\documentstyle[12pt,epsf,epsfig,amsmath]{article}
\newfont{\thiplo}{msbm10 scaled\magstep 2}
\newfont{\gothic}{eufb10 scaled\magstep 2}
\newfont{\unc}{eurb10} 
\newskip\humongous \humongous=0pt plus 1000pt minus 1000pt
\def\caja{\mathsurround=0pt}
\def\eqalign#1{\,\vcenter{\openup1\jot \caja
        \ialign{\strut \hfil$\displaystyle{##}$&$
        \displaystyle{{}##}$\hfil\crcr#1\crcr}}\,}
\newif\ifdtup

\def\eqright #1\cr{\noalign{\hfill$\displaystyle{{}#1}$}}
\def\eqleft #1\cr{\noalign{\noindent$\displaystyle{{}#1}$\hfill}}

\def\oldreffmt#1{\rlap{[#1]} \hbox to 2\parindent{}}

\def\figfmt#1{\rlap{Figure {#1}} \hbox to 1in{}}

\def\sectioneq{\def\theequation{\thesection.\arabic{equation}}{\let
\holdsection=\section\def\section{\setcounter{equation}{0}\holdsection}}}%

\newcounter{holdequation}

\def\auto{\eqno(\refstepcounter{equation}\theequation)}
\def\begineq #1\endeq{$$ \refstepcounter{equation}\eqalign{#1}\eqno
	(\theequation) $$}
\def\contlimit{\,{\hbox{$\longrightarrow$}\kern-1.8em\lower1ex
\hbox{${\scriptstyle (a\rightarrow0)}$}}\,}
\def\centeron#1#2{{\setbox0=\hbox{#1}\setbox1=\hbox{#2}\ifdim
\wd1>\wd0\kern.5\wd1\kern-.5\wd0\fi
\copy0\kern-.5\wd0\kern-.5\wd1\copy1\ifdim\wd0>\wd1
\kern.5\wd0\kern-.5\wd1\fi}}
\def\centerover#1#2{\centeron{#1}{\setbox0=\hbox{#1}\setbox
1=\hbox{#2}\raise\ht0\hbox{\raise\dp1\hbox{\copy1}}}}
\def\centerunder#1#2{\centeron{#1}{\setbox0=\hbox{#1}\setbox
1=\hbox{#2}\lower\dp0\hbox{\lower\ht1\hbox{\copy1}}}}
\def\lsim{\;\centeron{\raise.35ex\hbox{$<$}}{\lower.65ex\hbox
{$\sim$}}\;}
\def\gsim{\;\centeron{\raise.35ex\hbox{$>$}}{\lower.65ex\hbox
{$\sim$}}\;}
\def\st#1{\centeron{$#1$}{$/$}}

\def\super#1{\ifmmode \hbox{\textsuper{#1}}\else\textsuper{#1}\fi}
\def\textsuper#1{\newcount\holdspacefactor\holdspacefactor=\spacefactor
$^{#1}$\spacefactor=\holdspacefactor}

\def\getcite#1,{\advance\citenumber by1
\def\getcitearg{#1}\def\lastarg{@}
\ifnum\citenumber=1
\ref{#1}\let\next=\getcite\else\ifx\getcitearg\lastarg\let\next=\relax
\else ,\ref{#1}\let\next=\getcite\fi\fi\next}

\def\pom{{\rm P\kern -0.53em\llap I\,}}
\def\spom{{\rm P\kern -0.36em\llap \small I\,}}
\def\sspom{{\rm P\kern -0.33em\llap \footnotesize I\,}}

\relax
\def\contlimit{\,{\hbox{$\longrightarrow$}\kern-1.8em\lower1ex
\hbox{${\scriptstyle (a\rightarrow0)}$}}\,}
\def\upon #1/#2 {{\textstyle{#1\over #2}}}
\relax
\renewcommand{\thefootnote}{\fnsymbol{footnote}} 

\def\mainhead#1{\setcounter{equation}{0}\addtocounter{section}{1}
  \vbox{\begin{center}\large\bf #1\end{center}}\nobreak\par}
\sectioneq
\def\subhead#1{\bigskip\vbox{\noindent\bf #1}\nobreak\par}

\def\til#1{\centeron{\hbox{$#1$}}{\lower 2ex\hbox{$\char'176$}}}
\def\tild#1{\centeron{\hbox{$\,#1$}}{\lower 2.5ex\hbox{$\char'176$}}}
\def\sumtil{\centeron{\hbox{$\displaystyle\sum$}}{\lower
-1.5ex\hbox{$\widetilde{\phantom{xx}}$}}}

\newcommand{\bit}{\begin{itemize}}
\newcommand{\eit}{\end{itemize}}

\newcommand{\beq}{\begin{equation}}
\newcommand{\eeq}{\end{equation}}
\newcommand{\beqa}{\begin{eqnarray}}
\newcommand{\eeqa}{\end{eqnarray}}

\begin{document} 

\begin{titlepage} 

\rightline{\vbox{\halign{&#\hfil\cr
&\today\cr}}} 
\vspace{0.25in} 

\begin{center} 
  
{\large\bf THE PHYSICS OF A SEXTET QUARK SECTOR}\footnote{Work 
supported by the U.S.
Department of Energy under Contract
W-31-109-ENG-38} 

\medskip

Alan. R. White\footnote{arw@hep.anl.gov }

\vskip 0.6cm

\centerline{Argonne National Laboratory}
\centerline{9700 South Cass, Il 60439, USA.}
\vspace{0.5cm}

\end{center}

\begin{abstract} 

Electroweak symmetry breaking may be a consequence of
color sextet quark chiral symmetry breaking. 
A special solution of QCD is involved, with 
a high-energy S-Matrix that can be constructed ``semi-perturbatively'' 
via the chiral anomaly and reggeon diagrams. An infra-red fixed point 
and color superconductivity are crucial components of the 
construction.
Infinite momentum physical states contain both quarks and  
a universal ``anomalous wee gluon'' component, and the spectrum is more limited
than is required by confinement and chiral symmetry breaking. 
The pomeron is approximately
a regge pole and the Critical Pomeron describes asymptotic 
cross-sections. 

The strong coupling of the pomeron to the electroweak sector
could produce large $x$ and $Q^2$ events at HERA, and   
vector boson pairs at Fermilab. Further evidence for the sextet sector  
at Fermilab would be a large $E_T$ jet excess, due in part to 
the non-evolution of ${\alpha}_s$, and other
phenomena related to the possibility that top quark production is due
to the $\eta_6$.

The sextet proton and neutron are the only new baryonic states.
Sextet states dominate high energy hadronic cross-sections and 
stable sextet neutrons could produce both dark matter 
and ultra high energy cosmic rays. The
cosmic ray spectrum knee suggests the effective sextet threshold 
is between Fermilab and LHC energies, with
large cross-section effects expected at the LHC. Jet and vector
boson cross-sections will be very 
much larger than expected, and sextet baryons should also be produced. 
Double pomeron produced states could provide definitive evidence for the 
existence of the sextet sector in the initial low luminosity 
running.

\end{abstract}

\renewcommand{\thefootnote}{\arabic{footnote}} \end{titlepage}

\mainhead{1. INTRODUCTION}

The initial pursuit\cite{arw77,arw78}, nearly thirty years ago, of a 
particular solution of supercritical Reggeon Field Theory (RFT) has led us
to first associate the Critical Pomeron\cite{cri} with a special high-energy 
S-Matrix solution of QCD, then to connect this QCD solution to a very particular 
form of electroweak symmetry breaking\cite{wm,bww}. If this is the 
symmetry breaking and solution of QCD chosen by nature then, as 
outlined in \cite{arw04}, \cite{arw041} and \cite{mga}, 
we anticipate that there is a major change in the strong 
interaction above the electroweak scale. A new color sextet 
sector appears, with electroweak scale masses, that at high enough energies
should become responsible for the major part of the total cross-section. The
existence of this sector offers a natural explanation for the dominance 
of dark matter and
in fact, an interaction change of just this kind 
could be responsible for the apparent ``knee'' in the cosmic ray spectrum 
that occurs just above the Tevatron energy.  Other 
cosmic ray phenomena, that occur above the knee energy, 
also appear to be clear evidence for the same interaction change. 
That the knee is associated with the 
effective energy threshold for the sextet sector would be natural if 
inclusive pomeron exchange has to be involved when 
sextet states are produced, with large cross-section, from initial
triplet states.

We should emphasize that 
it could be (even though we consider it unlikely)
that the knee is not associated with sextet physics.  
If it is, however, then large cross-section effects have to appear very rapidly as 
the energy increases and they should be apparent  
at the LHC, with  dramatic and exciting physics involved.
In particular, jet cross-sections
and electroweak vector boson cross-sections will be overwhelmingly large,
with pomeron exchange cross-sections containing the most
distinctive signals.
Some indication of this physics could be observable at the Tevatron,
or even at HERA. Hints of what is to come, that may already have been seen, could be
the large $E_T$ jet excess at the Tevatron and large $x$ and $Q^2$ events at HERA.

We will use $QCD_S$ to denote\footnote{The suffix can be thought of as denoting 
``special'', or ``sextet'', or ``saturated'' - the asymptotic freedom constraint is
``saturated''. The ``special'' nature of the S-Matrix will become evident.} 
the S-Matrix solution of 
QCD with six color triplet and two color sextet quarks that we will describe.
Within $QCD_S$, sextet chiral symmetry breaking gives a triplet 
of ``sextet pions'' ($\Pi^{\pm}$, $\Pi^0$) and also, at first sight,
a ``higgs-like'' particle - the $\eta_6$. When the electroweak sector of 
the Standard Model is added, the ``sextet higgs mechanism'' takes place. 
By ``eating'' the $\Pi$'s, the $W^{\pm}$ and $Z^0$ aquire masses
that are a manifestation of the QCD sextet chiral scale. 
Thus, electroweak symmetry breaking is accomplished without any 
new interaction being added to the established 
$SU(3)\otimes SU(2)\otimes U(1)$ 
gauge interactions of the Standard Model. (We will only briefly
discuss how an $SU(2)\otimes U(1)$ 
anomaly is avoided since a special 
unification, requiring additional discussion, 
is most likely involved\cite{kw,arw05}.)
Furthermore, the electroweak scale is a new QCD scale and 
the symmetry breaking is connected
with the major change in the strong interaction discussed above. 

We obtain the $QCD_S$ high-energy S-Matrix via the powerful technology of reggeon 
diagrams\cite{fkl}-\cite{arw93}. While this S-Matrix
has some important distinctive properties relative to conventional QCD, we believe
that it is consistent with all the (experimentally established) 
properties of QCD below the electroweak scale.
A crucial distinctive property is, however,
the limitation on the spectrum of states compared
to what would be anticipated from just color confinement and chiral
symmetry breaking. As we will describe, the S-Matrix is constructed as a
reggeon critical phenomenon by starting within a ``color 
superconducting'' phase of $QCD_S$
(in which SU(3) color is broken to SU(2)). This starting 
point introduces reggeon ``anomaly interactions'' that are a key physical ingredient. 
These interactions produce divergences which have the consequence that
the physical states of $QCD_S$ are directly related to the chiral
Goldstone bosons of the superconducting theory. This implies that, 
in the normal hadronic sector, both glueballs and quark 
resonances (such as the $\rho$) are directly excluded as asymptotic states.

In general, because of the central role played by anomaly couplings, only a very 
limited sub-set of the gluon degrees of freedom contribute to the 
$QCD_S$ high-energy S-Matrix.
(Presumably, there is a corresponding limitation in the finite energy S-Matrix.)
As a result, there is no BFKL pomeron and no odderon. 
We are not aware of any experimental evidence against this. Rather,
strong experimental evidence that this should be the case is, surely, 
provided by the (almost total) absence of glueballs in the resonance
spectrum, the absence of the odderon\cite{ce} in experiments at HERA,
and the lack\cite{ce} of any definitive evidence for the BFKL pomeron.

The spectrum of states involving sextet quarks is, perhaps, the deepest 
consequence of the construction of the spectrum of $QCD_S$ via the anomaly
interactions of the superconducting phase.
Because there are no chiral symmetries linking the sextet and triplet 
quarks, there are no hybrid sextet/triplet states and the only new sextet states,
in addition to the sextet pions and the $\eta_6$, 
are a ``sextet proton'' (the $P_6$) 
and the ``sextet neutron'' (the $N_6$), both of which will have electroweak
scale masses that could be, we will suggest,
as low as $500 ~GeV$. Because of the conservation of 
sextet quark baryon number, one of the sextet nucleons must be absolutely stable.
The absence of sextet current quark masses (that is necessary for 
electroweak symmetry breaking) implies that the stable state
must be the $N_6$. Therefore, at the ultra-high energies relevant for the early
universe, the production of stable, neutral, sextet neutrons 
will dominate over the production of stable, charged, triplet protons. 
Consequently, we have a very natural explanation for the dominance of dark matter
- formed (as nuclei, clumps, etc.) from sextet neutrons. Furthermore,  
because neutral, massive, $N_6$'s will avoid the GZK cut-off they could also 
be the mysterious, ultra-high energy, cosmic rays. 
Since they would simply be very high energy dark matter their origin would, 
presumably, be much less of a mystery than is currently believed.
 
The purpose of this paper is two-fold. Firstly, we want to lay out 
what we believe we know about $QCD_S$ and why we think  we know it. 
Secondly, we will outline experimental consequences that we expect
from the combination of $QCD_S$ with the sextet higgs mechanism. 
While we have discussed high-energy phenomena that 
$QCD_S$ could produce in the past\cite{arw94,arw97}, 
we did not have the detailed understanding that we now have 
of how the chiral anomaly produces high-energy states and amplitudes. As a result,
the emphasis in this paper will be very different to that of our earlier 
papers\footnote{Most notably we 
believe our discussion of instanton interactions and dynamical
masses is irrelevant in the, infinite momentum, 
S-Matrix formulation within which we now 
work.}. Particularly important will be 
the strong coupling of the pomeron to sextet states that follows
from the anomaly pole method that we develop to estimate cross-sections
for hard diffraction. Predictions can then be made for soft diffraction
by combining the hard diffractive estimates with pomeron regge theory.

If pomeron exchange amplitudes are large, then
cut-pomeron amplitudes should also be large.
This leads to the prediction of large inclusive cross-sections
for sextet states (multiple $W$'s and $Z$'s, in particular)
across most of the rapidity axis, that we 
expect to be the major characteristic of $QCD_S$ physics above the
electroweak scale. 
There will also be ``non-diffractive'' consequences of the sextet sector 
that we will discuss. At current energies, these include 
the non-evolution of ${\alpha}_s$ above the electroweak scale and 
the possibility that top production is due to the $\eta_6$.

While our papers have suggested a link for some time, we believe that
the arguments presented in this paper make it clear
that the sextet higgs mechanism is inextricably tied to the  
pomeron and infinite momentum hadron states that have 
emerged from our work on the regge limit of $QCD_S$.
If this were not the case then, as we discuss again below, 
the $\eta_6$ would be\cite{cllr} a light axion-like state
that is not seen experimentally and the sextet Higg's mechanism would be
ruled out as a realistic possibility. 
We will emphasize (see Appendix C in particular) the 
likelihood that the left-handed vector nature of the electroweak sector 
of the Standard Model plays an important role with respect to 
inducing the special $QCD_S$ S-Matrix. 

That the high-energy behavior can be constructed by starting from 
the reggeon diagrams of $CSQCD_S$ (``color superconducting'' $QCD_S$) is the 
most crucial property of $QCD_S$. The original motivation for this starting point
came from a correspondence between supercritical pomeron 
RFT and $CSQCD_S$. This correspondence is referred to indirectly above and the 
arguments for it 
are described in Appendix C - where we outline our full multi-regge
program. There is, however, an important
technical reason why the construction can be carried through.
$CSQCD_S$ can be obtained from $QCD_S$ by introducing 
an asymptotically free scalar field. (This would not be possible if the number of 
quarks was any fewer!) Asymptotic freedom implies that 
this field can be smoothly decoupled in the ultra-violet region.
In the infra-red region the only remnant of 
the decoupling is the ``anomaly contribution'' of unphysical 
longitudinal wee gluons that provides the all important 
mechanism that produces a non-perturbative spectrum out of perturbative 
diagrams, as we discuss next. 

The presence of  massive gluons in $CSQCD_S$ produces\cite{arw03}-\cite{arw021}
triangle diagram anomalies
in the effective vertices of reggeon diagrams. The contribution of the anomalies
is (not surprisingly) strongly dependendent on 
ultra-violet and infra-red cut-offs and so
different ``solutions'' of the theory can be obtained, depending on how 
such cut-offs are handled. The essential part of our reggeon diagram 
analysis (described in Appendix C) is 
the initial imposition of a transverse momentum cut-off.
This cut-off produces a violation of gauge invariance Ward identities for 
the anomaly vertices. As a result, infra-red transverse momentum
divergences appear which, when the quarks involved are massless,
produce residue amplitudes that contain ``anomaly poles'' resulting from
infra-red chirality transitions. (An anomaly pole is produced, 
in part, by a pinching of massless particle
and antiparticle poles in the same zero momentum propagator and so, automatically,
involves a chirality transition.) The identification of anomaly poles  
as chiral Goldstone boson particle poles provides a crucial
mechanism for a bound-state, confining and chiral symmetry breaking, spectrum 
(and the appropriate amplitudes) to appear 
via the contribution of anomalies and transverse momentum infra-red divergences.

Because our starting point is perturbative reggeon 
diagrams, the final amplitudes we obtain are not very far from perturbation theory. 
Very complicated multiparticle diagrams are involved and
there is an elaborate phenomenon of cut-off dependent
infra-red divergences coupled to triangle diagram anomalies. Nevertheless, 
both confinement 
and chiral symmetry breaking have a diagrammatic description. The primary
reason that the physics involved stays perturbative 
is the existence of an infra-red fixed point due to the large 
number of quarks. By preventing the infra-red growth of $\alpha_s$,
the infra-red fixed point also
produces infra-red scaling properties for reggeon interaction kernels
that are vital for the emergence of physical scattering amplitudes
via infra-red divergences. 

Because both the infra-red fixed-point and 
infra-red effects of the chiral anomaly are crucial, it is essential that 
all quarks, including the sextet sector, are massless (initially). 
In this paper, we will discuss only how 
vector boson masses are generated by the sextet higgs mechanism. 
This mass generation is responsible for raising all effects of the sextet
sector to momenta at or above the electroweak scale.  
This is necessary, of course, to obtain normal QCD at low energies since, within 
massless $QCD_S$, $\alpha_s$ remains less than it's fixed point value 
($\approx 1/34$). The familiar, larger, value of $\alpha_s$ is obtained
only after an effective low-energy theory is obtained by integrating
out the sextet sector. In addition, to be physically 
applicable, triplet quark effective masses must also be 
added to the S-Matrix of $QCD_S$. 
We will not discuss the origin of effective quark masses. 
This is related to the unification of $QCD_S$ and the electroweak sector
of the Standard Model in a larger theory\cite{kw} and we will discuss this
in forthcoming papers\cite{arw05}. Fortunately, for most of
our discussion in this paper, only vector boson masses are relevant and so the issue
can be avoided. 

The transition from $CSQCD_S$ to $QCD_S$ is to be 
achieved via supercritical RFT and the 
phase transition appearance of the Critical Pomeron\cite{cri}.
If this can be carried through in full detail, the regge behavior 
of $QCD_S$, together with the infinite momentum hadron states, 
will be obtained from the much simpler infra-red divergence and
anomaly structure that appears in $CSQCD_S$. In particular, 
within (infinite momentum) $QCD_S$,  
confinement and chiral symmetry breaking will be understood
as resulting from dynamical infra-red chirality transitions produced by wee gluon
interaction anomalies. However, as we already emphasized above,
the spectrum of physical states will be significantly 
limited compared to that normally anticipated. Only 
states that correspond to Goldstone bosons in $CSQCD_S$ will be present. 
These states (and only these) have, as a consequence of the flavor anomaly,
a wee gluon content that produces the infra-red divergent amplitudes
giving the, eventual, physical amplitudes.
Pions and nucleons are included amongst such states, but flavor singlet Goldstone
bosons, unstable resonances and glueballs, 
are all excluded (as asymptotic states). As we have already emphasized,
the absence of hybrid sextet/triplet baryons in $QCD_S$
is crucial for the stability of the $N_6$ and, hence, for our 
explanation of the origin of dark matter. 

In conventional QCD, the only non-conserved axial U(1) charge 
is that coupling to the short-distance topological 
anomaly. If this were the case in $QCD_S$, the 
U(1) symmetry (essentially the sextet symmetry)
associated with the $\eta_6$ would be unbroken. In addition to being the
analog of the usual higgs scalar, the $\eta_6$ would be
a light axion of the kind that is ruled out experimentally. 
In our solution of $QCD_S$ the anomaly vertices that are initially obtained 
by imposing a cut-off, and that are responsible for
the dynamical ``wee gluon'' component of infinite momentum physical states,
break both the sextet and triplet U(1) symmetries and so there is no light axion. 
Consequently, although the $\eta_6$ appears, at first, to be a Goldstone 
boson of the  appropriate kind to appear as a physical state, there is
a multigluon regge exchange (a daughter of the pomeron) that 
mixes with it. This mixing, presumably, generates a large (electroweak scale) mass
for the $\eta_6$. The $\eta_6$ also couples to the triplet sector via the gluon 
intermediate state and if it has an electroweak scale mass 
the mixing will be primarily with the $t\bar{t}$ state. Consequently,
as we will briefly discuss, the $\eta_6$ could actually be responsible for top 
production at the Tevatron. 

Clearly, that the infra-red anomaly contributions persist,
via longitudinal wee gluons, after the removal
of the large $k_{\perp}$ cut-off and the restoration of SU(3) gauge symmetry,
is a central element of our construction of $QCD_S$.
It is well-known that the contribution of longitudinal wee gluons is an, 
a priori unresolved, ambiguity in the infinite
momentum quantization of $QCD$ which is closely related to
the well-known Gribov problem\cite{gm} and, therefore, to
the choice of vacuum at finite momentum. In effect, therefore, 
we resolve this ambiguity 
in $QCD_S$ by constructing the high-energy behavior via $CSQCD_S$.

It is well-known that 
both $s$-channel and $t$-channel unitarity (via reggeon unitarity) 
impose very strong constraints on the behavior of a theory
in  multi-regge limits. 
A solution of QCD in all such limits necessarily determines how
unitarity, the physical spectrum, and the validity of perturbation theory
all coexist. Obtaining such a solution is, therefore, likely to be almost
as difficult as solving the full theory. As we have said, 
according to our arguments the  
multi-regge limits of $QCD_S$ are described by 
the Critical Pomeron\cite{cri}, which is known
to satisfy all unitarity requirements. In addition,
we are able to give a diagrammatic construction
in which the connection between perturbation theory, the pomeron, 
and the physical bound state spectrum is clear. 
If everything goes through as we describe, it will be apparent that
$QCD_S$ is a  version of QCD that, perhaps uniquely, 
satisfies all general requirements.

On the lattice, it would be very difficult, if not impossible, 
to introduce the co-ordinated infra-red dynamical fluctuations of 
longitudinal wee gluons and the Dirac sea that provide the anomaly couplings, 
and consequent infra-red divergences that lead to the
infinite momentum $QCD_S$ S-Matrix. Not surprisingly, perhaps, within the lattice 
framework, the infra-red fixed-point that we have discussed 
is generally believed\cite{irfp} to be associated with a non-confining 
continuum theory and there is no sign of the confining 
``anomaly-driven'' S-Matrix that we have discovered. 

Similarly, there are general arguments\cite{asv} 
that the infra-red fixed-point in $QCD_S$ will produce
Green's functions that are conformally invariant in the infra-red region 
and do not contain any particle-like physical states. In fact this is,
essentially, the infra-red scaling property of reggeon kernels which
plays a central role in our analysis. Clearly, it is a subtle challenge to find
the asymptotic states and S-Matrix amplitudes that emerge from our construction.
They do not appear within quark or gluon Green's functions. Indeed,
their existence depends crucially on S-Matrix fermion anomalies that also
do not appear in off-shell Green's functions. For the reasons that we 
elaborate on in Appendix C, 
it may be necessary to consider the (on-shell) scattering of vector bosons
with left-handed couplings to quarks, to see the emergence of the
desired amplitudes.  

Section 2 is devoted to the high-energy solution of $CSQCD_S$. Our essential aim
is to focus on the physics that underlies this solution. To this end, we 
keep the discussion at a fairly broad level and 
supplement it with Appendices. In Appendix A
we describe the formal infra-red and ultra-violet $\beta$-function properties that
are needed to connect $CSQCD_S$ to $QCD_S$. We do not use (in Section 2)
the full multi-regge theory that is necessary to actually derive the solution
that we describe.
Instead, we use the anomaly-pole vertex method developed in \cite{arw02}. 
Needed properties of the triangle anomaly and the contribution of 
the anomaly pole are described in Appendix B.
In Appendix C we outline our full multi-regge program and,
as part of our description, we include (very briefly) the historical development
which led to our association of the Critical Pomeron with $QCD_S$. Since
many of the 
details of how the transition from $CSQCD_S$ to $QCD_S$ is described by the
Critical Pomeron have still to be worked out we give, in Section 3, only a 
brief outline of the features that are relevant for the purposes of this paper.

We begin the process of combining the electroweak sector with $QCD_S$
in Section 4. In particular, we show how masses for the electroweak bosons
are generated by anomaly interactions that result from the presence of 
wee gluons in infinite momentum physical states. This is the infinite
momentum S-Matrix analog of vacuum generation of the masses. Most importantly, we see
that the mass scale is determined by the coupling of wee gluons to sextet 
quarks. We can then carry the knowledge of this coupling over to the coupling of
the pomeron to sextet quark states and, in particular, to multiple 
$Z^0$ and $W^{\pm}$ states.

In Sections 5 and 6 we discuss processes that might 
be seen (or may have already been seen) at 
current accelerators and could provide evidence for the existence of the sextet 
sector. In Section 5 we discuss diffractive deep-inelastic scattering and suggest 
that the most dramatic 
large $x$ and $Q^2$ event presented\cite{ZEUS} by ZEUS, may have been diffractive
production of a $Z^0$.
Sextet quark physics that might be seen at the Tevatron is the focus of 
Section 6. We describe a number of small cross-section effects that might
be seen in diffractive, and diffractive related, processes involving 
$W^{\pm}$ and $Z^0$ vector bosons.  We also suggest 
that $t\bar{t}$ production could  
originate from the $\eta_6$, even though this process
can be understood perturbatively. The interpretation of the top quark mass 
would be different and non-perturbative decay modes should also be seen, at some level. 
A jet excess at large $E_T$  would provide 
supporting evidence for this proposal since, in this case,
$\alpha_s$ evolution should stop at $E_T \sim ~m_{top}$.

If the sextet sector exists, the LHC will most probably 
be the discovery machine. Sections 7 and 8 are devoted to explaining why we expect
that dramatic effects will be seen. In Section 7 we discuss 
dark matter and the cosmic ray phenomena that tell us that the sextet sector
could appear at the LHC.
We discuss the specifics of what we expect to see at the LHC in Section 8. 
While jet cross-sections and cross-sections
for multiple vector boson production will be orders of magnitude  
larger than expected,
the double pomeron cross-section for electroweak vector boson pairs,
which can be studied (in part) during the initial low luminosity running, 
may well be the most definitive early evidence that is seen.
There could be spectacular events 
in which the forward protons are tagged and only
large $E_T$ leptons are seen in the central detector. 
``Dark matter'', in the form of sextet neutron/antineutron pairs, 
should have significant inclusive cross-sections and may even 
be produced in double pomeron exchange. If so, this would be really
dramatic!

\newpage

\mainhead{2. COLOR SUPERCONDUCTING $QCD_S$}

\subhead{2.1 Symmetry Breaking, Reggeization, and Infinite Momentum States} 

The breaking of the SU(3) color symmetry of $QCD_S$ to SU(2)
can be achieved with an asymptotically free, complex color triplet, scalar field.
(This is discussed in more detail in Appendix A.)
As a consequence of the symmetry breaking, $CSQCD_S$ 
contains an SU(2) triplet of massless gluons,
plus two SU(2) doublets (with mass $\frac{2}{\sqrt{3}}M$) 
and one singlet (with mass $M$) of massive gluons.  
Each SU(3) triplet quark gives one complex SU(2) doublet and one singlet quark.
Each SU(3) sextet quark gives one complex SU(2)
triplet, one complex doublet, and one singlet quark. Reflecting the absence of any
corresponding chiral symmetry in $QCD_S$, there is obviously
no chiral symmetry relating the, sextet originating, SU(2) complex triplet to either  
of the SU(3) triplet originating representations.

All quarks and gluons (massive or not) are reggeized,
but only the SU(2) singlets have infra-red finite regge trajectory functions. 
The infra-red scaling behavior of various
``transverse momentum kernels'' that describe the interactions 
of reggeized quarks and gluons will be an essential ingredient of the following 
analysis.
The scalar particle produced by the scalar field does not reggeize and
so at the non-leading power level $CSQCD_S$ is, presumably, a non-unitary 
theory - implying  that only the leading
high-energy behavior of $QCD_S$ can be constructed via $CSQCD_S$.

The status of the full program that we have developed to construct
the multi-regge behavior of $CSQCD_S$ is outlined in 
Appendix C. We believe that this program, as it is now formulated, 
would give the high-energy behavior of $QCD_S$ unambiguously
if pursued to completion. However, we can arrive much more 
simply at the physics involved if we utilise
the approach that we developed in \cite{arw02}. In that paper, we introduced 
a procedure that was designed to bypass the multi-regge construction and 
instead obtain directly the $CSQCD_S$ scattering amplitudes
for infinite momentum states. This procedure is what we now describe.

We note, before we start, that if high-energy states and amplitudes  
can be derived from perturbative reggeon diagrams, then the 
parton model must have a broad validity, well
beyond leading-twist perturbation theory. For this to be the case,
the ``naive'' validity of the 
perturbative vacuum at infinite momentum must hold for deeper reasons. This  
can be so if infinite momentum
states have a universal ``wee parton'' component that carries the  
finite momentum ``properties of the vacuum''. (Note that, although it is not 
directly relevant at this point, regge pole factorization
properties for the pomeron are, most probably, a pre-requisite for
a universal wee parton distribution in hadrons.) As we shall see, it is indeed
a universal wee gluon component of infinite momentum states that determines our  
solution of $CSQCD_S$. 

\subhead{2.2 Pion Anomaly Pole Vertices} 

The primary assumption 
in \cite{arw02} was that the wee gluon properties of the physical states 
could be obtained\footnote{We expect this 
to be an outcome of the full multi-regge program 
and we emphasized in \cite{arw02} that if 
the assumptions made appeared to be ad-hoc this was, in large part, 
because of our deliberate efforts to avoid the full complexity of multi-regge theory.}
from properties of the chiral anomaly and ``anomaly
pole'' vertices. It is well-known that an anomaly pole appears, in particular 
kinematic circumstances, in a three-point vertex of 
local currents when the triangle diagram anomaly is present and when the fermions
producing the anomaly are massless. When
the vertex involves an axial current that is the generator of a chiral symmetry
that is spontaneously-broken, this pole can be directly interpreted 
as a Goldstone boson particle pole associated with the symmetry breaking. 

The invariant functions of a triangle 
diagram depend on the invariants $k_1^2$, $k_2^2$, and $q^2$, 
where, as shown in Fig.~1(a), $k_1, k_2$ and $q$ are the momenta 
entering at each of the vertices.    
\begin{center}
\epsfxsize=5.5in
\epsffile{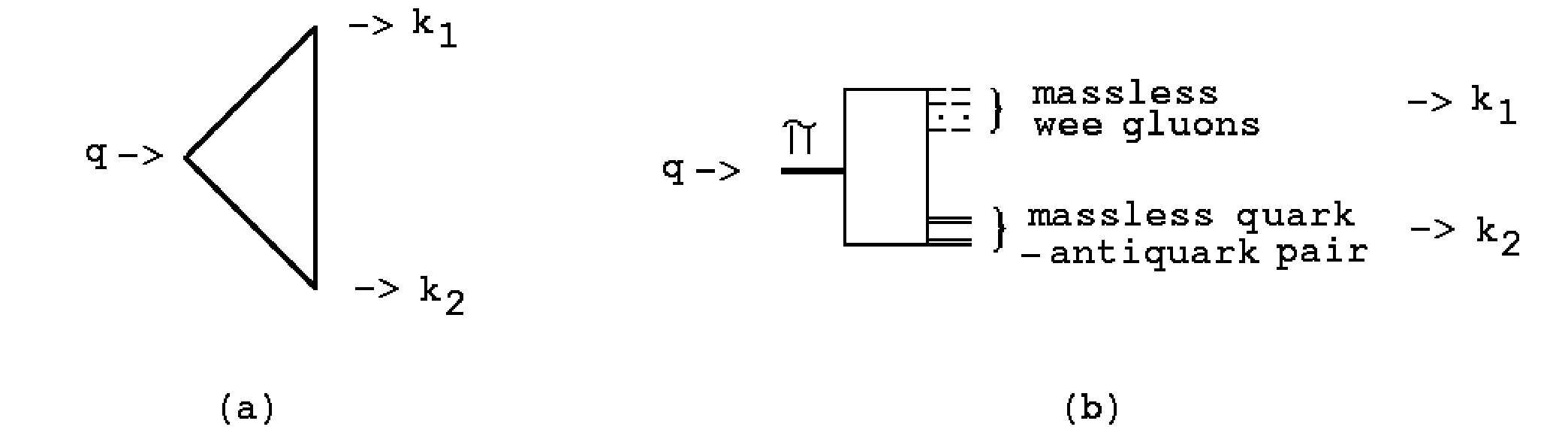}

Fig.~1 (a) Triangle momenta (b) How wee gluons give a pion anomaly pole.
\end{center}
The pole is present when either 
$$
k_1^2~=~k_2^2~=~0, ~~~q^2 ~\to ~0~, ~~~~~~~ \hbox{or} 
~~~~~~~ k_1~=~0, ~~~k_2^2~=~q^2 ~\to ~0
\auto\label{akin}
$$
and the residue is determined by the anomaly. 
(Some details of how the pole is generated are given in Appendix B.)

We anticipate that the occurrence of anomaly poles in
regge limit effective triangle diagrams will be a widespread phenomenon in the
full multi-regge analysis of $CSQCD_S$. 
They appear whenever components of the relevant currents
(not the full currents) appear as effective vertices in a triangle diagram.
Poles associated with a flavor anomaly current component are Goldstone
boson particle poles that are, in effect, dynamically generated.
As illustrated in Fig.~1(b), the kinematics
producing a Goldstone (pion) pole can occur when a set of wee gluons produces a 
divergence at $k_1^2=0$ and couples via an effective triangle diagram to 
a quark-antiquark pair that carries a light-like momentum $k_2$. 
In this Section,
we will refer to all quark/antiquark (triplet or sextet) Goldstone bosons in
$CSQCD_S$ as ``pions'' and, when we need to, will refer to quark/quark or 
antiquark/antiquark Goldstone bosons\cite{kog} 
as ``nucleons''. Effectively, all of our 
discussion of pions will also apply to nucleons, even though we will not 
usually say so explicitly. Poles
associated with the U(1) anomaly do not contribute as particle poles but instead 
contribute as $\delta$-functions that conserve wee gluon transverse momenta
during an interaction.

The underlying calculations needed 
to demonstrate the existence of the initial anomaly pole vertices we require
can now be found in \cite{arw03}. 
In calculations carried out  after \cite{arw02} was published,
we showed explicitly how, 
in the scattering of electroweak vector bosons, 
effective vertices containing a 
triangle diagram are generated by the contraction of larger loop diagrams, 
in the channel with pion exchange quantum numbers. 
As a result, we can anticipate that in general
scattering processes involving an infinite momentum vector boson, if a transverse
momentum cut-off is imposed, a pion anomaly pole will indeed appear with the wee gluon
couplings we assumed to exist. This should be sufficient to show that 
a massless on-shell pion carrying light-cone momentum $k_+$ has a coupling
to wee gluons (carrying total light-cone momentum $k_-$, with $k_-/k_+ \to 0$)
given by the anomaly pole residue of a triangle diagram
that is generated as illustrated in Fig.~2. (The use of vector boson scattering
states is explained in Appendix C.)  
\begin{center}
\epsfxsize=4.8in
\epsffile{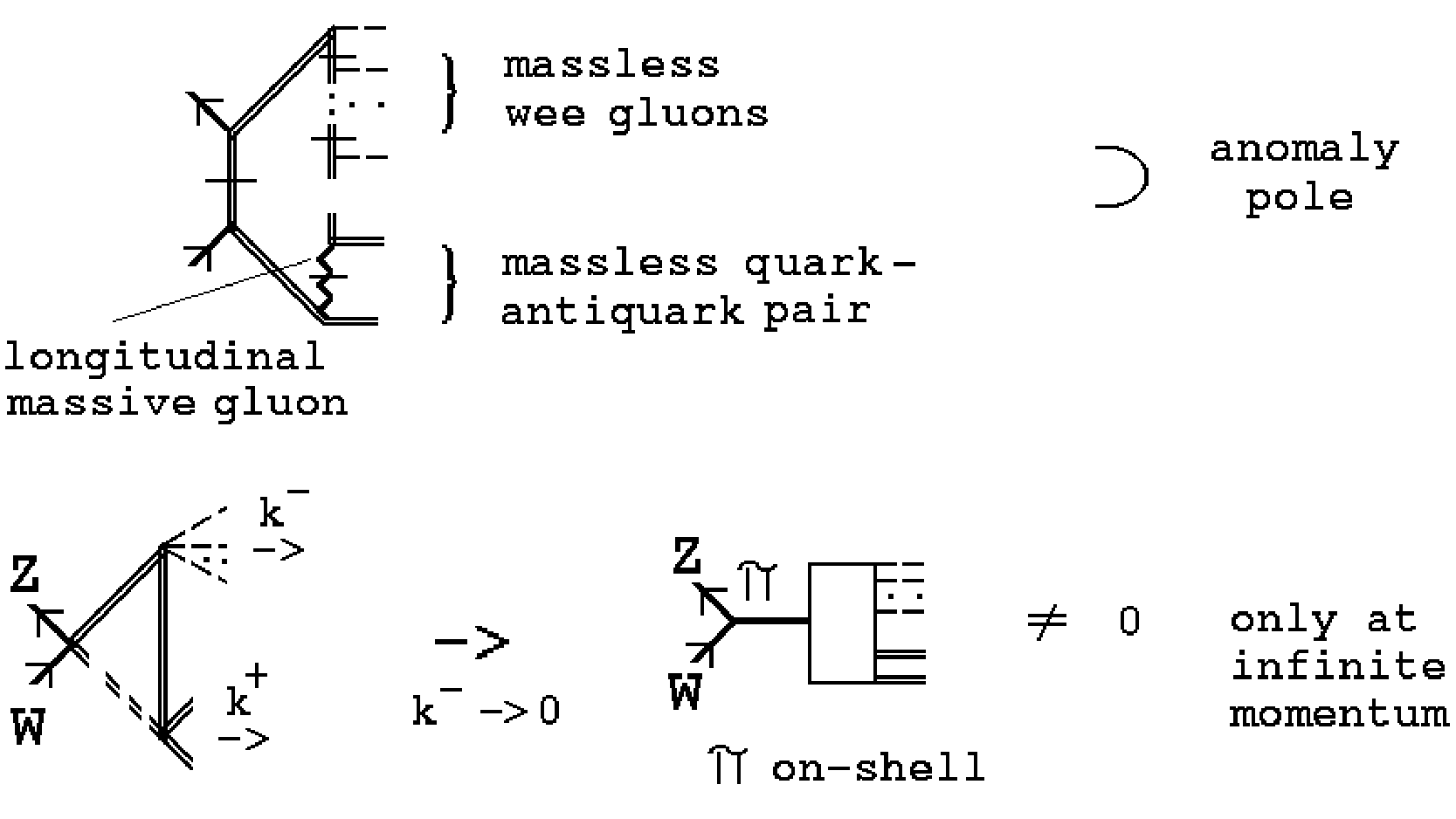}

Fig.~2 Anomaly pole generation in an effective triangle diagram.
(The hatched lines are on mass-shell.)
\end{center} 

The coupling shown in Fig.~2 involves a massless quark-antiquark pair 
that has a vector-like helicity
and any number of ``wee gluons'', that are also in a vector-like state.
The dashed line in the triangle diagram is a zero momentum quark propagator 
that, as discussed in Appendix B, 
generates the anomaly pole and also produces a chirality transition.
According to (\ref{imf}), in an ``infinite momentum'' 
frame reached via a boost $a_3(\zeta)$, the momentum dependence of the 
anomaly pole coupling is
$$
[~k_+ k_- \sinh{\zeta}~] 
\auto\label{ifc}
$$ 
which is finite when $k_- \to 0$, if $ k_- \cosh{\zeta}~$ is kept finite.
It is important that (as we will discuss further later)
it is the longitudinal component of the massive gluon that 
is responsible for the quark/antiquark vertex of the triangle diagram.

As we develop a complete dynamical picture in the following, 
we will introduce
a variety of anomaly pole effective vertices whose existence is a
natural extrapolation of existing vertices but, for which, the underlying
(very complicated) multi-regge calculations still need to be performed.

\subhead{2.3 Transverse Momentum Kernels and Infra-Red Divergences}

In \cite{arw02} we also argued that, because the anomaly pole is generated by
a light-cone internal momentum region within the triangle diagram, we could  
use transverse momentum diagrams to discuss wee gluon
interactions within the infinite momentum pion state. (Again, this should, 
straightforwardly, be the case in the multi-regge framework of Appendix C.)  
The coupling (\ref{ifc}) is defined at $k_{\perp}=0$, where $k_{\perp}$ is
the transverse momentum of the wee gluons. That it is non-zero is correlated  
with the fact that, for $k_{\perp} \neq 0$, the  
anomaly pole contribution to the effective triangle diagram
violates the wee gluon Ward identity (for reasons discussed in Appendix B).
A direct consequence is that infra-red divergences appear in the transverse 
momentum diagrams and dominate the physical
pion scattering amplitude. 
(In the multi-regge framework, a transverse momentum cut-off is initially 
responsible for the failure of Ward identities
that then leads to the occurrence of divergences and the correlated
appearance of anomaly pole couplings.)

To describe the infra-red divergences that occur, we must first describe
the infra-red properties of the transverse momentum kernels that are 
involved. These kernels are defined in more detail in Section IIIB of
\cite{arw02}, where a detailed review of elastic scattering
reggeon diagrams is also given.
We begin with the kernels $K^I_N(\underline{k},\underline{k}')$ 
that involve only the SU(2) triplet of massless gluons.
($I$ denotes SU(2) color.) When the  
color of the multigluon state is non-zero, infra-red divergences 
give (in a sense explained in \cite{arw02})
\newline \parbox{0.5in}{$~$}
\parbox{1.8in}{ 
\begin{center}
\epsfxsize=1.4in 
\epsffile{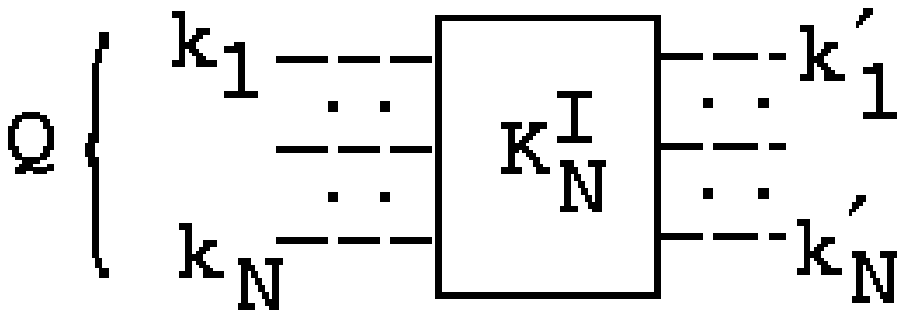}
\end{center}
}
\parbox{3.6in}{
$$ 
=~K^I_N(\underline{k},\underline{k}')~\to ~\infty~,~~~Q^2, I \neq ~0
~~~~~~~~~~~~~~~~
\auto\label{kin}
$$}
As a result,  
the sum of all gluon transverse momentum diagrams in any colored 
channel exponentiates to zero. 

When $I=0$ and $Q^2 ~\neq 0$, the kernels $ ~K^0_N(\underline{k},\underline{k}')$ 
are finite and have an important scaling property, as described in \cite{arw02}. 
As a result, there is no  
exponentiation of divergences in color zero gluon channels. However, 
the disappearance of all colored multigluon states is not 
confinement since gluon poles remain in the color zero diagrams.
Confinement is produced when the remaining $Q^2 = 0$ singularity in color zero
channels is absorbed into a ``condensate'', as we describe below. 

The most 
important contribution of the $K^0_N$ kernels comes when a color zero set of
massless gluons accompanies another SU(2) color
zero transverse momentum state, as can be the case
in states produced by the pion anomaly 
pole couplings. In Fig.~3 we show the kernel 
$K_R(\hat{k},\underline{k},\hat{k}',\underline{k}')$
describing the interactions of
massless gluons with the massive (SU(2) singlet) reggeized gluon and the kernel 
$~K_Q(\hat{k},\underline{k},\hat{k}',\underline{k}')$ 
describing the analagous interaction with
an SU(2) singlet quark-antiquark pair.
\begin{center}
\epsfxsize=4in
\epsffile{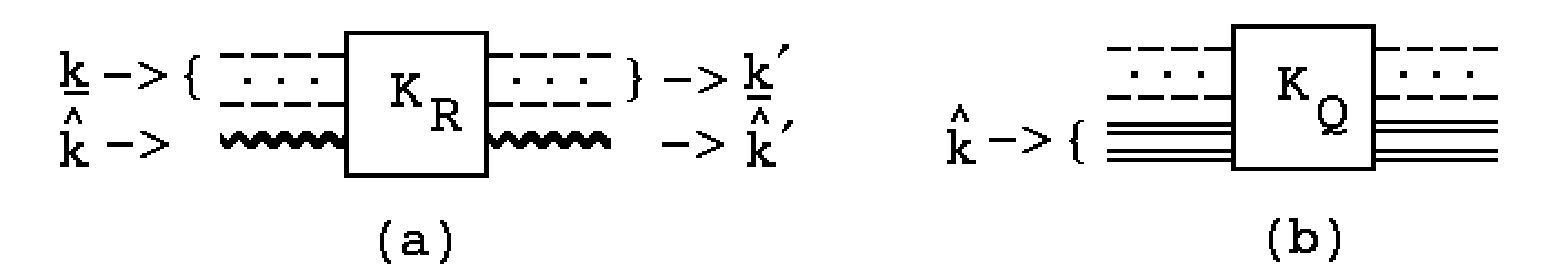}

Fig.~3 Kernels for massless gluon interactions with (a) a massive reggeized gluon
(b) a quark-antiquark pair.
\end{center}
Ward identities require that both $K_R$ and $K_Q$ vanish when either 
$\underline{k} \to 0$ with $\underline{k}'$ fixed 
or when $\underline{k}' \to 0$ with $\underline{k}$ fixed. 
But, because these kernels have a dimension of $[$momentum$]^2$ and 
additional non-zero mass and momentum scales (i.e. $M^2$
and $\hat{k}^2$) are present, 
we expect that these kernels neither vanish, nor have an infra-red scaling property,
when $\underline{k} \sim \underline{k}' \to 0$.
As a result, whenever the interactions of Fig.~3  exist, infra-red 
divergences again cause the sum of all diagrams to exponentiate to zero. 

However, as illustrated in Fig.~4(a), 
\begin{center}
\epsfxsize=5.5in
\epsffile{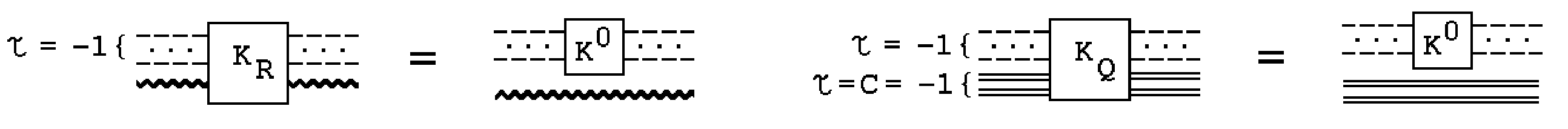}

(a)\hspace{2.7in}(b)

Fig.~4 Reggeon states without interaction kernels. 
\end{center}
because of helicity conservation in 
the massless quark and gluon sector, there is no transverse momentum
kernel describing the interaction of negative signature, color zero, 
massless gluons with the massive reggeized gluon. This is because 
a multigluon state containing an odd number of gluons and carrying SU(2) color 
zero necessarily has ``anomalous color charge parity'', i.e. the color charge
parity is necessarily positive and can not be equal to the negative signature. 
Similarly, as illustrated
in Fig.~4(b), for a massless quark-antiquark state
that carries negative signature, color zero, 
and normal color charge parity, there is also no interaction. 

Related to the lack of interactions, transverse momentum states of the kind 
shown in Fig.~4 will couple only 
through anomalies. As a result, there will be no exponentiation of divergences
in reggeon channels with these quantum numbers. Instead, the scaling property
of the massless gluon kernels leads to an overall divergence. 

\subhead{2.4 Pion Scattering Amplitudes Via Infra-Red Divergences}

In \cite{arw02} we considered feynman diagram contributions to
the particular transverse momentum diagram shown 
in Fig.~5, in which there are three wee gluons in each of the pion 
channels and also in the pomeron channel. 
\begin{center}
\epsfxsize=3.3in
\epsffile{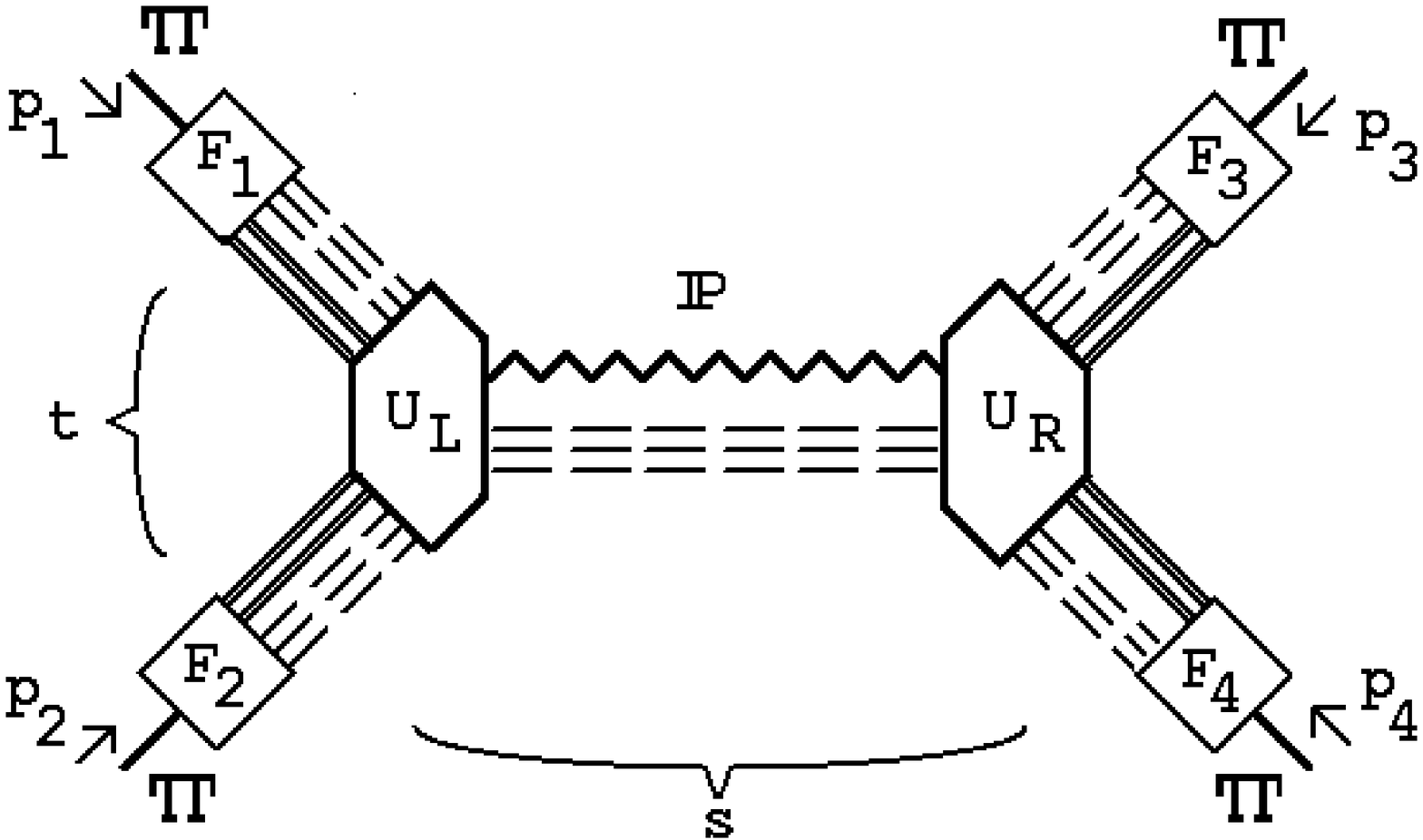}

Fig.~5 A transverse momentum diagram for pion scattering.
\end{center}
(The notation in Fig.~5 is the same as for Fig.~2.) 
Because of the foregoing discussion, this diagram is amongst the
simplest, describing pion scattering, that contain a transverse momentum
divergence that does not exponentiate to zero. In \cite{arw02}
we carried out a detailed infra-red analysis to extract the resulting amplitude. 

We will not reproduce the analysis of \cite{arw02} here
but, rather, will elaborate on features of the underlying physics that we did not
discuss in \cite{arw02}. For this purpose we need to describe, briefly,
the kinematics involved in the analysis. The kinematics were chosen 
so that each of the initial and final state pions
was in an infinite momentum frame, reached by an appropriate boost, such that
an anomaly pole residue corresponding to (\ref{ifc}) could give 
the contribution of each of the four external pion couplings $F_i$. 
To also produce internal triple-regge anomaly 
interactions, the wee gluons in the outgoing pions were associated with 
light-cones whose space direction is orthogonal to that of the incoming
wee gluon light-cones. We, therefore, introduced distinct Lorentz frames as follows. 
We calculated the left-hand part of Fig.~5 in a 
``left-hand finite momentum frame'' in which $p_1$ and $p_2$ have the  
form\footnote{The notation is straightforward in that ${k}^{1^+}$ is 
a vector with raised index component along the light-cone defined by the
positive $\{1\}$ - axis (and all other othogonal components are 
zero). Simiilarly ${q}^{1^-}$ is a vector with raised
index component along the light-cone defined by the
negative $\{1\}$ - axis. 
The same vectors can be labeled via lowered index components as usual.}
$$
\eqalign{p_1 ~&=~{k}^{1^+}~+~{q}^{1^-} ~~=~{k}_{1^-}~+~{q}_{1^+} \cr
&=({k \over \sqrt{2}},{k \over \sqrt{2}},0,0)~
+~({q \over \sqrt{2}},- {q \over \sqrt{2}},0,0 )}
\auto\label{kin1l}
$$ 
$$
\eqalign{p_2 ~&=~- ~{k}^{2^+}~-~{q}^{2^-} ~~=~- ~{k}_{2^-}~-~{q}_{2^+} \cr
 &= ~- ({k \over \sqrt{2}},0,{k \over \sqrt{2}},0)
~-~({q \over \sqrt{2}},0,-{q \over \sqrt{2}},0)}
\auto\label{kin2l}
$$ 
where ${q}^{1^-}$ and ${q}^{2^-}$ are, respectively, the wee gluon momenta in
$F_1$ and $F_2$. For simplicity, we took the scale of 
the light-cone momenta for all on-shell pions to be $k$ and the scale of
all wee gluon (longitudinal) 
momenta to be $q$ although, as we discuss further below, this is
clearly not essential. Since
$$
p_1^2 ~= ~p_2^2~=~2kq 
\auto\label{kin021l}
$$
$q$ is both the wee gluon scale and the scale which puts pions on-shell
as it vanishes. 

The right-hand part of Fig.~5 was calculated 
in a ``right-hand finite momentum frame'' in which  
$$
\eqalign{p_3 ~&=~{k}^{2^+}~+~{q}^{2^-} \cr
 &=({k \over \sqrt{2}},0,{k \over \sqrt{2}},0)
~+~({q \over \sqrt{2}},0,{- q \over \sqrt{2}},0)}
\auto\label{kin5}
$$ 
$$
\eqalign{p_4 ~&=~-~{k}^{1^+}~-~{q}^{1^-} \cr
 &=~- ({k \over \sqrt{2}},{k \over \sqrt{2}},0,0)
~-~({q \over \sqrt{2}},-{q \over \sqrt{2}},0,0)}
\auto\label{kin6}
$$ 
and so we also have 
$$
p_3^2 ~= ~p_4^2~=~2kq
\auto\label{kin21r}
$$

The full scattering amplitude for Fig.~5 was calculated
in the ``infinite momentum frame''  in which 
$$
\eqalign{p_1 ~&=~ \bigl(~C ~{k +q \over \sqrt{2}},~{k -q\over 
\sqrt{2}},~0,~S~{k+q \over \sqrt{2}}~\bigr) \cr
p_2 ~&=~ - ~\bigl(~C~ {k +q  \over \sqrt{2}},~0, 
~{k -q \over \sqrt{2}},~S~{k+q \over \sqrt{2}}~\bigr) \cr
p_3 ~&=~ \bigl(~C ~{k +q \over \sqrt{2}},~0,~{k-q  \over \sqrt{2}},
~-S~{k+q \over \sqrt{2}}~\bigr) \cr
p_4 ~&=~-~ \bigl(~C~ {k +q \over \sqrt{2}},~{k-q \over \sqrt{2}},~0, 
~-S~{k+q \over \sqrt{2}}~\bigr)}
\auto\label{kin4r}
$$
where $C=cosh~ \zeta$, $S=sinh~\zeta$, and so 
$$
\eqalign{s~&=~(p_1+p_3)^2  
~\centerunder{$\longrightarrow$}{\raisebox{-5mm}{
$q \to 0$}}
~~(C^2 + S^2)k^2 ~
\centerunder{$\sim$}{\raisebox{-5mm}{
$C \to \infty$}} ~2C^2 k^2 \cr 
t~&=~(p_1+p_2)^2 ~\centerunder{$\longrightarrow$}{\raisebox{-5mm}{
$q \to 0$}}
~ ~- k^2 }
\auto\label{kin6r}
$$
We combined the mass-shell limit $q\to 0$ and the regge limit
$ s/t \to \infty$ by taking 
$$ 
q ~\sim ~1 / C ~\to ~0 
\auto\label{sca4}
$$
Note that, as is apparent from (\ref{kin4r}), the wee gluon momentum $q$ 
is exchanged only as a zero transverse momentum contribution in the infinite
momentum frame. 
  
The internal couplings $U_L$ and $U_R$ appearing in Fig.~5
are anomaly pole contributions
from effective vertices of the form shown in Fig.~6

\begin{center}
\epsfxsize=2.2in
\epsffile{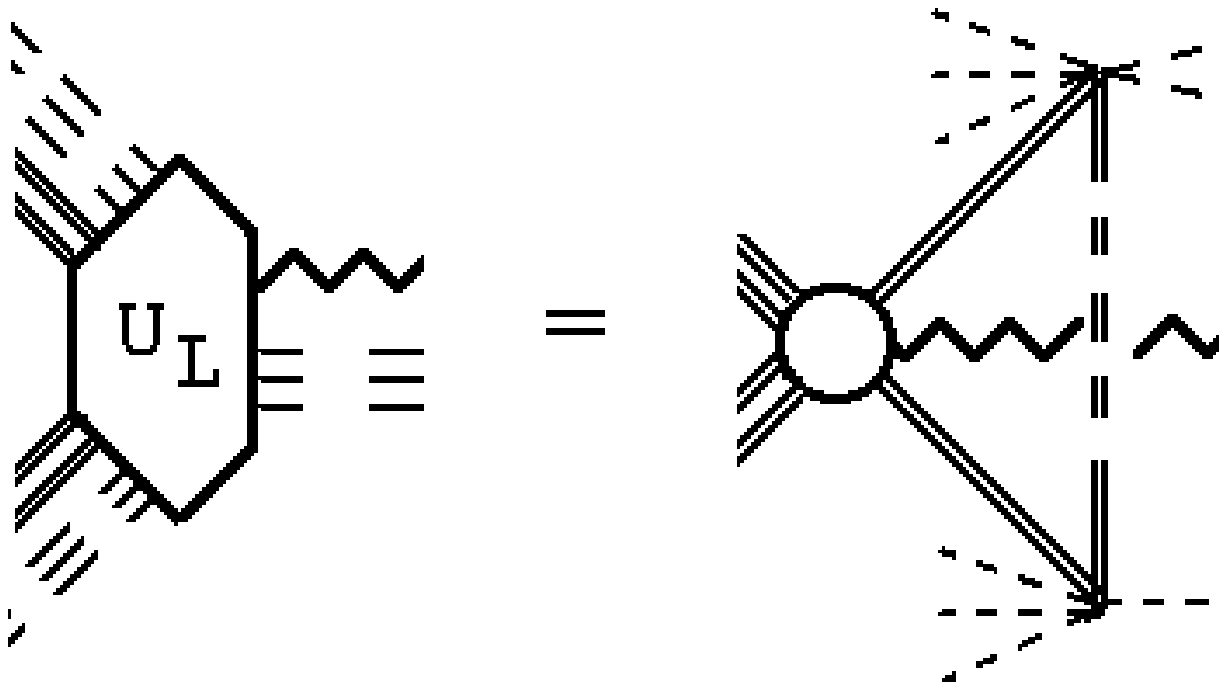}
\hspace{0.6in}
\epsfxsize=1.7in
\epsffile{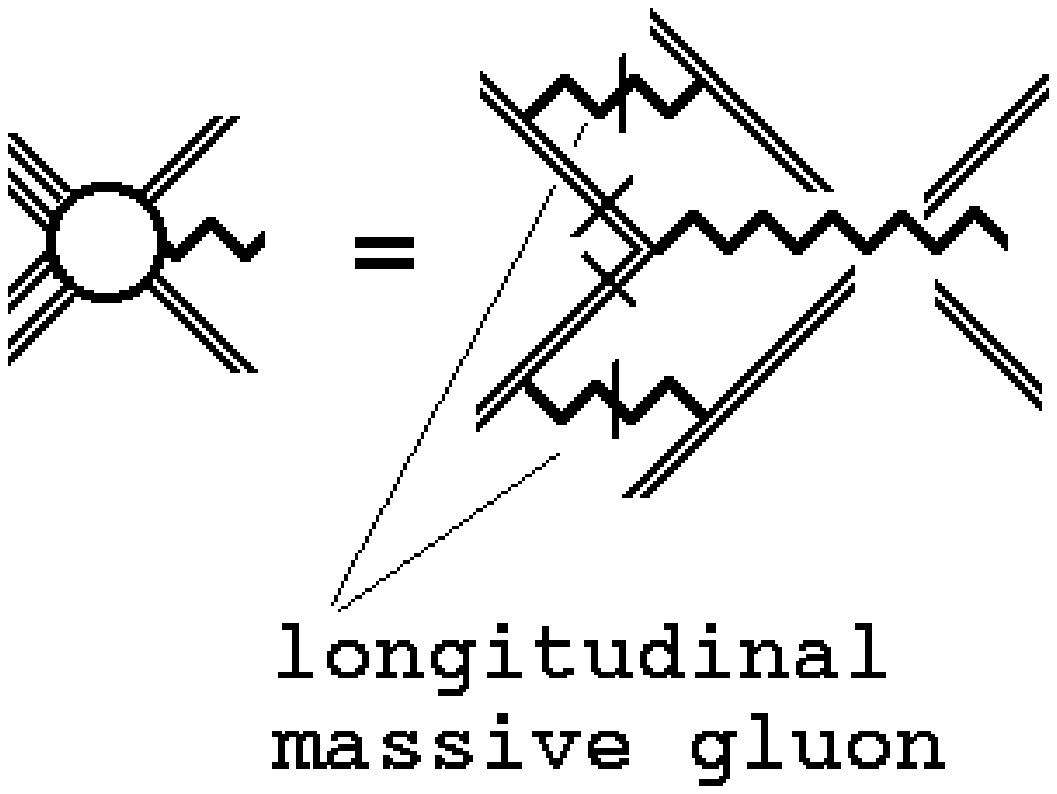}

Fig.~6 An anomaly pole coupling.
\end{center}
(These vertices are illustrated in more detail in Fig.~C6). 
Because the anomaly poles are integrated over, they 
contribute as ``anomaly $\delta$-functions'' that produce a separate 
conservation of transverse momentum
for the massless gluon interactions. This separate momentum
conservation allows these interactions 
to be factorized off from the remaining ``hard interaction''. 
As a result, the diagram of Fig.~5 has an overall logarithmic divergence 
from the region where the transverse momenta of all massless gluons are
scaled uniformly to zero. After this divergence is 
factorized off (as a zero transverse momentum 
``reggeon condensate'') and the pion poles in each channel are also extracted, 
the amplitude obtained has the form 
$$
\eqalign{ A_{\pi\pi\pi\pi} ~\sim&~\prod_i\{F_i~ \hbox{anomaly pole coupling}\}~
\{\hbox {quark $k_{i\perp}$ integrals}\}\cr 
& \times ~~\prod_{j=L,R} \{U_j~ \hbox{anomaly amplitude} \}
~\{\hbox{massive gluon propagator}\} \cr
\sim&~ \biggl\{~{k~ C~q \over M^2}~\biggr\}^4~
\biggl\{~{(k C)\over M^2}~{(kCq ) \over M^2 }~\biggr\}^2~ 
~\biggl\{~C~q ~\biggr\}^4~ 
~~\biggl\{~\frac{1}{t + M^2 }~\biggr\} }
\auto\label{phamp0}
$$
Writing $t \sim k^2$ and $s \sim C^2 k^2 ~$,
(\ref{phamp0}) can be rearranged to give  
$$
A_{\pi\pi\pi\pi} ~\sim ~\bigl[~{C~q \over M}~ \bigr]^8 ~~
\bigl[~{s~q^2 \over M^4}~ \bigr]~~
\bigl[~{t \over M^2}~ \bigr]^2~~
\bigl[~{s \over t+M^2}~ \bigr]
\auto\label{phamp}
$$

Since the first two square brackets in (\ref{phamp}) are finite constants 
when the limit (\ref{sca4}) is taken, the kinematic
structure of the pion 
scattering amplitude we obtain is, essentially, that of massive gluon 
exchange, i.e.
$$
A_{\pi\pi\pi\pi}(s,t)~=~ \bigl[~{t \over M^2}~ \bigr]^2~~
\bigl[~{s \over t+M^2}~ \bigr]
\auto\label{psca}
$$
(Note that this result is obtained for $t >> M^2$.)
In higher-orders the massive gluon will 
reggeize, with an infra-red finite trajectory $\alpha_g(t)$ that satisfies 
$\alpha_g(M^2)=1$. But, since the exchange of
four reggeized gluons is involved, as we add all diagrams and go to
higher-orders, only the even signature amplitude will survive. As a result,
reggeization of the massive gluon will give
$$
\bigl[~{s \over t+M^2}~ \bigr] ~\to~ \bigl[~{s^{\alpha_g(t)} ~+~
(-s)^{\alpha_g(t)}
 \over t+M^2}~ \bigr]
\auto\label{hor}
$$
That is, reggeized gluon
exchange will provide the leading contribution to the pomeron but 
there will be no gluon pole at $-t=M^2$.

\subhead{2.5 Momentum Flows and Wee Gluon Couplings}

The general dynamical structure of the diagrammatic 
contributions to $A_{\pi\pi\pi\pi}$ is illustrated in 
Fig.~7. Where there is a broken quark line (and a $T$)
there is a chirality transition of a zero momentum massless quark. 
Wee gluon couplings, that we will discuss shortly, are denoted
by a circle containing a $w$.
\begin{center}
\epsfxsize=3.7in
\epsffile{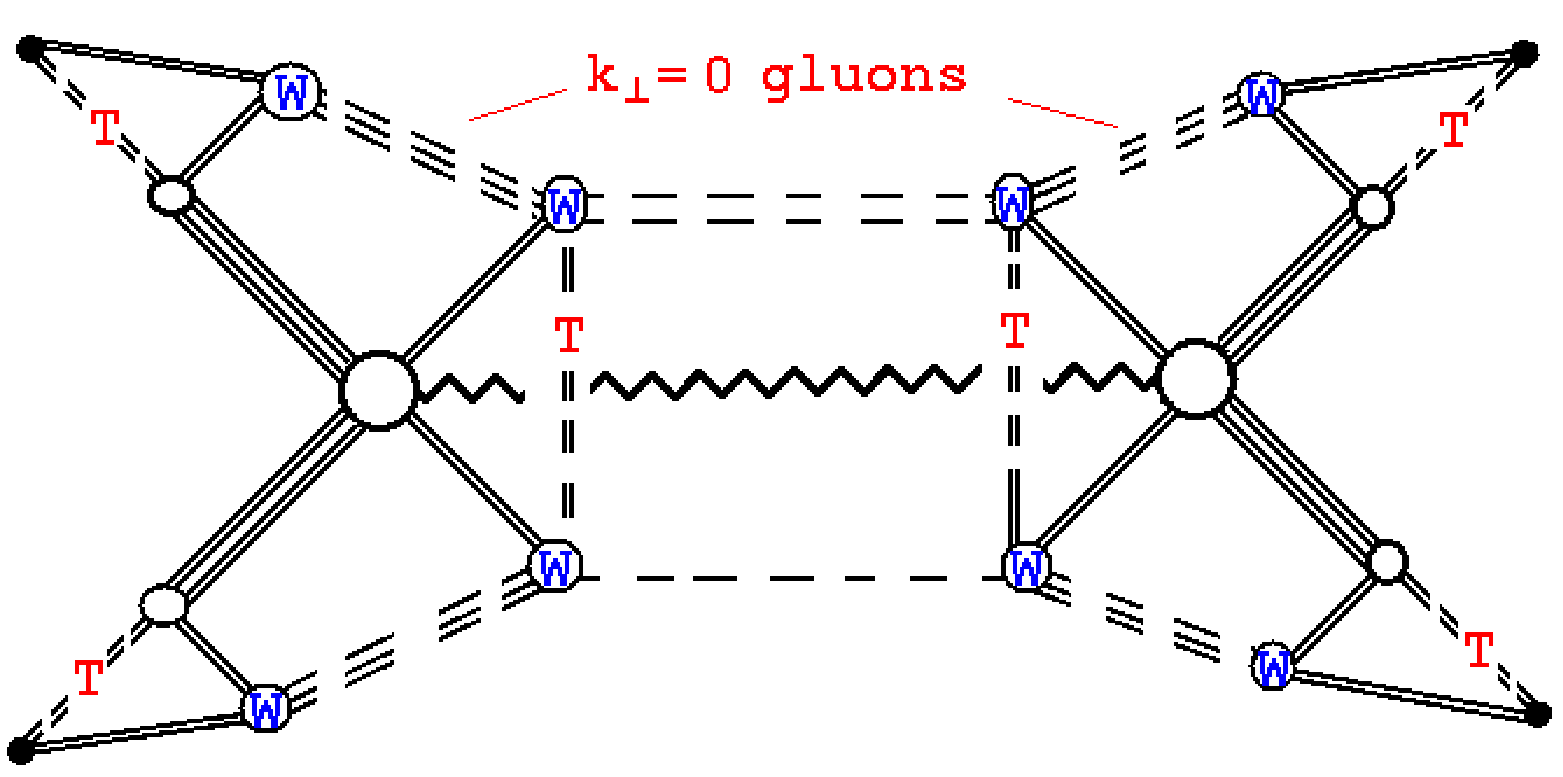}

Fig.~7 Dynamical structure of the scattering amplitude.
\end{center}
Using the origin of the anomaly pole described in Appendix B,
the scattering process can be interpreted as follows. 
A ``pion'' is created by the product of a physical
quark field and a zero momentum ``unphysical'' antiquark
field in which the Dirac sea is shifted.
The antiquark becomes physical, via a chirality transition, that introduces
an accompanying ``semiclassical'' anomalous wee gluon field (condensate)
that effectively
moves the sea back to it's perturbative location. 
In the scattering process, the wee gluon field of an incoming pion is transformed
into that of the outgoing pion by an anomaly coupling that 
involves a further rearrangement of the Dirac sea. The final state pions are  
created via a final shift of the Dirac sea that absorbs the anomalous wee 
gluon field.

The flow of large momentum ($\sim k$ in the finite momentum frame) 
through the left side of Fig.~7 is shown in Fig.~8(a), 
\begin{center}
\epsfxsize=4.5in
\epsffile{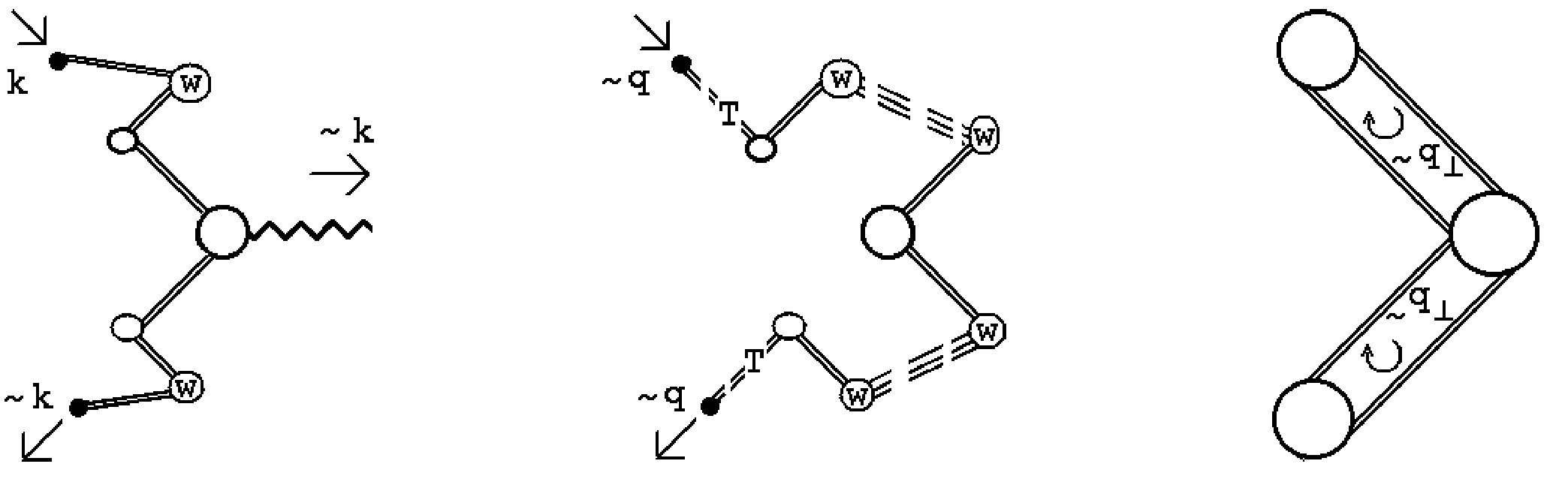} 

(a)\hspace{2.2in}(b)\hspace{2.2in}(c)

Fig.~8 Momentum flows.
\end{center}
while the flow
of wee gluon longitudinal momentum ($\sim Cq$ in the infinite momentum frame) 
is, as shown in Fig.~8(b), along an (almost) orthogonal set of lines. Note that
the large momentum flows along either the quark or the antiquark, but not both. 
The remaining momentum scale is the relative transverse momentum ($\sim q_{\perp}$)
of the quark-antiquark pair which simply flows around a loop, as illustrated in
Fig.~8(c). In the finite momentum frame (``inside the pion'')
the wee gluon limit $q \to 0$ gives the 
zero momentum required for the first and last 
chirality transitions. In the infinite momentum frame  
$Cq$ provides the light-cone momentum flowing around the
triangle diagram giving the anomaly $\delta$-function. The ``zero momentum'' line
in the $\delta$-function triangle 
therefore has momenta much smaller than $q$ in the finite momentum frame.

There are eight wee gluon couplings that originate from the 
chirality transitions. As we already noted, they are denoted 
by a circle containing a $w$ in Fig.~7. A factor of $Cq$ 
for each wee gluon coupling gives
the factor of $[Cq/M]^8$ in (\ref{phamp}). The other two factors in (\ref{phamp}),
apart from (\ref{hor}), arise from the integrations over the 
quark-antiquark relative transverse momenta.
All of the factors in (\ref{phamp}), apart from (\ref{hor}), 
are scaled by the vector boson mass $M$. 
The overall factor of $M^{-16}$ 
can be traced back to the eight contributions of longitudinal massive gluon 
exchange. Four appear via anomaly pole vertices of the form appearing
in Fig.~2, and are represented by small circles in Fig.~7. 
The other four appear in the two vertices, of the form shown in Fig.~6, 
represented by large circles in Fig.~7. In each case the 
longitudinal contribution of the on-shell massive gluon gives a contribution 
of the form 
\newline \parbox{5.5in}{ 
{\Large $\raisebox{2mm}{``}$
$~\frac{k_{\mu}k_{\nu}}{M^2}~~$$\raisebox{2mm}{''}$} $\leftrightarrow$
[wee~gluon~momentum/M~$]_{\mu}$~[quark~transverse~momentum/M~$]_{\nu}$}
\parbox{0.4in}{$~$
\newline $~$
$$
~
\auto\label{lint}$$}
\newline $~$
\newline The existence of the amplitude (\ref{phamp0})
depends entirely on this interaction which, it is important to note,
couples wee gluon related chirality transitions and 
small transverse momentum quark dynamics. Also, 
the appearance of a wee gluon momentum
scale in the amplitude is 
crucially dependent on the presence of such transitions.

Clearly we need not have taken the wee gluon momentum scales of both scattering
pions to be equal. In general the factor of 
$[Cq/M]^8$ in (\ref{phamp}) would be replaced by a separate factor 
of $[(Cq)^2/M^2]^2$ for each scattering pion. Furthermore, we anticipate
that if we were to carry through the complete multi-regge calculation 
of Appendix C, the wee gluon factor for each pion would be replaced by the 
(integrated) contribution of a wee gluon distribution $w(Cq/M)$ so that, in the
pion amplitude,
$$
\bigl[{C q \over M} \bigr]^8\to \prod_{i=1,2} 
~\biggl[\int~d(Cq_i/M)~(Cq_i/M)~w(C q_i/M)\biggr]^2 
\equiv~\bigl[\frac{C}{M}\bigr]^8~ \prod_{i=1,2}~
\biggl[\int dq_i~q_i~w(q_i)\biggr]^2 
\auto\label{wgd}
$$

\subhead{2.6 Higher-Order Diagrams}

Consider now the higher-order diagrams that will add to that of Fig.~5. 
As we noted
above, and discuss in more detail in \cite{arw02}, adding interactions amongst the
wee gluons will not change the nature of the overall divergence.    
Similarly there will be no change if
the three wee gluons in the pomeron, and
in each pion channel, are replaced by infinite sums
over arbitrary (allowably different in each channel), odd, 
numbers of massless gluons that similarly have zero transverse momentum,
carry overall SU(2) color zero, and have 
(anomalous) positive color charge parity. Again such wee gluons 
will have self-interactions but will not interact with
the quark/antiquark pairs in the pions, or the 
SU(2) singlet reggeized, massive, gluon in the pomeron. The same discussion
would also apply if the single massive gluon is replaced by 
any number of massive gluons (giving multiple pomeron exchange).

\subhead{2.7 Pomeron Production Vertices}

In the remainder of this Section, 
and the following Sections, we will go far beyond 
the explicit calculations of \cite{arw03} and \cite{arw02}. We will introduce 
effective vertices for which the underlying (in general, multi-regge)
calculations have not, as yet,
been carried out but whose existence is a natural extrapolation of the 
vertices that we have already discussed. 
We begin by considering, briefly, a set of 
effective vertices which are responsible for 
the vacuum production of pomerons that is one of the defining features of
supercritical RFT. 

A priori, it might appear that the anomaly 
$\delta$ function vertex of Fig.~6 could give rise to simple 
``vacuum production'' of
massive reggeized gluon pairs by wee gluons, as illustrated in Fig.~9(a). 
\begin{center}
\parbox{1.8in}{
\epsfxsize=1.65in
\epsffile{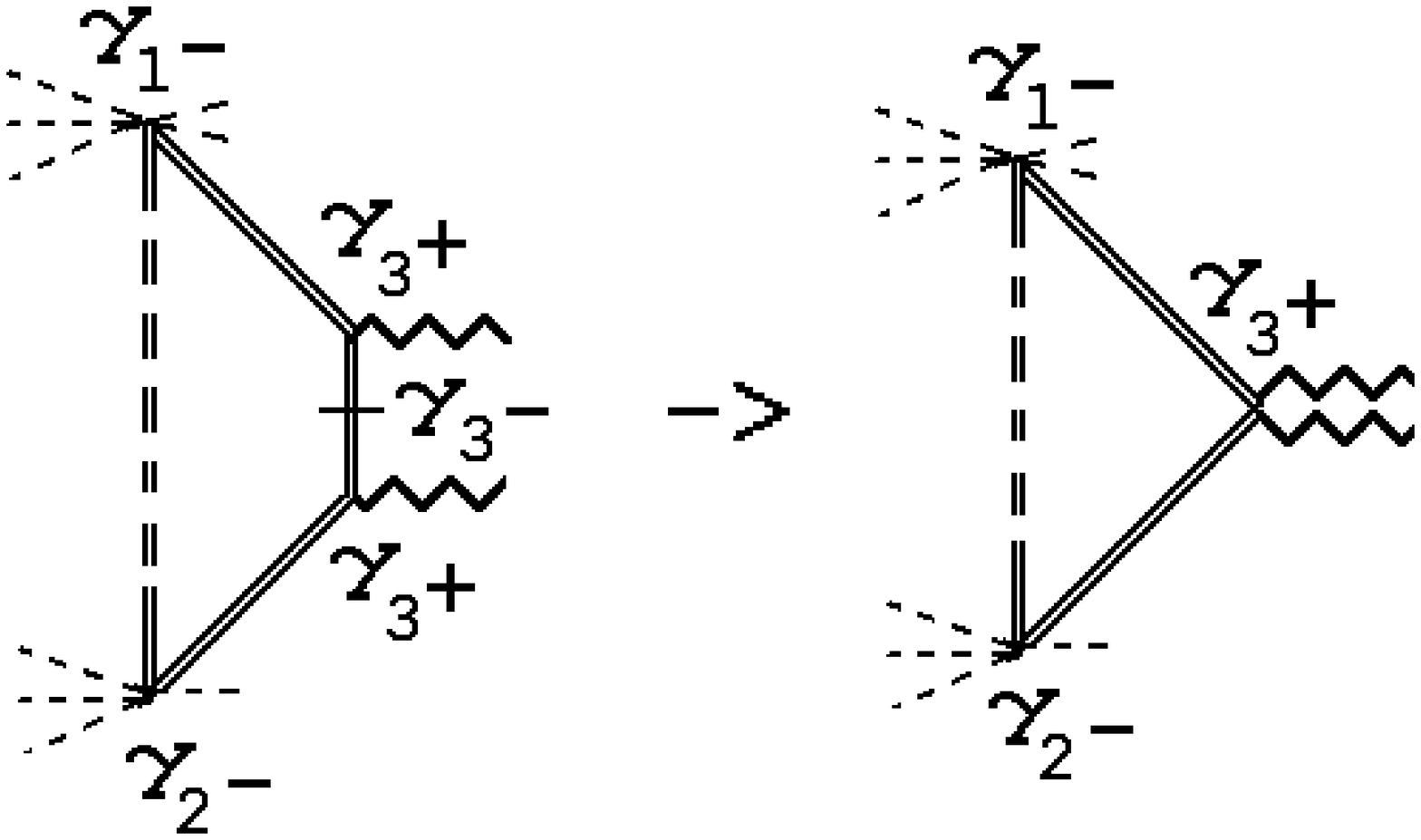}}\parbox{1in}{$$
\st{\to}~ \hbox{anomaly}$$}\hspace{0.3in}
\parbox{1.8in}{
\epsfxsize=1.65in
\epsffile{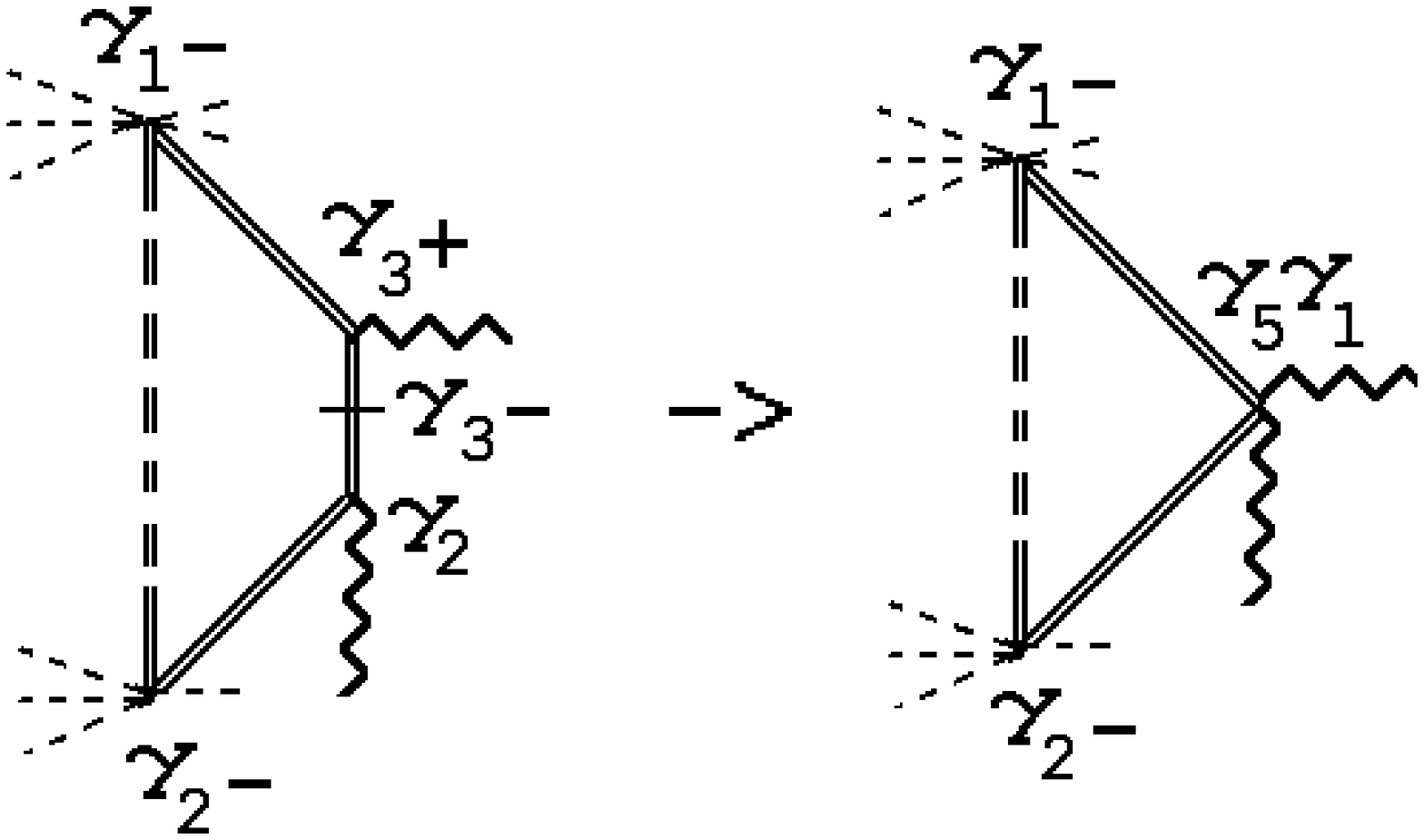}}\parbox{1in}{$$
\to~ \hbox{anomaly}$$}

(a)$~~~~~~$\hspace{3in}(b)

Fig.~9 Wee gluon vertices (a) that do not give an anomaly (b) that give
an anomaly.
\end{center}
In fact, to have the axial vector structure for the anomaly, 
both gluons can not have the polarization needed  
to be exchanged in the scattering process. Instead, as illustrated in Fig.~9(b)
one gluon must have a different polarization. Since the  
interaction can, nevertheless, take place some distance across the rapidity
axis it leads to particle pole interactions within pomeron 
vertices.

The most general pomeron vacuum production vertices are generated as illustrated
in Fig.~10. When these vertices are included, 
we reproduce the complete range of pomeron vertices that arise from
the ``vacuum production of pomerons'' due to the pomeron condensate
in the supercritical pomeron phase\cite{arw91}. 
A more detailed study is needed to determine that
 the non-exchanged massive gluon in Figs.~9 and 10 is 
longitudinal.
\begin{center}
\epsfxsize=3.5in
\epsffile{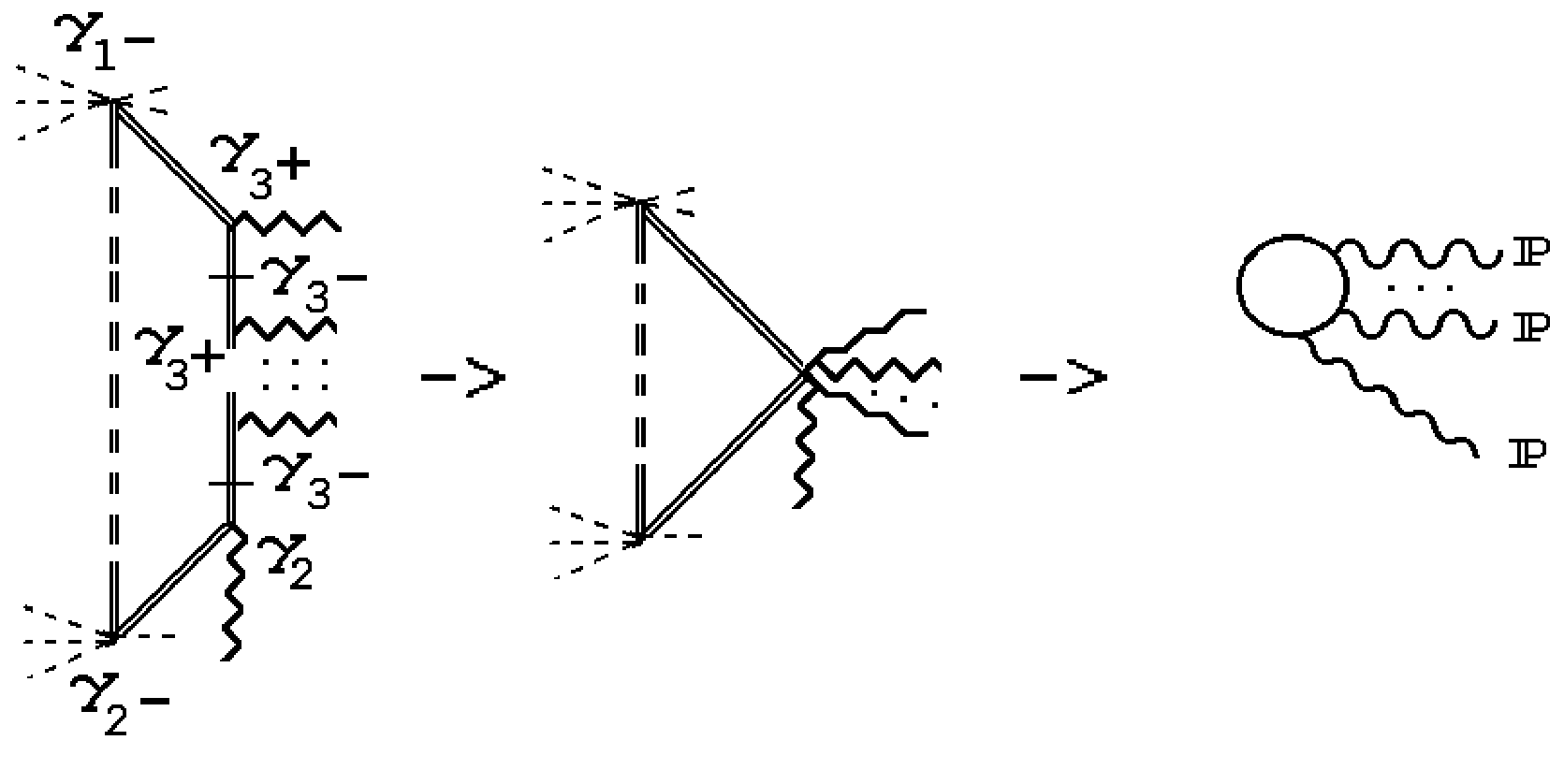}

Fig.~10 Generation of pomeron vacuum vertices.
\end{center}

\subhead{2.8 The Complete Set of Amplitudes and States}

While it remains to be shown that the high-energy behavior of $CSQCD_S$ maps 
completely on to supercritical RFT, we will assume,
in this paper, that the connection is established. Our major purpose, here,
has been to elaborate the physics that is involved. As we have seen,
the essential physics of $CSQCD_S$ is 
that a wee gluon condensate is produced by chirality transitions that are part 
of anomaly interactions introduced by the massive vector mesons. 
We can view the condensate as originating 
from a shift of the Dirac sea that 
produces states, and an 
S-Matrix, in which
SU(2) color confinement and chiral symmetry breaking
completely determine the spectrum.  
The wee gluon condensate has no connection with instantons. It is a 
``semi-classical'' infra-red effect that, as we discuss in the next
Section, becomes a dynamical effect in $QCD_S$. Note also that,
since the anomalous wee gluons in a pion
can not be produced from the perturbative quark/antiquark component
by normal perturbative interactions (without an anomaly-related 
chirality transition), we can say that there is no simple quark/antiquark component
in the infinite momentum pion ``wave function''.

We expect the complete set of (infinite momentum)
physical scattering 
amplitudes in $CSQCD_S$ to be produced via a logarithmic divergence, as in our 
discussion of the amplitude obtained from Fig.~5. If this is the case, then any
physical amplitude must involve initial and final scattering states that 
contain anomalous wee gluons. If such gluons appear only via 
anomaly pole vertices then, according to our discussion, all physical states must be
color zero Goldstone bosons. Unfortunately, we 
have only been able to study on-shell pion 
amplitudes. If we were to carry through the multi-regge program of Appendix C,
then we would obtain amplitudes for off mass-shell reggeized pions to scatter. 
This would give us much more information about how a pion appears as 
an anomaly pole 
and would, perhaps, allow us to determine the role played by chiral symmetry in 
ensuring that such a pole is present. For the present we assume that an
anomaly pole occurs if and only if there is a chiral symmetry that can be 
broken spontaneously. We also assume that the anomaly pole 
mechanism provides the only possibility for the dynamical formation of bound state
Goldstone bosons.

We can refer to the Goldstone bosons as created by a product of quark/antiquark 
operators alone provided we 
remember that the wee gluon component can be eliminated only by a shift of the 
Dirac sea in one of the operators. If we denote SU(3)
color triplet quarks, generically, 
by {\large $q$} and SU(3) color sextet quarks, generically, by {\large $Q$},
the Goldstone boson states of $CSQCD_S$ obviously include 
all flavor non-neutral {\large $q\bar{q}$} and {\large $Q\bar{Q}$} 
pseudoscalar mesons. (There will be two separate $Q\bar{Q}$ states 
formed from SU(2) color triplets and doublets.) In Section 5,
we will discuss how the flavor neutral mesons (the $\eta_6$ and the $\eta_3$) 
mix with pure gluon states and, hence, do not appear as Goldstone bosons. 
Because of the equivalence of quark and antiquark representations 
when the gauge symmetry is SU(2), there are also {\large $qq$}, 
{\large $\bar{q}\bar{q}$}, {\large $QQ$}, 
and {\large $\bar{Q}\bar{Q}$} states that are Goldstone boson mesons
in $CSQCD_S$ but will become baryons, by aquiring an additional quark (or
antiquark) in $QCD_S$. Such states reflect real chiral symmetries\cite{kog} 
of $CSQCD_S$. (Again, the $QQ$ states will appear as separate states formed 
from SU(2) color triplets and doublets.) 

We will not 
discuss the dynamics of baryon formation in this paper, 
although we will briefly discuss the spectrum
in the next Section. To discuss dynamics we need to know the full role
of the SU(2) singlet quarks and gluons in $CSQCD_S$. According to the 
above argument, since they are not Goldstone bosons, they can not be
physical states. If they are, nevertheless, ``physical'',
it must be that they appear as regge exchanges, without
producing physical states. 
For example, within $CSQCD_S$ there can be a regge 
exchange involving the combination of a 
Goldstone boson ``nucleon'' and an SU(2) reggeized quark 
that can become a normal, reggeized, nucleon 
in $CSQCD$, as SU(3) color is restored.

\subhead{2.9 Background Wee Gluon Interactions}

A more subtle question is the role played by the SU(2) singlet gluon. In particular,
is there an odd-signature amplitude involving only exchange of the SU(2)
singlet gluon reggeon? A divergent amplitude can be 
produced by background wee gluon anomaly interactions, as illustrated 
in Fig.~11.  
\begin{center}
\epsfxsize=3.5in
\epsffile{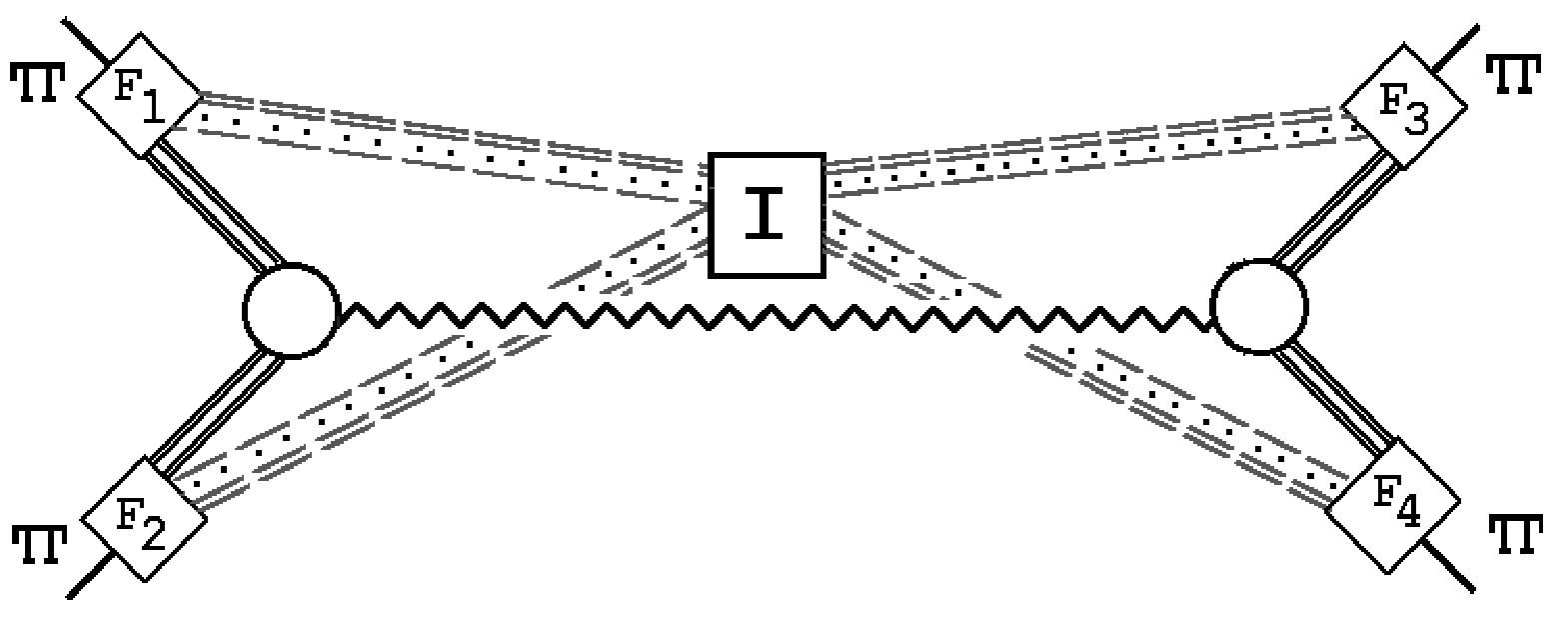}

Fig.~11 Background Wee Gluon Interactions Accompanying Reggeon Exchange
\end{center}
In general, we would expect that there should be (multiple)
chirality violating interactions that involve just wee gluons, 
accompanying all interactions and contributing to the overall divergence. 
As we will see in Section 5, the existence of wee gluon interactions
of this kind is essential
for adding the electroweak sector of the Standard Model to $CSQCD_S$. Unfortunately,
to establish the existence and nature of such interactions requires 
elaborate multi-regge calculations that have yet to be carried out.
The interaction of Fig.~11
must contain anomaly effective vertices generated by the orthogonality of the 
$\gamma$-matrices involved, as illustrated in Fig.~12.
\begin{center}
\epsfxsize=3.6in
\epsffile{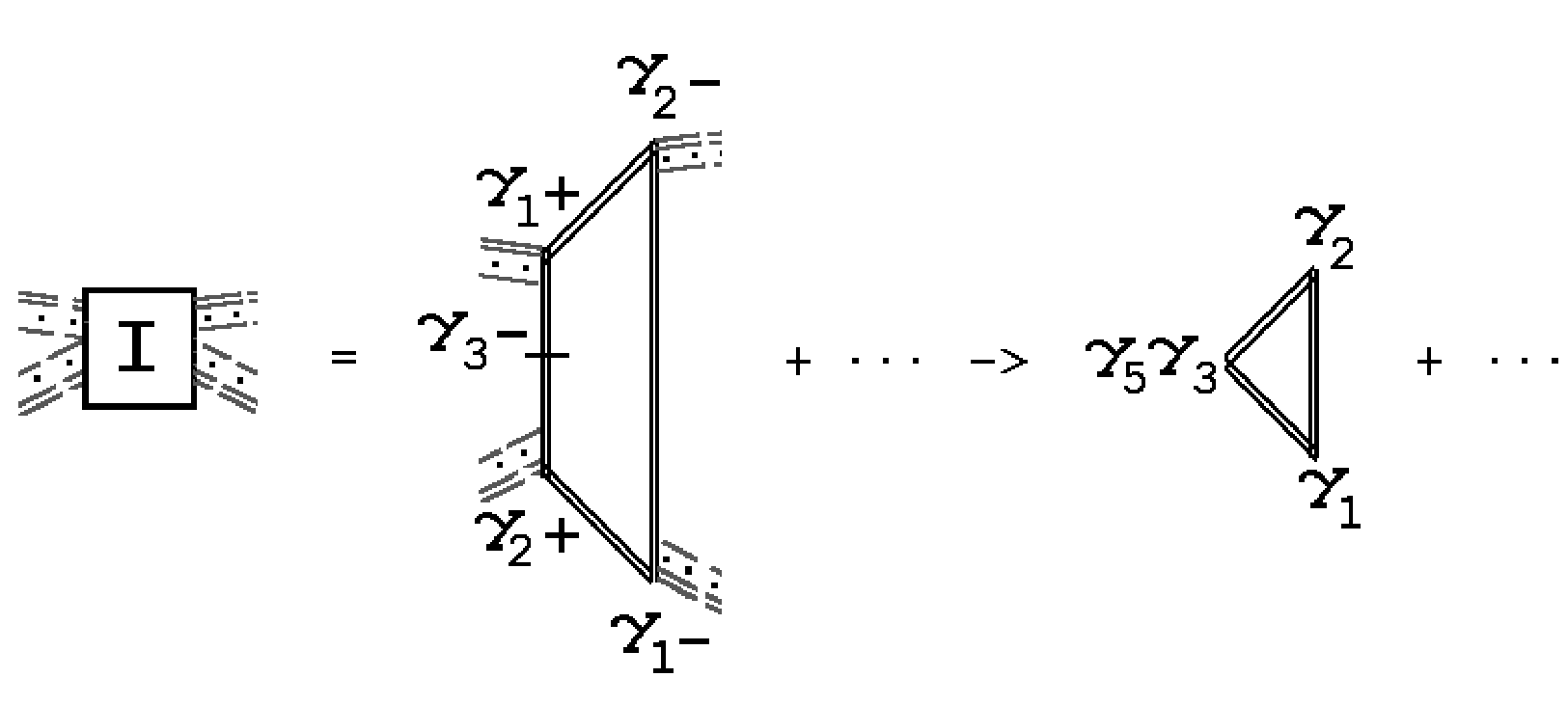}

Fig.~12 A Background Effective Vertex Containing an Anomaly 
\end{center}
If there is no anomaly,
there will be an exponentiation of the divergences via even signature (BFKL)
gluon interactions with the reggeon, as illustrated in
Fig.~13, that will produce a zero amplitude. The anomaly vertex of Fig.~12
necessarily couples directly to the wee gluons 
in the scattering state, and so avoids the exponentiation. 
\begin{center}
\epsfxsize=3.5in
\epsffile{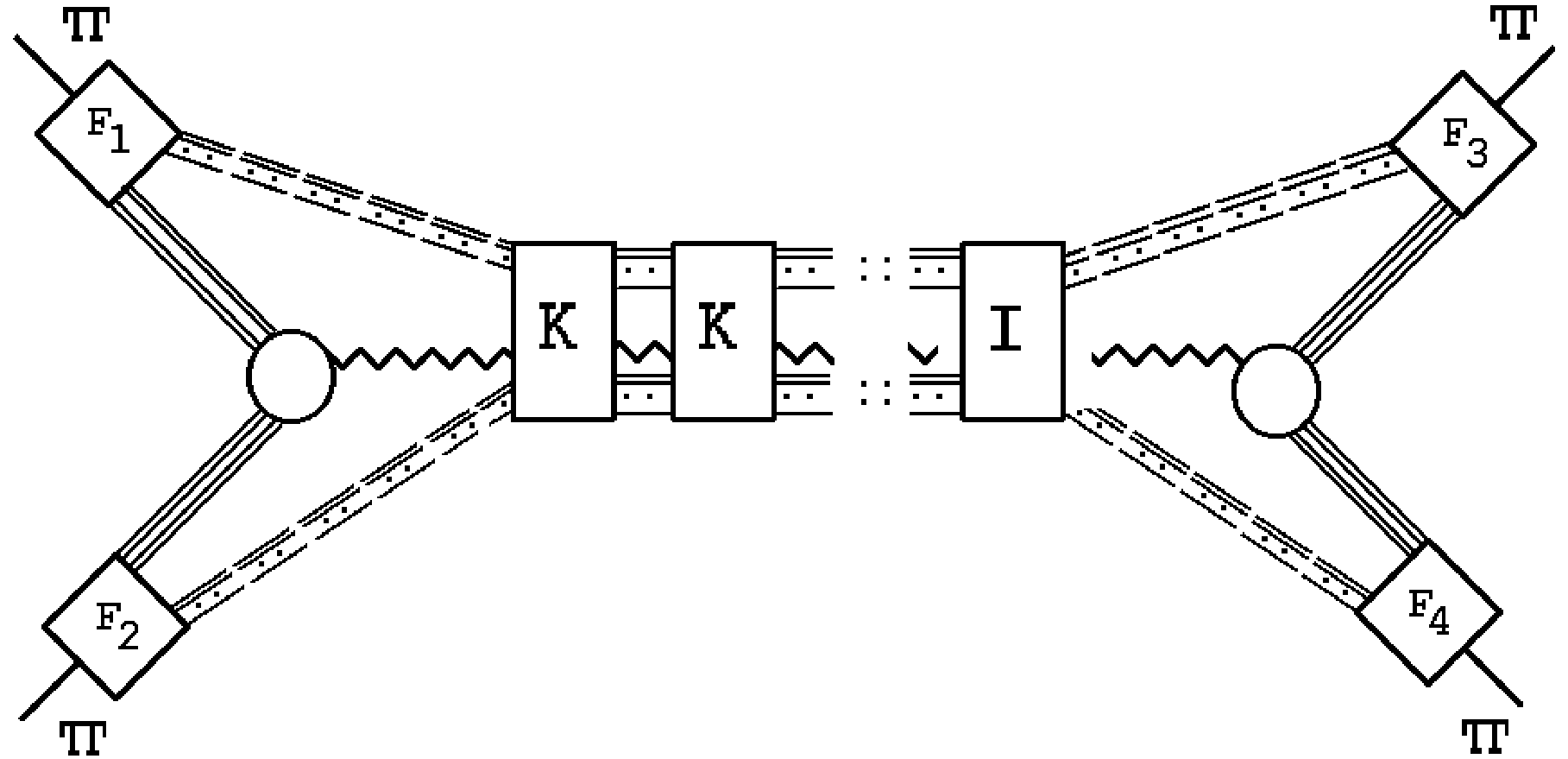}

Fig.~13 Potential Exponentiation of the Wee Gluon Interaction
\end{center}

As SU(3) symmetry is restored, the background wee gluon interaction should 
become SU(3) symmetric. As a result, the non-zero SU(3)
color of the reggeon in Fig.~11
should lead to the vanishing of this amplitude. However, when the reggeon is 
replaced by
an electroweak vector boson which does not carry color, as we discuss in Section
5, the corresponding amplitude will not vanish. 

\newpage 

\mainhead{3. THE CRITICAL POMERON IN $QCD_S$}

If the high-energy behavior of 
$CSQCD_S$ is mapped onto supercritical RFT, as discussed in the last 
Section (and in Appendix C), SU(3) color will be restored 
via the Critical Pomeron phase transition. 
As part of this transition, the SU(2) singlet gluon will become massless and 
decouple. Simultaneously, the wee gluon condensate will disappear 
and a corresponding dynamical degree of freedom will appear. That is, the 
shifting of the Dirac sea will become dynamical.
Dynamical, gauge-invariant, infinite  
number),wee gluon combinations  carrying octet color, will produce the chirality 
transitions illustrated in Fig.~7 (and many more). 
For this to happen, the longitudinal
vector meson interactions, which at first sight should decouple as the 
color symmetry breaking is removed, must still be present - at zero light-cone
momentum. 

In fact, the role
of zero light-cone momentum, longitudinal, gluons is a major ambiguity of
light-cone quantization\cite{arw84}. When we discuss wee gluons in a pion, as we 
did in the previous Section, we are essentially invoking light-cone quantization
in a frame in which the pion carries light-cone momentum 
$k_+$. For the dynamical wee 
gluon processes that we are discussing to be present the longitudinal, zero
light-cone momentum, gluons must provide the interactions, of the form 
of (\ref{lint}), that are responsible for the occurrence of the 
chirality transitions (and anomaly poles) in Fig.~7. There is, of course, 
no vector gluon mass ``$M$'' in $QCD_S$. Consequently, there must be
an intrinsic momentum scale $\mu$ that is generated as part of the symmetry 
restoration process that will
provide the scale for dynamical wee gluon contributions in a hadron.  
Whether, or not, this scale should simply be identified with the normal
dynamical scale of $QCD_S$ remains to be determined. In any case,
by constructing the high-energy behavior of $QCD_S$ via $CSQCD_S$ we are, 
effectively, fixing the ambiguity of the 
role of zero light-cone momentum, longitudinal gluons.

The dynamical shifting of the Dirac sea produced by wee gluon interactions will, 
as we said above, no longer correspond to the introduction of a semi-classical
gauge field, or condensate, in a fixed direction of the SU(3) color group. Rather,
the chirality transitions, which will be many in any scattering process, will 
correspond to random gauge field fluctuations within the color group. The 
transition from a fixed ``magnetization'' for the gauge field associated
with Dirac sea shifts to a randomized, fluctuating, field, characterizes the
nature of the ``critical phenomenon'' that is associated with the high-energy 
behavior of $QCD_S$. The shifting of the Dirac sea is the ``order parameter'' of
the transition. In the supercritical phase this degree of freedom is ordered into
a single, semi-classical, wee gluon gauge field contribution, 
while in the sub-critical phase it is random. 

It is obviously essential for the quarks to be massless if the 
physics of the Critical Pomeron is to be as we have just described it. 
The chirality transitions can take place in 
a ``perturbative manner'' (i.e. within effective vertex
triangle diagrams) only if the quarks are massless. We would expect, however, that 
the high-energy behavior is independent of the physical states aquiring masses and
therefore would expect that the Critical Pomeron remains, at high-energy, even when 
effective quark masses are added to $QCD_S$. To add such masses and preserve the
physics involved would appear, nevertheless, to be non-trivial. It would appear
that the Dirac sea would have to undergo major shifts
(as envisaged by Gribov\cite{dd}) in a random 
dynamical manner, as part of any scattering process and as part of the creation
of asymptotic states. In fact, it now seems likely that the solution  
to this obviously complex problem is provided by the
embedding of $QCD_S$ and the electroweak sector of the Standard Model in
``very special'' unified theory\cite{kw,arw05}. This unified theory should also 
answer the question of how the short-distance electroweak anomaly due to 
the sextet quarks is canceled.

The large transverse momentum (``short distance'')
pomeron will be the least sensitive to the wee gluon phase transition. At large 
transverse momentum, therefore, the $QCD_S$ pomeron will be approximately
a short-distance (gauge-invariant) reggeized  
gluon combined with a color compensating dynamical, anomalous,
wee gluon contribution. Also, at large transverse momentum, 
both triplet and sextet pions will have a
wee gluon component that is the same as the pomeron, 
but with a short-distance quark-antiquark pair replacing the 
reggeized gluon. It can be shown that
the quark-antiquark state in a pion reggeizes and so becomes 
gauge-invariant, like the reggeized gluon in the pomeron, but we will not
discuss it in this paper. (Note that we expect that at large transverse momentum
the quark and antiquark in a reggeized 
pion have equal dynamical status while, in an on-shell
pion one or the other carries, essentially, all of the corresponding 
light-like momentum.)

As we said in the last Section, we also 
will not attempt to follow the formation of
baryons as SU(3) color is restored. However, there is one very importent 
feature of baryon formation which is clear. Namely,
there are no ``hybrid states'' formed, for example, by a sextet quark 
{\large $Q$} combining with
a {\large $\bar{q}\bar{q}$} triplet 
state that is a ``nucleon'' in $CSQCD_S$. This combination 
is possible in principle, but the Goldstone boson nucleon will have the wrong
symmetry properties to combine with the SU(2) singlet component of a sextet quark.
In addition, for the complete SU(3) invariant 
state to be formed it would be necessary to also have a {\large $\bar{q}Q$} 
state in $CSQCD_S$ combining with an SU(2) singlet {\large $\bar{q}$} 
(as the symmetry is restored) and, as is clear from the previous Section, this
is prevented by the complex SU(2) triplet component of the $Q$. 
We conclude, therefore, that the only new baryon states formed by the sextet
sector are the sextet proton - the $P_6$, and the sextet neutron - the $N_6$. 
The importance of this conclusion will become apparent in later Sections. 

We can enumerate the formation of the asymptotic states 
of $QCD_S$ from those of $CSQCD_S$, as follows. 
\begin{enumerate}
\item{``pions'' $\leftrightarrow$
\{{\large $q\bar{q}$} + wee gluons\}  $\to$ normal meson spectrum in $QCD_S$}
\item{``Pions'' $\leftrightarrow$ 
\{{\large $Q\bar{Q}$} + wee gluons\}  $\to$ $\Pi^{\pm},\Pi^0,$ in  $QCD_S$ }
\item{``nucleons'' $\leftrightarrow$ 
\{{\large $qq~/~\bar{q}\bar{q}$} + wee gluons\} + \{{\large $q~/~\bar{q}$}\}, 
$\to$ SU(3) color singlet 
\newline $\to$ normal nucleon spectrum in $QCD_S$ }
\item{``Nucleons'' $\leftrightarrow$
\{{\large $QQ~/~\bar{Q}\bar{Q}$} + wee gluons\} + \{{\large $Q~/~\bar{Q}$}\}, 
$\to$ $N_6$, $P^{\pm}_6$ in $QCD_S$ }
\end{enumerate}

In Section 5 we will discuss hard diffractive interactions of the pomeron
with either a photon or an electroweak vector boson.
In these interactions the wee gluon component has only a limited role and,
most importantly, there are no wee gluon interactions. In these circumstances,
we can continue to represent the wee gluon component as a 
zero transverse momentum ``condensate''. Even though, in reality, it is
a much more complicated dynamical contribution of wee gluons 
over a range of infra-red transverse momenta. As we will see, 
the effective vertices involved will 
not contain a longitudinal vector interaction and so, as a consequence, 
the scale of wee gluon couplings will be an important effect. It will be crucial that,
as we determine from the electroweak mass scale in the next Section, 
the wee gluon couplings 
for triplet and sextet quarks are very different. This will 
be represented by distinct condensate couplings for triplets and sextets.

With the wee gluons treated as semi-classical, we will be able to use
the anomaly pole mechanism to obtain a
limited understanding of the production 
of sextet pions and the resultant production of $W$'s and $Z$'s in
hard diffractive processes. Not surprisingly, 
the minimal representation of the dynamics of the
wee gluon component will have major limitations. 
Most significantly, we will be able to apply the ``condensate anomaly mechanism'' 
only at large $k_{\perp}$ and then, directly, only 
to the production of an ``on-shell'' sextet pion carrying light-like momentum. 
Dynamical wee gluons can, presumably, produce
sextet pions at both small $k_{\perp}$ and off-shell, 
but we will not try to discuss this explicitly.
Instead, we first use the kinematic form given directly by 
the anomaly amplitude to go ``off-shell''. This leads to 
rough order-of-magnitude
estimates and (some) qualitative kinematic features 
of hard diffractive phenomena. We can then combine the knowledge 
of hard diffraction that we obtain, with regge theory, 
to discuss expectations for soft
diffraction. We will argue that, at the LHC, 
the most immediate place to see that new physics is in evidence is
likely to be the double pomeron exchange cross-section!

\newpage

\mainhead{4. ELECTROWEAK VECTOR BOSONS AND THE SEXTET QCD SCALE}

We consider, now, the addition of the electroweak vector boson sector 
to $QCD_S$. We first add a triplet $\{W^{\pm},W^0\}$ of 
massless SU(2) gauge bosons with Standard Model left-handed 
couplings (with coupling constant $g_w$) to both triplet and sextet quarks. Later 
we will add a 
massless hypercharge gauge field $Y$ (with coupling constant $g_y$) that 
also has Standard Model couplings to all quarks. We define ``Standard Model'' 
couplings for sextet quarks by recognizing that sextet antiquarks have the 
same SU(3) triality as triplet quarks. It is natural, therefore, for  
sextet antiquarks (quarks) to have 
the same electroweak couplings as triplet quarks (antiquarks). In 
fact, this is also what occurs when both kinds of quarks originate from an 
underlying unified theory\cite{kw}. In massless $QCD_S$ there will be three
flavor doublets of color triplet quarks that each produce a triplet of 
``pions'' that have the quantum numbers to couple directly to the $W$'s.
The triplet of Pions produced by the single sextet doublet 
similarly has the quantum numbers to couple directly to the $W$'s.  
We begin in $CSQCD_S$, however, because
this will enable us to understand the generation of a vector boson mass in terms 
of anomaly pole pions and Pions. We will see how the 
wee gluon component of a scattering, infinite momentum,
pion generates a mass for an exchanged
vector boson, as we would expect if the universal wee gluon component of 
infinite momentum states is able to reproduce vacuum properties.

\subhead{4.1 Background Wee Gluon Interactions}

To obtain an infra-red divergent scattering amplitude involving W exchange, 
there must be a wee gluon 
exchange accompanying (but not interacting with) 
the $W$, as illustrated in Fig.~14.
\begin{center}
\epsfxsize=2.5in
\epsffile{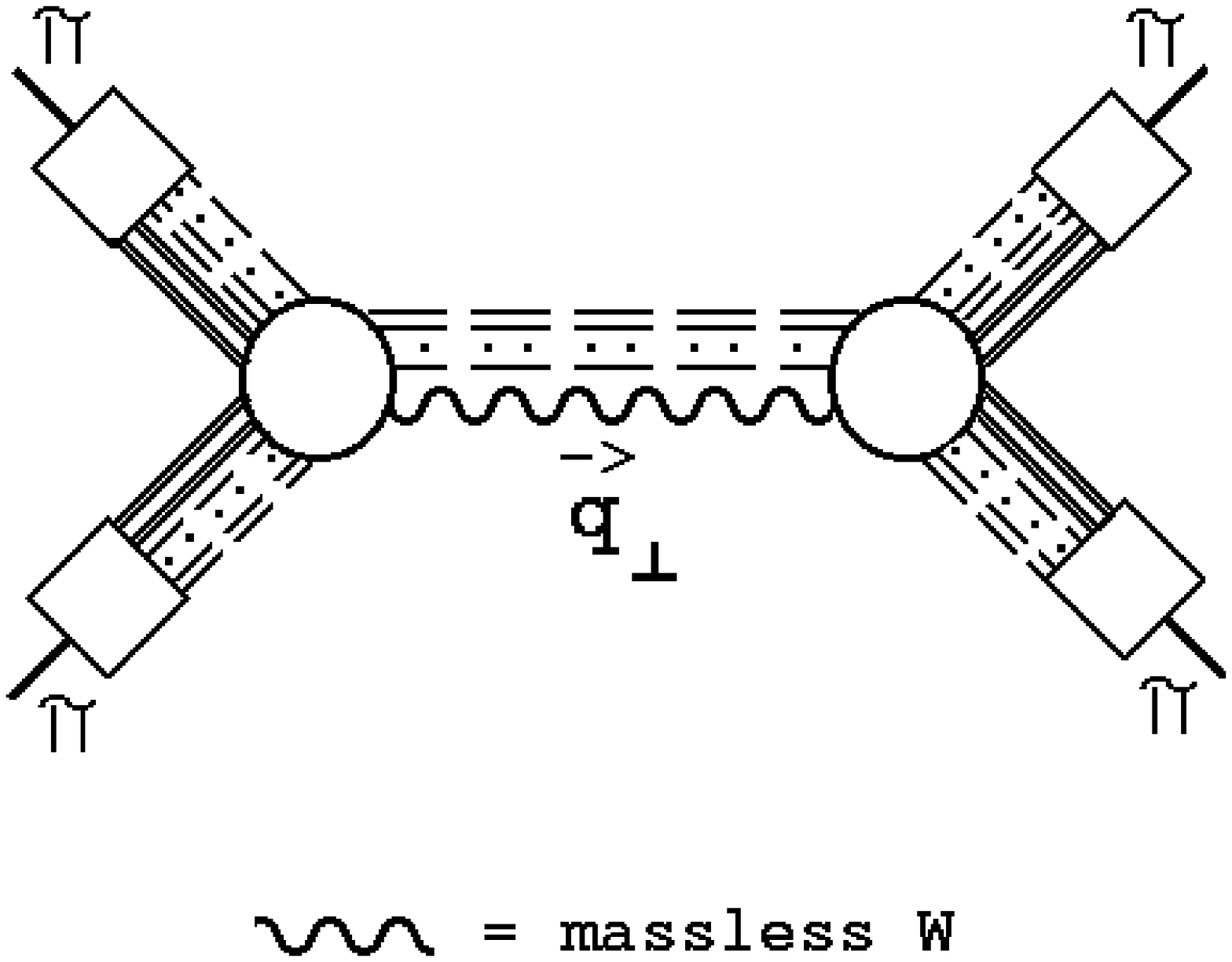}

Fig.~14 Scattering via $W$ exchange.
\end{center}
(Apart from the exchanged vector boson, the notation is the same as in Section 2.)
However, because of the left-handed coupling, there will be interaction kernels,
analagous to that of Fig.~3, for the $W$ to interact with multi-gluon states that 
carry both normal and anomalous color charge parity. As a consequence, all infra-red
divergent amplitdes will be exponentiated to zero, except for those produced by
background wee gluons. Although the underlying multi-regge calculations remain
to be carried out, we expect that
there will be amplitudes analagous to that of Fig.~11, but 
with the gluon reggeon replaced by a (reggeized) vector boson. In this case, 
we expect that the full
anomaly vertices, of the kind illustrated in Fig.~12, will not survive the 
exponentiation analagous to Fig.~13. Instead, 
the left-handed component of the axial-vector
coupling shown in Fig.~12 will, because of the left-handed $W$ coupling, contribute
to an exponentiation of the form of Fig.~13. Implying that perturbative $W$ exchange
will be accompanied by a background, ``right-handed'', wee gluon interaction. In
$QCD_S$, with SU(3) color restored, this background interaction will be SU(3)
symmetric.

\subhead{4.2 $W$ Mass Generation}

A priori, we anticipate that the existence of Goldstone boson $\Pi$'s  
will lead to the $W$'s aquiring a mass via the mixing 
illustrated schematically in Fig.~15. 
\begin{center}
\epsfxsize=4in
\epsffile{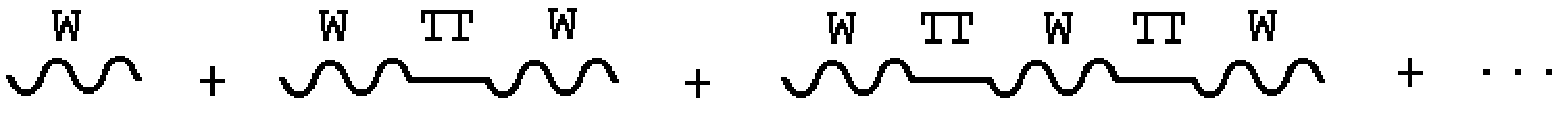}

Fig.~15 The Anticipated Mass Generation.
\end{center}
We will show that, in the regge limit, the first interaction term is produced
(when $q^2_{\perp} \to 0$ ) by wee gluons in one, or the other,  
of the scattering pions. The wee gluon anomaly interactions 
involved are illustrated in 
Fig.~16. 
\begin{center}
\epsfxsize=2in
\epsffile{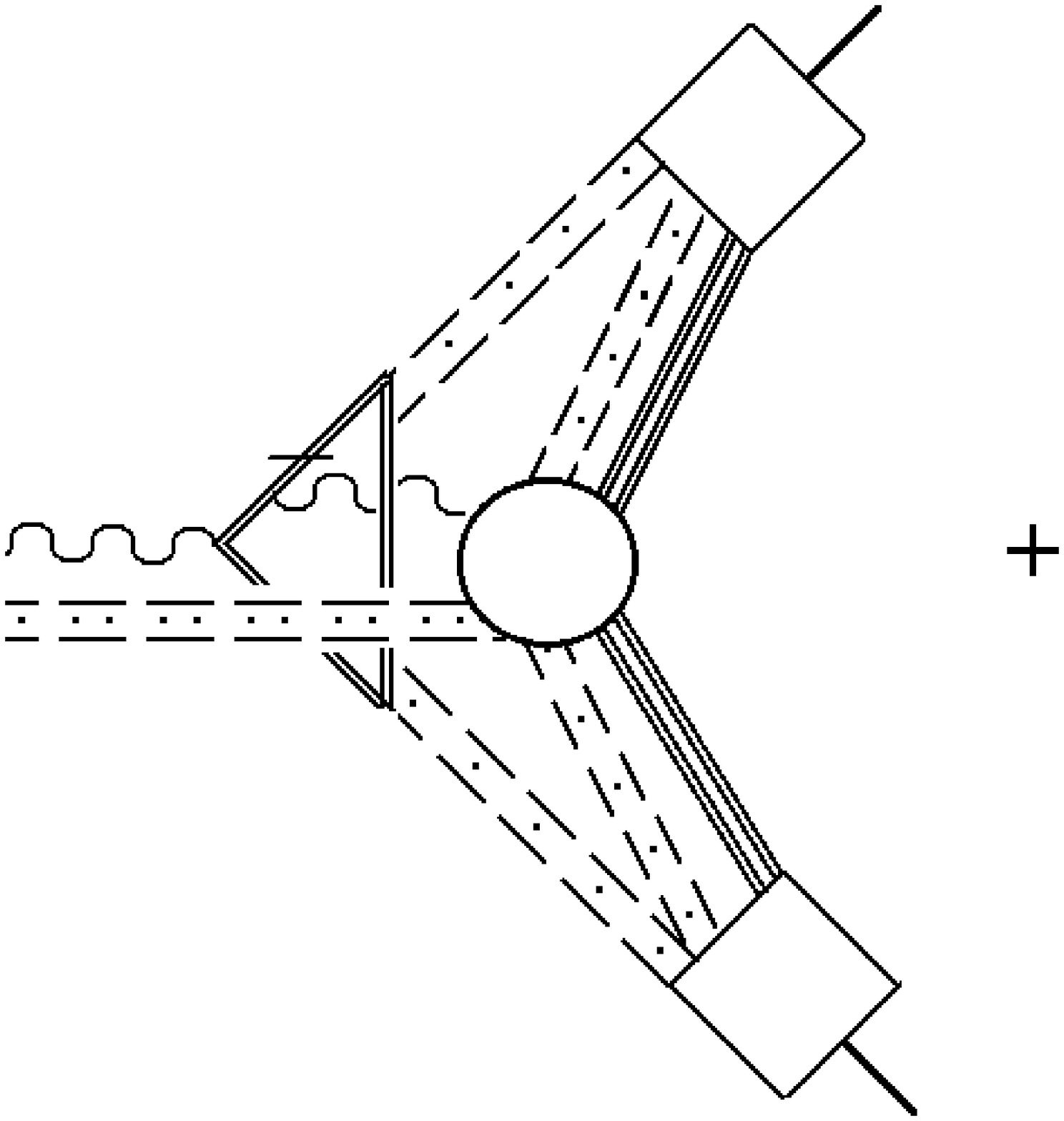}
\epsfxsize=2in
\epsffile{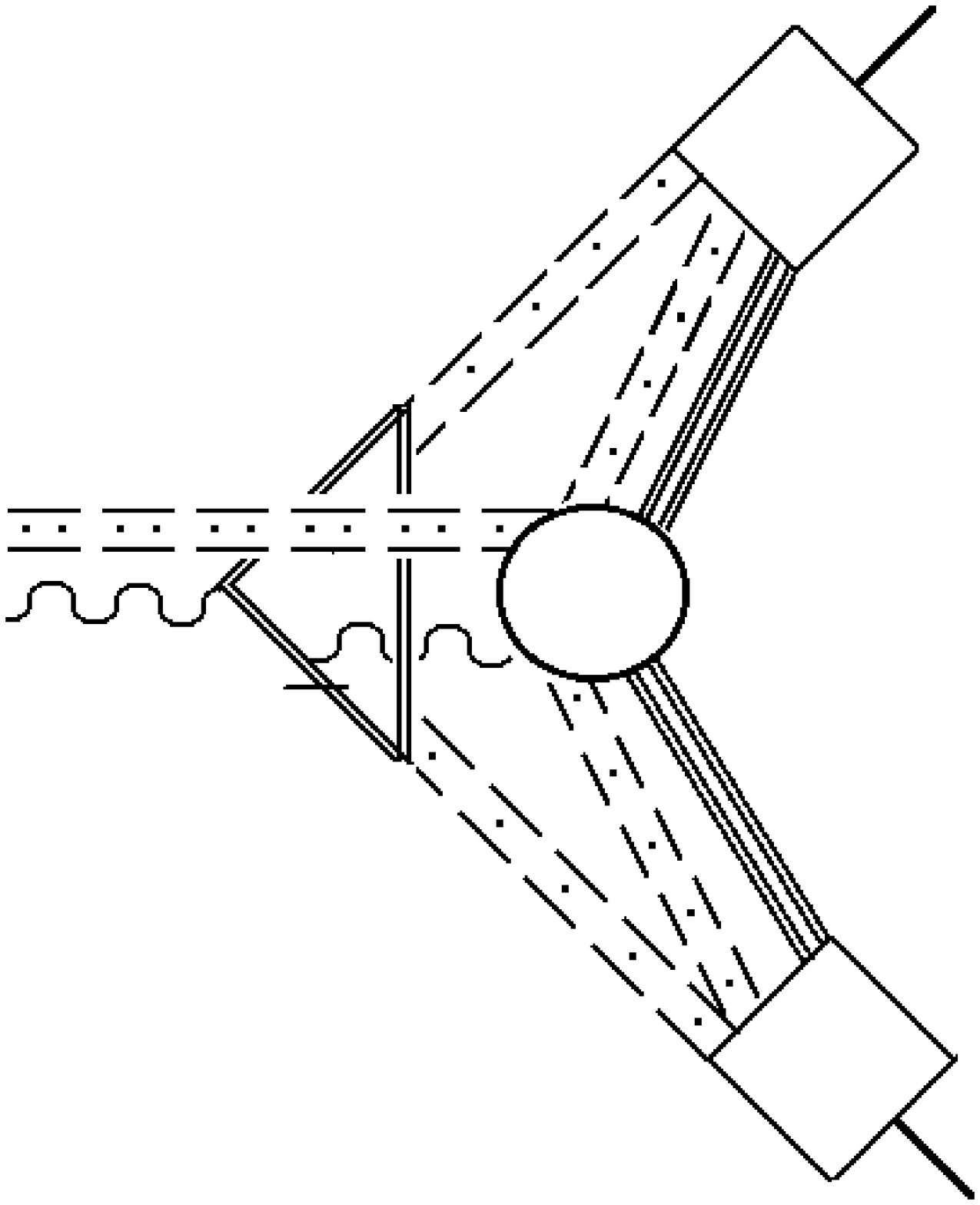}

Fig.~16 Anomaly interactions.
\end{center}
For the moment, the quark loop involved 
can be either sextet or triplet. We will not attempt to identify the higher-order
terms in Fig.~15. Identifying the first term will give us sufficient information
for our purposes.

With the wee gluon kinematics used in Section 2, the first interaction 
in Fig.~16 gives, as $q^2_{\perp} \to 0$, 
the anomaly pole contribution shown in Fig.~17.
\begin{center}
\epsfxsize=3.5in 
\epsffile{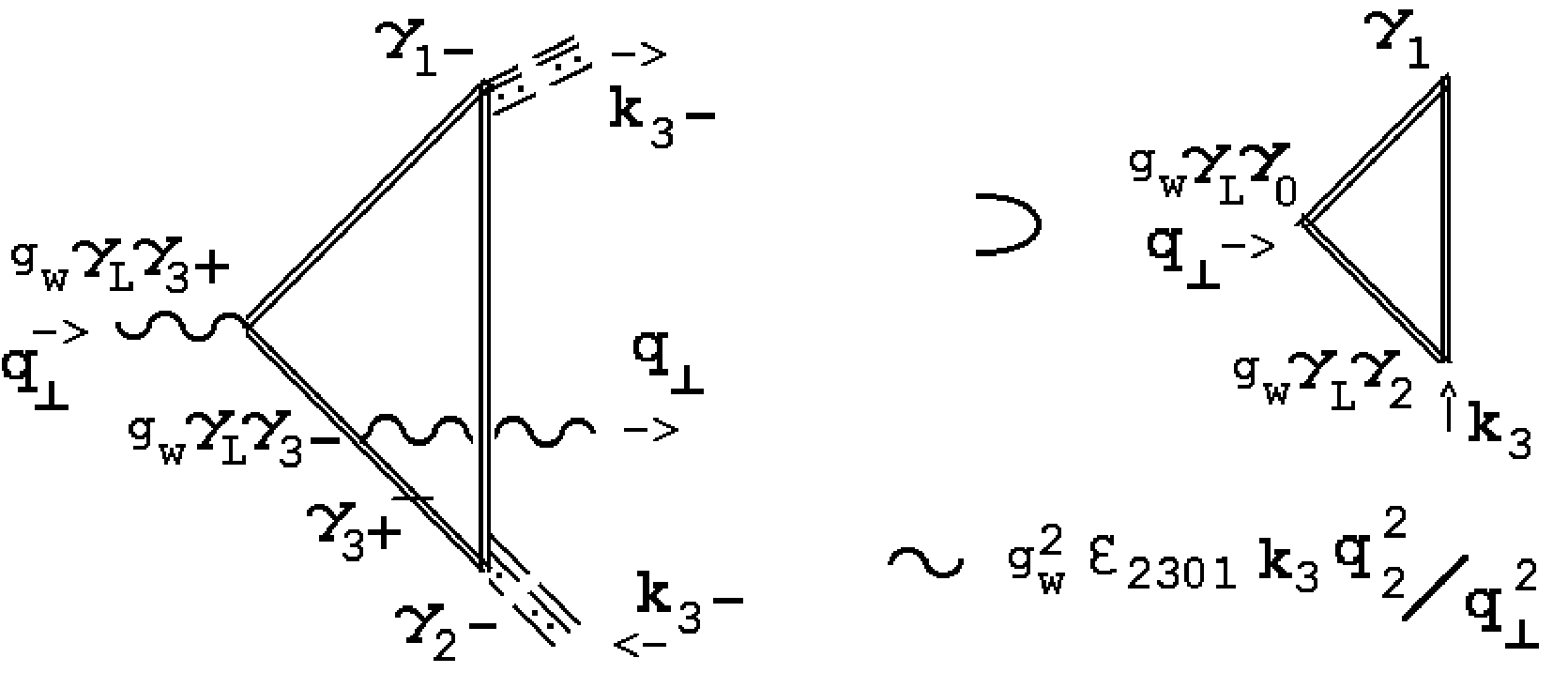}

Fig.~17 The anomaly pole contribution. 
\end{center}
(Again the notation is the same 
as in previous diagrams, except that we have introduced $\gamma_L = 1 + \gamma_5$.)
The $g_w \gamma_L\gamma_{3^{\pm}}$ couplings appear because 
the $W$ is exchanged over a large rapidity interval. The $\gamma_{1^-}$
and $\gamma_{2^-}$ couplings are similarly determined by the wee gluon
kinematics.

If we add the two diagrams shown in Fig.~16, and integrate over the
wee gluon momentum $k_3$, we produce a $W$ mass of the form
$$
M_W^2 ~\sim~ \frac{q_1^2 + q_2^2}{q_{\perp}^2} ~ g_W^2 ~\int dk_3 ~k_3
~=~ g_W^2 ~\int dk_3 ~k_3
\auto\label{mw}
$$
That there is actually no pole at $q_{\perp}^2 = 0$ is consistent with
our argument in Section 2 (and Appendix B) that the on-shell residue of an
anomaly pole is finite only in an infinite momentum frame. Nevertheless, the 
quantum numbers at each vertex of the triangle diagram producing the denominator 
pole are identical to those of the 
effective triangle diagrams discussed in Section 2.
Therefore, the masss generation can be interpreted as due to the direct coupling 
of a $W$ to a Pion (or pion) just as anticipated in Fig.~15. 

As discussed in the previous two Sections, wee gluon momentum factors
are generally scaled by a mass factor 
($M$ in $CSQCD_S$ or $\mu$ in $QCD_S$). However, 
because the diagrams of Fig.~16 contain only perturbative $W$
vertices in addition to the wee gluon couplings (with no 
longitudinal interaction of the form of (\ref{lint}) ), the wee gluon momentum 
factor produced by the coupling to the anomaly diagram is not scaled by 
such a mass factor. As a result, the mass (\ref{mw}) that is 
obtained is a direct reflection of the wee gluon momenta involved together
with an overall normalization factor that will be
determined by the color factors associated 
with the sum over all wee gluon couplings to the quark loop involved.
Since this color factor will be
different for triplets and sextets, we can write the mass obtained
from all quark loop interactions of the form of Fig.~16 as 
$$
M_W~=~ g_w^2~F_{\Pi}^2 ~+~ \Sigma_{\pi 's} ~ g_w^2~F_{\pi}^2
\auto\label{mass}
$$
and consider this to be a definition, for our purposes, of 
both $F_{\Pi}$ and $F_{\pi}$.

We will discuss the relative magnitude of $F_{\Pi}$ and $F_{\pi}$ shortly.
First, however, we note that the mass (\ref{mass}) appears only
for vector bosons with a purely left-handed coupling. 
The ``LLV'' structure of the triangle
diagram in Fig.~17 gives an anomaly, whereas if 
the $W$ couplings were purely vector we would have 
a ``VVV'' structure and no 
anomaly. Similarly, if the coupling were purely axial vector we would have an 
``AAV'' structure and, again, no anomaly. Hence, 
if we now introduce the Standard Model hypercharge gauge field $Y$,
with couplings as discussed above, 
the above mass generation mechanism will apply also to 
the left-handed component of $Y$. We, therefore, obtain exactly the mass 
generation pattern of the Standard Model and there is no mass for the photon.
(Note that photon exchange exchange will be accompanied by a background 
axial vector wee gluon interaction.)

To discuss the contribution of wee gluon color 
factors to $F_{\Pi}$ and $F_{\pi}$, it will be simpler to 
make the transition from $CSQCD_S$ to $QCD_S$. As we have discussed in the previous
Section,
the wee gluons will no longer be a simple condensate and the $W$ mass generated
by wee gluon interactions will be a much more complicated dynamical effect.
Nevertheless, we can continue to define $F_{\Pi}$ and 
$F_{\pi}$ by (\ref{mass}).

The large sextet color factors  surely imply that $F_{\Pi}$ is much larger 
than $F_{\pi}$. A common expectation,
based on Feynman graph color factors, is that 
triplet and sextet quark momentum scales for gluon interactions
will be related (approximately) by the ``Casimir Scaling'' rule. 
This rule would say that $F_{\Pi}$ and $F_{\pi}$ 
should be related by 
$$
C(6)~\alpha_s (F_{\Pi}^2)~\sim ~C(3) ~\alpha_s(F_{\pi}^2)
\auto\label{cas}
$$
where $C(3)$ and $C(6)$ are Casimirs for triplet and sextet quarks respectively.
For SU(3) there are two Casimir operators which are (representation
dependent) multiples of the identity.
In terms of the generators $G_a$, these operators can be 
written as 
$$
C_2~=~G^2~ \sim ~f_{abc}G_aG_bG_c ~,~~~~ C_3~\sim~d_{abc}G_aG_bG_c
\auto\label{cas1}
$$
and since 
$$
C_2(6)/C_2(3)~=~ 5/2~, ~~~~ C_3(6)/C_3(3)~=~ 7/2
\auto\label{cas2}
$$
we can say
$$
C(6)/C(3) ~\approx~ 3
\auto\label{cas3}
$$

To apply (\ref{cas}) to the real world
we must use the physical $\alpha_s$ that is defined via ``low-energy''
QCD, with the sextet sector 
integrated out and with the physical quark masses in place.
In this case, if $\alpha_s$ evolves as slowly as is commonly believed 
(e.g. $\alpha_s (F_{\pi}^2)
\sim 0.4~$), the order of magnitude of $F_{\Pi}$ will indeed
be the electroweak scale! We conclude, also, that the sextet quark Pions will 
dominate the mass generation for $W$ bosons, as anticipated in Fig.~15, 
and we can effectively ignore the triplet quark contribution.

We can look at the Casimir scaling rule (\ref{cas}) in two complimentary ways.
We can use it, as we just did, to obtain directly the relative magnitude of triplet 
and sextet factors with a momentum dimension, on the basis that this is entirely
controlled by the evolution of $\alpha_s$. More directly,
we can say that in going
from triplet to sextet graphical 
contributions, $\alpha_s$ is effectively replaced by 
$\{C(6)/C(3)\}~ \alpha_s$. (An explicit example of this is provided by the
$\beta$-function calculations described in Appendix A.) In this case, we 
can say that the large
factor of $F_{\Pi}^2$ that appears in the $W$ mass results from the color
factors associated with the product of the two
wee gluon couplings, in the diagrams of Fig.~16, 
to the sextet quark loop involved. 
Since, essentially, the same color factors and wee gluon interactions will
be involved, we conclude that the wee gluon coupling
that provides the coupling of  
the wee gluon component of the pomeron to a sextet quark loop (in an
anomaly pole amplitude) similarly,
has the order of magnitude of $F_{\Pi}$. This tells us,
as we shall see explicitly in the following, that the pomeron will couple
very strongly to the electroweak sector, even though the states are very massive.

\newpage

\mainhead{5. SEXTET PIONS AT HERA} 

We begin our discussion of 
the hard diffractive production of vector bosons ($W$'s and $Z$'s) via sextet
pions by discussing deep-inelastic diffractive scattering in this Section.
As we anticipated in the previous Section, the strong coupling
of the sextet sector to wee gluons will be directly evident in the coupling of this
sector to the pomeron and, as we show below, we can begin 
to estimate cross-sections by utilising the generation
of a sextet pion via an anomaly pole. Because the produced vector
boson carries a large longitudinal momentum and it
is longitudinally polarized it has, as we will explain below, 
an enhanced probability (compared to a 
transversely polarized vector boson) for decay to a jet pair 
that are sufficiently close together, in phase space, to appear
as a single massive jet.
In the kinematical situation at HERA, this is particularly 
difficult to detect unambiguously. We are encouraged, nevertheless, by the  
fact that the features of the most dramatic 
large $x$ and $Q^2$ event presented\cite{ZEUS} by ZEUS, in the original paper
highlighting such events, are such that we are able to argue that 
a $Z^0$ may, indeed, have been produced.

\subhead{5.1 Diffractive Hard Interactions }
 
A sextet pion can be directly produced via a  
hard interaction of the pomeron with a color neutral $ \gamma$, $Z^0$ or $W^{\pm}$.
When no pomeron self-interactions are 
involved, it should be reasonable to treat the wee gluon component
of the pomeron as a condensate, as discussed in Section 3.
In this case, as illustrated in Fig.~18, the pomeron can provide directly 
the wee gluon component that is needed for the sextet pion to appear via 
an anomaly pole.
\begin{center}
\epsfxsize=4.7in
\epsffile{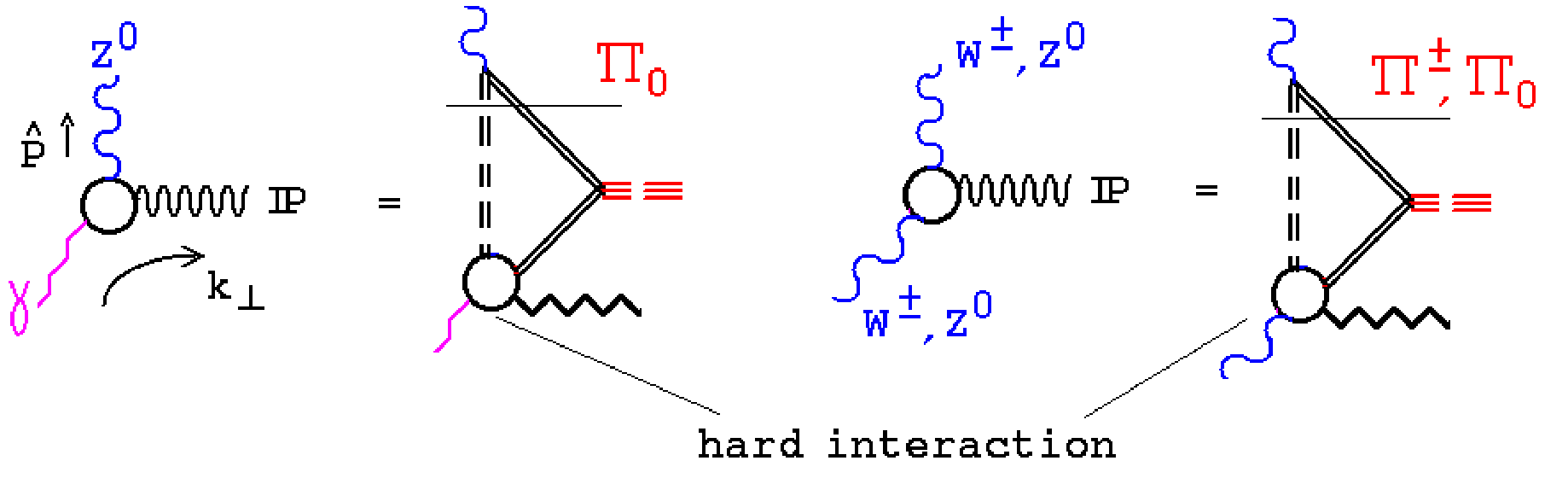}

Fig.~18 Hard diffractive interactions.
\end{center}
(We use the same diagrammatic notation as in Section 2.) 

At moderate and low $Q^2$, deep-inelastic scattering 
is dominated by photon exchange. 
To see the sextet pion process, we will require
large $x$ and $Q^2$ and, in fact, $Q^2$ will be sufficiently large that
$Z^0$ exchange, in the neutral current, and $W$ exchange, in the charged
current will give equally large (or even larger) contributions. 
In the following we will specifically discuss interactions initiated by a photon
and only occasionally refer to the fact that the photon could equally well be
a $W^{\pm}$ or a $Z^0$. 

The simplest photon interaction  
that is effectively pointlike at large $k_{\perp}$ and has the right
$\gamma$ - matrix structure to produce an anomaly pole, is shown in Fig.~19(a). 
\begin{center}
\epsfxsize=2.2in 
\epsffile{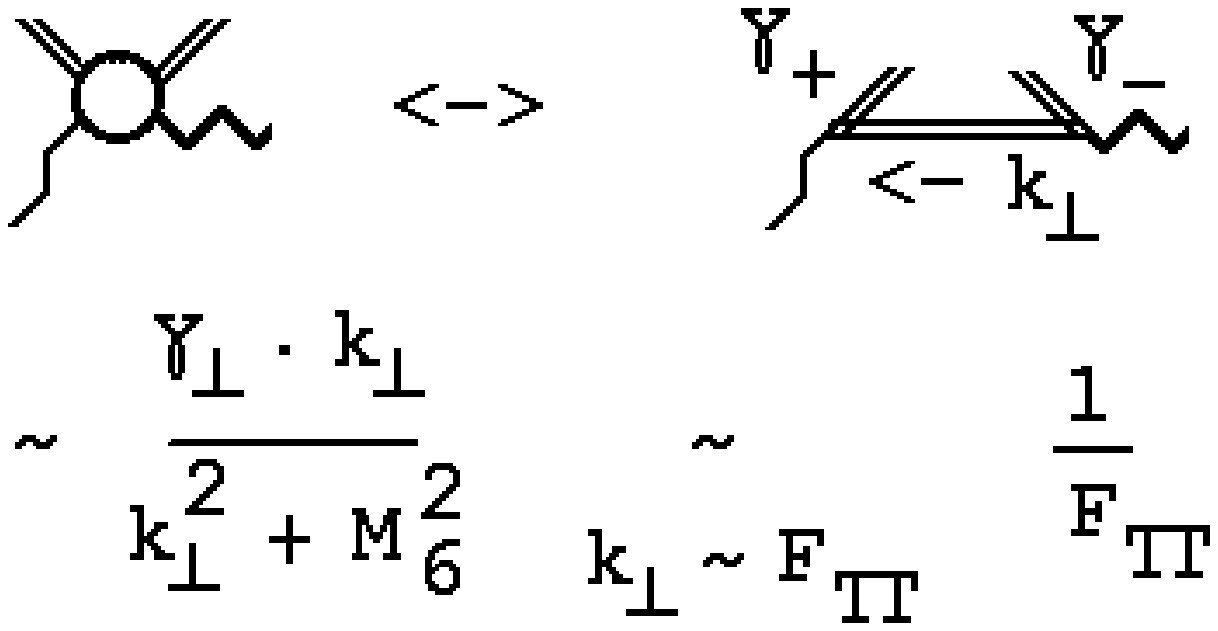}
\hspace{0.9in}
\epsfxsize=1.2in 
\epsffile{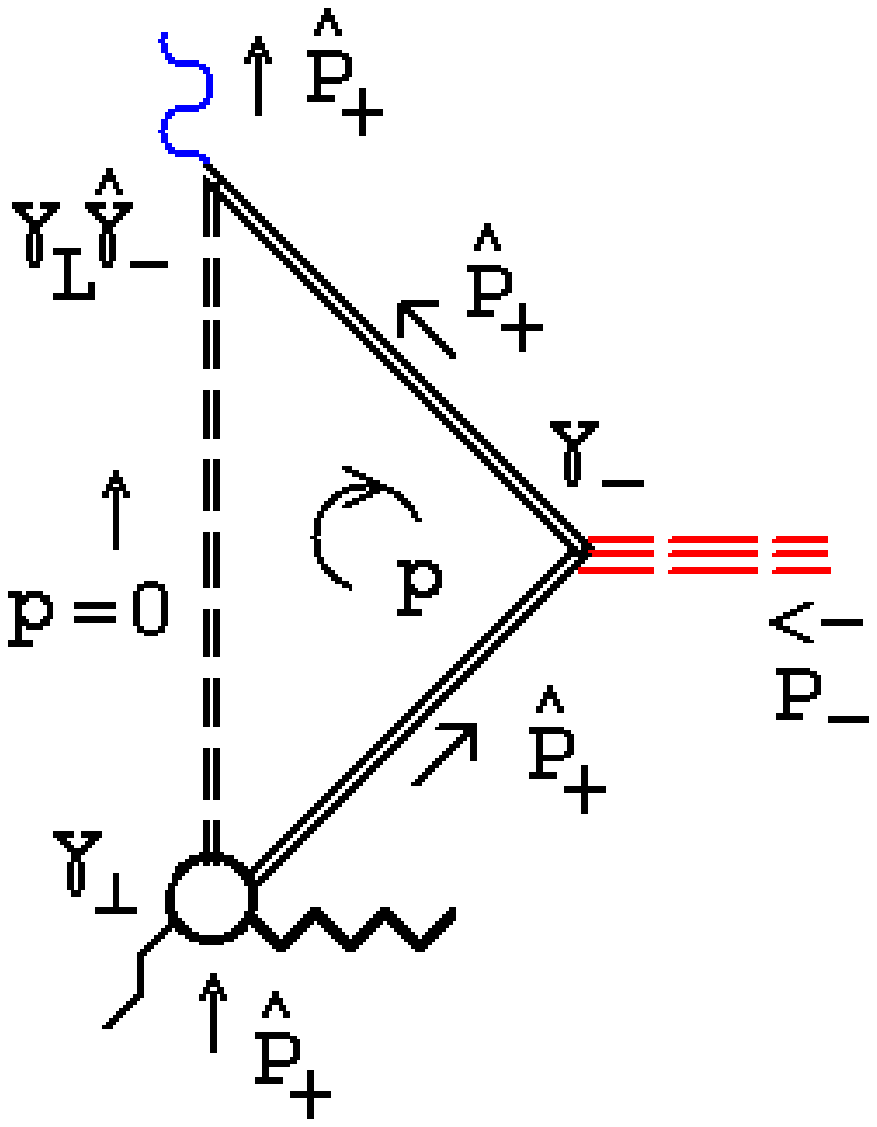}

(a)\hspace{2.5in}(b)

Fig.~19 (a) The hard interaction (b) the anomaly pole diagram.
\end{center}
($M_6 $ is a dynamical sextet quark mass that we take to be $\sim F_{\Pi}$.)
The resulting anomaly pole diagram is shown in Fig.~19(b). 
We will see that, in addition to the large $k_{\perp}$,
the hard gluon in the pomeron must also carry a large light-like momentum. 

To obtain an anomaly pole amplitude via a finite on-shell residue we 
should, in principle, go to the infinite momentum frame of the produced pion.
In addition, the anomaly pole description is valid only when
when the $\Pi$ is on mass-shell, with zero mass. However, to produce
a $Z^0$, and not a $\Pi$ on mass-shell, it should
be reasonable to use the finite 
momentum anomaly pole amplitude, initially defined close to the Pion mass-shell, 
and continue that towards the $Z^0$ pole.  

\subhead{5.2 Diffractive Deep-Inelastic Scattering}

The anomaly amplitude shown in Fig.~19(b) gives the contribution
to deep-inelastic diffractive jet production illustrated in Fig.~20.
\begin{center}
\epsfxsize=1.7in
\epsffile{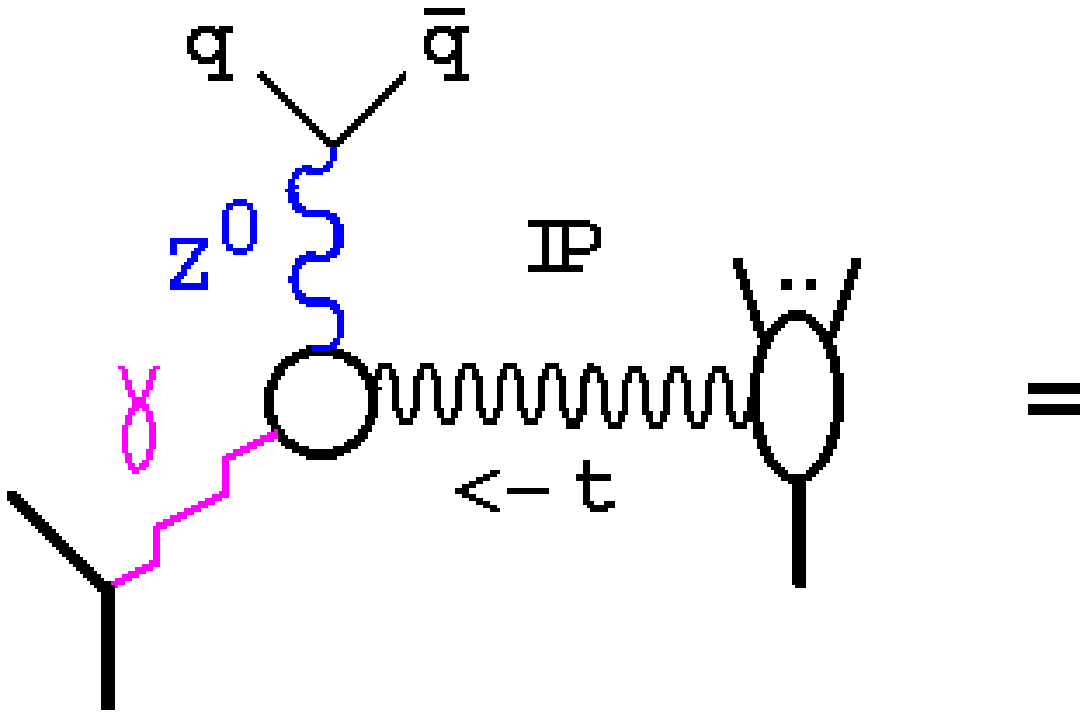}
\epsfxsize=1.7in
\epsffile{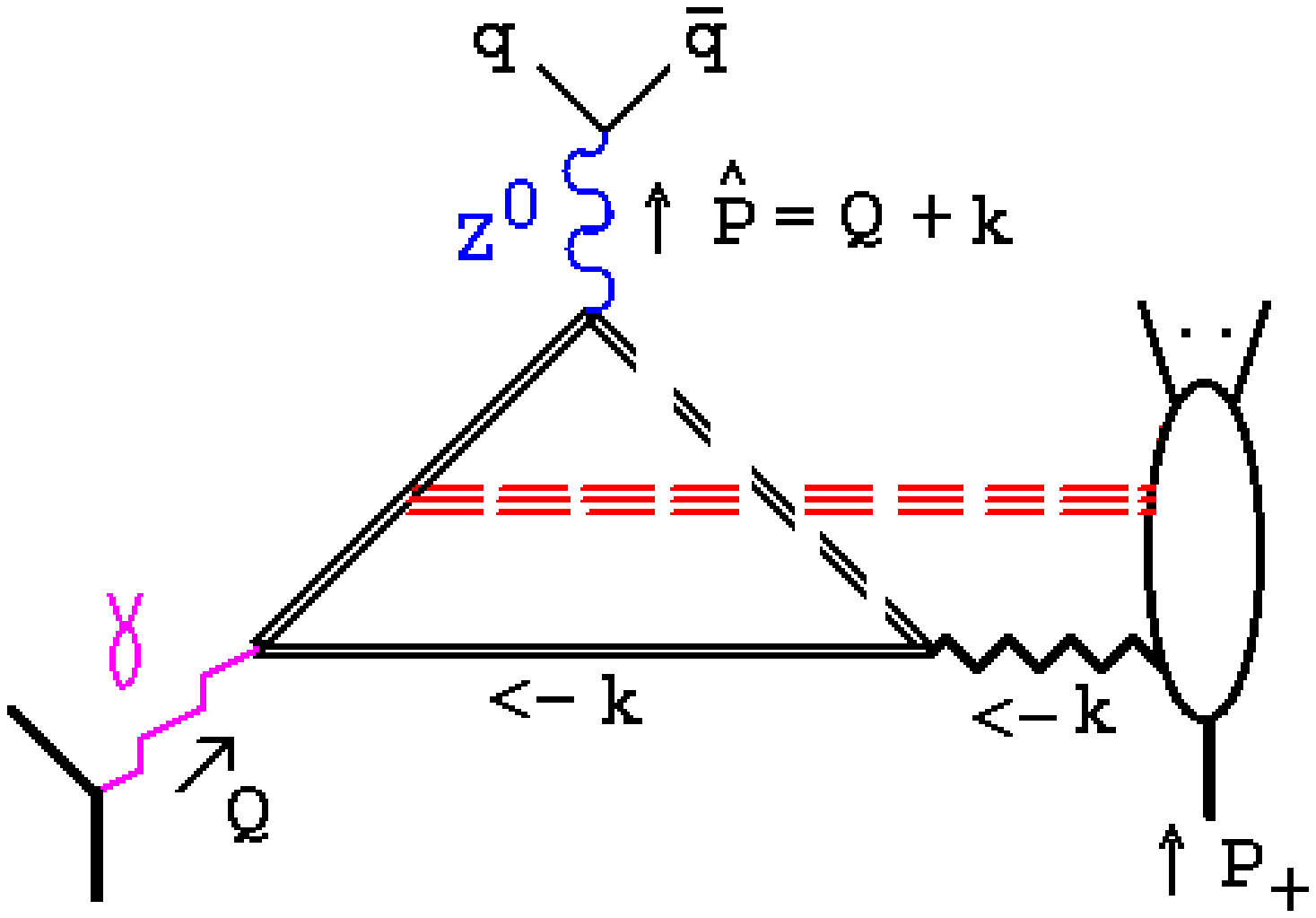}

Fig.~20 Deep-inelastic diffractive jet production.
\end{center}
Using the kinematic notation shown in Fig.~19(b) and Fig.~20, we initially
take $\hat{P}= Q+k$ to be light-like ($= \hat{P}_+$, as in Fig.~19(b) )
but very importantly, because $Q$ is spacelike, the light-cone is not parallel
to that defining $P_+$ and $P_-$. In this case, 
with the $\gamma$-matrix couplings appearing
in Fig.~19(b), the anomaly amplitude has a contribution with the kinematic form
$$
\Gamma_{\perp \hat{n}_- -} ~\sim ~ \frac{\hat{P}_+P_-\hat{P}_+}{P_-
\hat{P}_+} ~= ~\hat{P}+
\auto\label{spa}
$$
where $\hat{n}_-$ is the light-cone vector orthogonal (in the euclidean sense)
to $\hat{P}_+$. Again there is no anomaly pole. Instead, the
effect of this pole is that the amplitude is independent of the wee gluon
momentum $P_-$.  
Therefore, the anomaly pole wee gluon coupling 
will produce a simple integral over the wee gluon
distribution that, for the reasons discussed in the last Section,
we take to be $\sim ~ F_{\Pi}$. Combining (\ref{spa}) 
with this coupling 
and with the $Z^0$ propagator and vertices $g_w$, and extrapolating away from
$\hat{P}^2=0$ by introducing $\hat{P}_-$, as a component of $Q+k$, gives 
$$
F_{\Pi} 
\hat{P}_+ g_w^2~ \frac{(g_{-\nu} - \hat{P}_-\hat{P}_{\nu}/M^2)}{
(\hat{P}^2-M^2)}
~=~- \frac{\hat{P}_-}{F_{\Pi}}\frac{\hat{P}^2}{\hat{P}^2 -M^2}
\delta_{-, \nu}
~-  ~\frac{\hat{P}_+}{F_{\Pi}}\delta_{+, \nu}
\auto\label{fspa}
$$
where $M$ is now $M_{Z^0}$ (but would be $M_W$ if we were discussing $W$ production)
and we have used $M= g_w F_{\Pi}$. (All light-cone co-ordinates are
now defined relative to the $\hat{P}$ light-cone.)
The first term in (\ref{fspa}) is present as soon as $\hat{P}^2 \neq 0$.
It produces a physical, longitudinal, $Z^0$. 

The second term in (\ref{fspa}) has no pole, but it is 
of comparable magnitude away from the pole and (when $\hat{P}_-$ is small)
it gives a direct coupling 
to fermion final states that is proportional to their mass.
Note that there is no explicit $g_w$ dependence in 
(\ref{fspa}) and when $\hat{P}_+ \sim \hat{P}_- \sim F_{\Pi}$ both terms are
$O(1)$. Therefore, at the electroweak scale, the anomaly 
amplitude produces electroweak final states with no electroweak suppression.

\subhead{5.3 Comparison With a Jet Amplitude}

At first sight, as illustrated in Fig.~19(a), the hard interaction that 
helps produce the anomaly amplitude gives a suppression $O(1/F_{\Pi})$ 
at the electroweak scale. 
However, as we discuss now, this is the
natural order of magnitude for a normal two jet amplitude 
that does not involve sextet pion production. 

We consider the two jet amplitude involving gluon exchange shown in Fig.~21(a),
and consider the two production vertices shown in Fig.~21(b). 
(Once again the photon could be replaced by either a $Z^0$ or a $W^{\pm}$.) 
The first vertex shown is a lowest-order amplitude
involving quark exchange. The second is a loop amplitude that gives
the lowest-order triplet sector vertex for $Z^0$ production. 
Considering (\ref{fspa}) to simply give a factor that is O(1), if we compare the 
the quark exchange amplitude with the  
anomaly amplitude of Fig.~20, the first difference is that in the 
jet amplitude a triplet quark propagator,  
carrying momentum $P_j$, replaces the hard interaction
in the triangle diagram amplitude. However, provided $|P_j|\sim |k_{\perp}|$, 
this will simply give the ``natural'' order of magnitude for a jet amplitude that we
referred to above. 
\begin{center}
\epsfxsize=1.6in
\epsffile{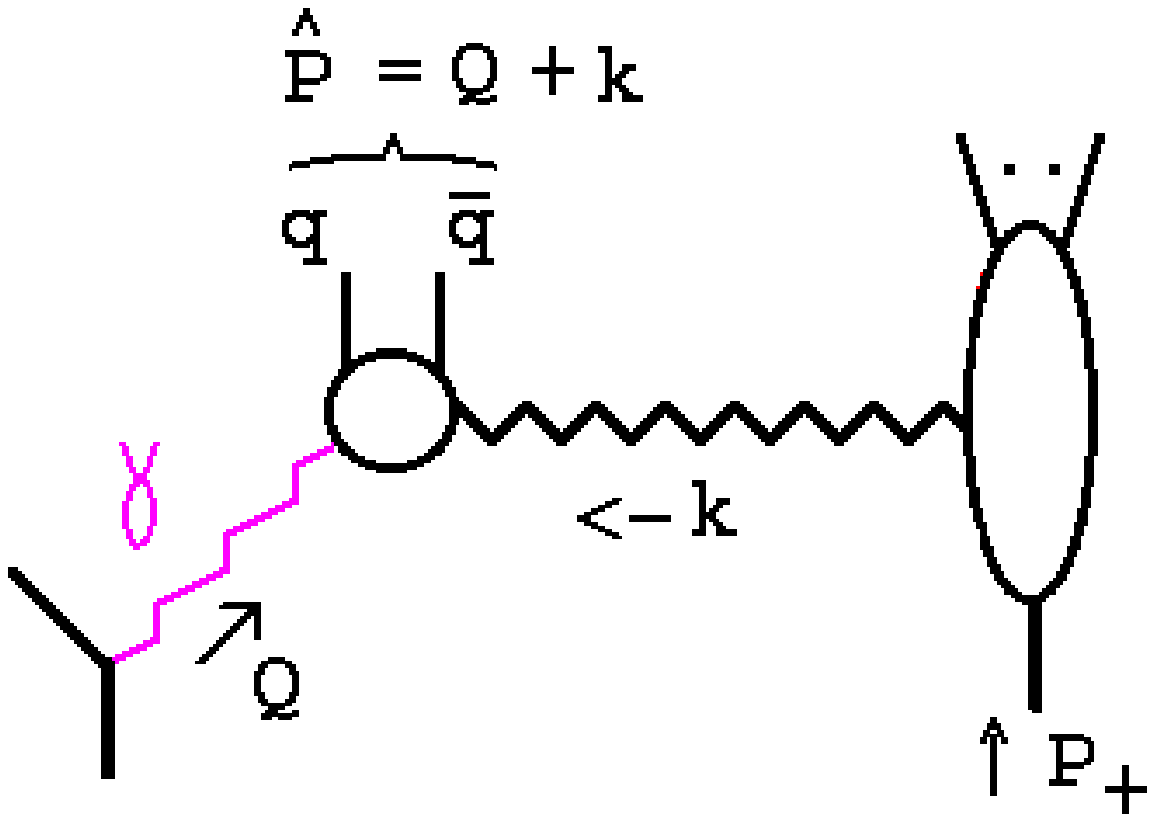}
\hspace{0.4in}
\epsfxsize=3.1in
\epsffile{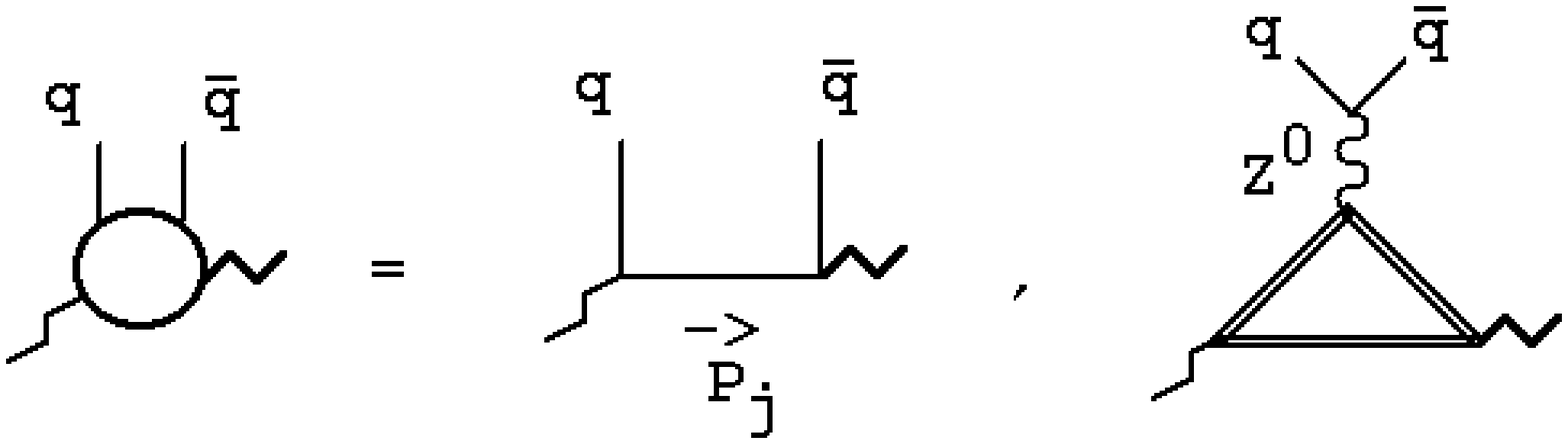}

(a)\hspace{2.5in}(b)$~~~~~~~~$

Fig.~21 (a) A two jet amplitude (b) production vertices
\end{center}

The second
difference between the jet amplitude and the anomaly 
amplitude is that a regular gluon (parton) vertex 
replaces the pomeron vertex (that is the reggeized gluon plus wee gluons vertex),
giving a reduction by a factor of $\approx 1/3$ in the amplitude. 
This will be compensated, in part, by the appearance of a
sextet quark coupling (relative to a triplet coupling). Therefore, when all the 
momenta involved are electroweak scale in magnitude, the diffractive production
of jet pairs via $Z^0$ production will give a comparable cross-section to that for
conventional (non-diffractive) two jet production.

To emphasize the (relatively) large magnitude of the diffractive production 
amplitude we are discussing, we 
consider corresponding cross-sections for $Z^0$ production when only the triplet
sector is present. First, we consider the anomaly pole mechanism. In this case,
the factor of $F_{\Pi}$ in the numerator of the left side of (\ref{fspa}) is 
replaced by $F_{\pi}$ and there is a reduction in cross-section of 
$~ \centerunder{\raisebox{1mm}{$ >$}}{$\sim$}
 ~O(10^6)$. If we instead consider
the one loop production vertex of Fig.~21(b), the
factor of $F_{\Pi}\hat{P}_+$ in the left side of (\ref{fspa}) is 
replaced by a factor of $\mu^2$, where $\mu$ is the triplet sector momentum
scale, leading to a much greater reduction of the cross-section.

\subhead{5.4 The Angular Distribution of Produced Jets and Leptons}

A high momentum longitudinal $Z^0$ (as, potentially, produced at HERA)
is more likely, than a transversely
polarized $Z^0$, to produce a jet or lepton pair 
that are sufficiently close together, in phase space, to appear
as a single massive jet. We will show this by 
comparing infinite momentum limits in the two cases.

If we denote the (four-) momentum vectors of the produced fermions 
by $X$ and $Y$, then if the $Z^0$ momentum is
$$
P_{Z^0}~=~(P_+ + P_-,~ P_+ - P_- ,~ 0,~ 0)~, ~~~~ where ~~~ 4P_+P_-~ =~M^2
\auto\label{mz0}
$$
the most general form for $X$ and $Y$ is
$$
\eqalign{X~&=~ (\lambda P_+ + (1-\lambda)P_-,~  
\lambda P_+ - (1-\lambda)P_-,~ p_{\perp},~0) \cr 
Y~&=~ ((1-\lambda) P_+ + \lambda P_-,~
(1- \lambda) P_+ - \lambda P_-,- p_{\perp},~0)
}
\auto\label{XY}
$$
where $0 \leq \lambda \leq 1$. The mass of both fermions is given by
$$
m_f^2~=~ 4 \lambda (1-\lambda) P_+ P_- ~- ~p_{\perp}^2
= \lambda (1-\lambda)~M^2 ~- ~p_{\perp}^2
\auto\label{mf}
$$  
This notation will be convenient for our purposes, even though it
obscures the fact that we could obtain all momenta via 
a boost from the rest frame of the $Z^0$. In this frame, the only variable would be
the angle between the transverse momenta of the fermion pair and the
direction in which the $Z^0$ is to be boosted. This is, of course, why
$p_{\perp}$ and $\lambda$ are related via (\ref{mf}).

We consider first a transverse coupling which, for the purpose of
$\gamma$-matrix manipulations, we write in the form
$$
<Y|n_{\perp}. \gamma_{\perp}|X>
\auto\label{XYp}
$$
where $n_{\perp}$ is a unit transverse vector. Suppose, first, that 
$P_+$ is so large that both $P_-$ and $p_{\perp}$ can be neglected.
Using the Dirac equation for $|X>$ then gives
$$
<Y|n_{\perp}.\gamma_{\perp}|X> ~ 
\sim~ \frac{\lambda}{1-\lambda}~<Y|n_{\perp}.\gamma_{\perp}|X>
\auto\label{XYp1}
$$
$$
\implies ~~<Y|n_{\perp}.\gamma_{\perp}|X> ~=~0
\auto\label{XYp2}
$$
except, possibly, when $\lambda= (1-\lambda) = 1/2 $.
Not surprisingly, we have to add transverse momentum in order to get 
substantial information about how a transversely polarized $Z^0$ will decay. 

If we repeat the above
manipulation keeping the transverse momentum dependence we obtain
$$
\eqalign{<Y|n_{\perp}.\gamma_{\perp}|X>&\sim ~ <Y|~n_{\perp}.\gamma_{\perp}
(\lambda \gamma_- P_+ + \gamma_{\perp}.p_{\perp})/m_f~|X> \cr
&\sim \frac{\lambda}{(1-\lambda)}<Y|n_{\perp}.\gamma_{\perp}|X>
+ \frac{(1-2\lambda)}{(1-\lambda)}<Y|p_{\perp}~n_{\perp}|X>/m_f + \cdots
}
\auto\label{XYp3}
$$
The additional terms cancel if we add the corresponding equation obtained by 
reversing the roles of $\lambda$ and $(1-\lambda)$. 
(Note that $(1-2\lambda)$ changes sign under 
$\lambda \leftrightarrow (1- \lambda)$ but, also, $m_f \leftrightarrow -m_f$.)
We then obtain the simple result
$$
<Y|n_{\perp}.\gamma_{\perp}|X> ~\sim ~-~
<Y|p_{\perp}.n_{\perp}|X>/m_f 
\auto\label{XYp4}
$$

We learn from (\ref{XYp4}) that a transverse $Z^0$ decays to fermions with
transverse momenta determined by the polarization. The amplitude is a maximum
when $|p_{\perp}|$ is a maximum which, from (\ref{mf}) occurs when 
$\lambda = 1/2$. In this case, the fermions symmetrically carry one half of the
light cone momenta of the $Z^0$. It is a smooth maximum, however, and so there 
is a significant probability that the $Z^0$ will decay into an asymmetric
configuration.

If we repeat the above discussion for the longitudinal polarization we obtain a
non-zero contribution already in the first manipulation, i.e.
$$
\eqalign{<Y|~\gamma_+~|X>~&\sim ~ <Y|~\gamma_+
\lambda \gamma_- P_+ /m_f~|X> \cr
& \sim~ \frac{\lambda}{1-\lambda}~<Y|~\gamma_+|X>
~- ~2\lambda ~<Y|~P_+~|X>/m_f
}
\auto\label{XYp5}
$$
giving
$$
<Y|~\gamma_-~|X>~\sim ~ - \frac{2\lambda (1-\lambda)}{m_f(1-2\lambda)} <Y|~P_+ ~|X>
\auto\label{XYp6}
$$
Now the symmetric case, with $\lambda =1/2$, is strongly enhanced. Although,
because terms that are non-leading as $P_+ \to \infty$ will also be singular as 
$\lambda \to 1/2$, we 
can use (\ref{XYp6}) only if we stay away from $\lambda =1/2$. It is,
nevertheless, sufficient for us to conclude that, at large momentum,
the symmetric configuration with two jets (or leptons) close together in 
phase space is enhanced for a longitudinal $Z^0$ decay, compared to the transverse
case. In general, the final result may often look 
like a broad single jet.

\subhead{5.5 HERA Kinematics}

For most of our discussion we will 
take both the proton and the positron to be massless.
We denote the momentum of the proton beam by $E_p$ and the momentum
of the positron beam by $E_e$. If we write the photon (or $Z^0$, or $W^{\pm}$)
momentum as 
$$
Q~=~(Q_+ + Q_-,~ Q_+ - Q_-,~ Q_{\perp})
\auto\label{Qcp}
$$
then  the light-cone components 
$Q_+$ and $Q_-$ are determined, at fixed $x$ and $Q^2$, 
by the mass-shell condition for the scattered positron, i.e.
$$
0~=~ 4 p_- Q_+ ~- ~ Q^2 
\auto\label{mse}
$$
and
$$
x~=~ \frac{Q^2}{4 P_+ Q_- }
\auto\label{xQ2}
$$
Solving for $Q_{\perp}^2$, we obtain
$$
Q_{\perp}^2~=~Q^2 ~-~\frac{Q^4}{Sx}~=~(1-y)Q^2
\auto\label{qpr}
$$
where $S = 4 P_+ p_-$ and $xyS=Q^2$. 

(\ref{qpr}) shows 
that large $Q_{\perp}$ requires both large  $x$ and  $Q^2$. 
With $E_p = 820$ GeV and $E_e = 27.5$ GeV (the original HERA values) 
we can obtain  $~ Q_{\perp} \sim ~ 100~ GeV$ with
$Q^2 ~\sim ~ 30,000~GeV^2$ and $~x~\sim ~ 0.5$.
However, if (in the notation of Fig.~20)
we also require that $k_{\perp} \sim~ 100~ GeV~$
and $~\hat{P}^2 ~\sim ~M_{Z^0}^2~ $ then, not surprisingly, it is
very difficult to have all conditions satisfied. First, it is necessary
for $k$ to have a very large light-cone component to put the
$Z^0$ on-shell. We then find that to keep the diffractively excited proton state
physical we must have
$ |k^2|~=~|t|~ \centerunder{$>$}{$\sim $} ~2k_{\perp}^2~ \sim~  20,000~GeV^2$.
In this case, the jet cross-section we are comparing with
will be far too small to be observable.

\subhead{5.6 Small $t$ Scattering}

We can extend the foregoing discussion with an argument that we will also
apply to other diffractive amplitudes in later Sections. According to
our analysis, the $QCD_S$ pomeron is essentially a regge pole and so has, 
approximately, the factorization properties of a regge pole all the way from
electroweak scale values of $|t|$ down to $|t| \sim 0$. The regge behavior is 
manifest at large $|t|$ via the reggeized gluon that gives the kinematic 
properties of
the hard pomeron that we have been discussing and this will match smoothly
with a soft pomeron regge pole as $|t|$ decreases. 
(Note that $t$ can be small
even though a large light-like momentum is exchanged.) Since we anticipate
that the ``non-perturbative'' 
$\gamma Z^0~ \pom~$ vertex is entirely due to electroweak scale dynamics it should
vary only slowly with $|t|$ (with a 
scale determined by $F_{\Pi}$). However, the proton/pomeron coupling will be the 
normal hadronic coupling and will increase exponentially fast as  
$|t|$ decreases. It is difficult to know how large this increase will be, since 
there are no measurements of this coupling for $|t| \sim ~20,000 ~GeV^2~ !$ 
We do know that
the cross-section for proton elastic scattering, which involves the square of
the coupling that we are interested in, 
decreases by five orders of magnitude between zero and 
$|t| \sim ~1~GeV^2$, and by another five orders of magnitude between 
$|t| \sim ~1~GeV^2$ and $|t| \sim ~10~GeV^2 $. 
The mass-shell condition for the proton to scatter elastically, with a large
longitudinal momentum exchanged, is $|t|~\centerunder{$<$}{$\sim$} ~2-3 ~GeV^2$.
With the 
increase by orders of magnitude as $|t|$ decreases, if we are also close 
to the $Z^0$ pole, the resulting cross-section may well be observable.

\subhead{5.7 What Can be Seen at HERA ?}

In the original ZEUS paper\cite{ZEUS} five events were highlighted which all had
relatively large $x$ and $Q^2$. 
We have identified an electroweak scale $|Q_{\perp}|$
as necessary for sextet pion $Z^0$ production and four events had $|Q_{\perp}|
\centerunder{$>$}{$\sim$} 100$ GeV. Although subsequent ZEUS data\cite{ZEUSa} 
appear to show that the $e^+p$ cross-section 
at large $x$ and $Q^2$ (up to, and including, $Q^2 = ~30,000~GeV^2$)
is not substantially above that predicted by the Standard Model, H1 data  
give a different impression.
The published H1 cross-section\cite{H1a} for the neutral current 
at $Q^2 = ~30,000~GeV^2$ (and the charged current at lower $Q^2$) seems to be 
significantly above the Standard Model value. Therefore,
it remains possible that some fraction 
of the original five ZEUS events (particularly at 
$Q^2> ~30,000~GeV^2 $) and, presumably, subsequently observed events, are due 
to a non Standard model process. In fact, as we now discuss,
only the largest $Q^2$ event clearly has a high 
probability to have resulted from $Z^0$ production. 

In each event there is a clear jet and the
essential question is whether it could have been 
a massive jet produced by a $Z^0$. We will make use of the fact 
that two, a priori independent, reconstruction 
methods are used to measure both $Q^2$ and $x$
and the results from both are quoted separately for each event. 
The first method,
called the ``double-angle'' (DA) method uses only the measured angles of the jet
($\gamma$) and the electron ($\theta_e$), together with the formulae
$$
 x_{DA}~=~ \frac{E_e}{E_p}\frac{sin \gamma}{(1-cos \gamma)}
\frac{sin \theta_e}{(1-cos \theta_e)}~,~~ 
y_{DA}~=~\frac{sin\theta_e (1-cos \gamma)}{sin \gamma 
+ sin \theta_e -sin(\gamma + \theta_e)}~,
\auto\label{dba}
$$
and $~Q^2_{DA}= s x_{DA} y_{DA}~$. 
The other ``positron'' method uses only the measured positron energy $E_e'$ and 
the angle $\theta_e$, together with the formulae
$$
x_e~=~ \frac{E_e}{E_p}\frac{E_e'(1 + cos_e)}{2E_e - E_e'(1-cos \theta_e)}~,
~~ y_e~=~1~-~ \frac{E_e' }{2E_e}~(1 - cos\theta_e)
\auto\label{psm}
$$
and, again, $~Q^2_e = s x_e y_e~$.
Although this second method is much more direct,
because of the difficulty of measuring $E_e'$ reliably, the double 
angle method is generally regarded as more reliable for discussing
large $Q^2$ deep-inelastic events. 

The double angle method is
predicated\cite{dam} 
on the assumption that the jet mass can be neglected. As a result, (\ref{dba})
correctly gives $x$ and $Q^2$ only when the jet is (at least approximately)
massless. Therefore, whether or not, there is agreement between the two methods
can be regarded as an indirect test of the smallness of
the jet mass. In fact, for all but the largest $Q^2$ event, there
is no significant disagreement.

\subhead{5.8 The Largest $Q^2$ Event}

This event is shown in Fig.~22. The jet is clearly very broad and, in fact, the results 
for $Q^2$ and $x$ obtained from the two reconstruction methods 
differ significantly, with the differences being outside
of the quoted errors. 
\begin{center}
\epsfxsize=6in
\epsffile{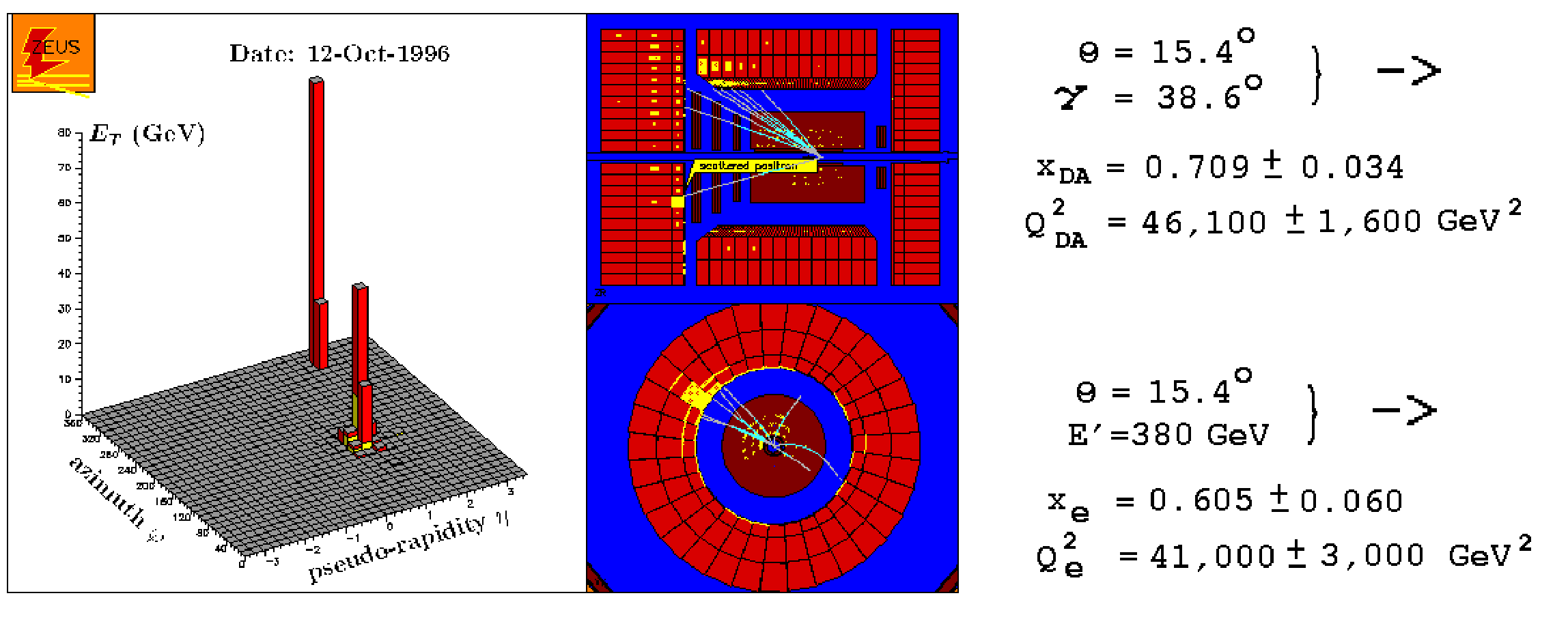}

Fig.~22 The largest $Q^2$ ZEUS event. 
\end{center}
If we reconstruct the full four-momentum $Q$
from (\ref{Qcp})-(\ref{qpr}) we obtain
$$
Q_{DA}~=~(-399,-439,-113,0)
\auto\label{qda1}
$$
and
$$
Q_e~=~(-352.5,-393.5,101,0)
\auto\label{qe1}
$$

We can regard the double-angle method as projecting the
experimentally measured calorimeter
energies and momenta onto the combination of a massless jet and an additional
momentum projected onto the measured direction of the positron. 
In effect, this is what is done by the process of 
eliminating the energy of the positron and assuming that only the angle is well
determined experimentally. The jet angle $\gamma$ is 
determined directly (under the assumption that the jet is massless). 
Following the procedure used in the ZEUS paper we determine the 
jet energy by using the fact that $p_{\perp}$ is approximately 
conserved (as is recorded in Fig.~22). 
As a result, the $p_{\perp}$ of the jet must balance that given by
$Q_e$. With $\gamma = 38.6^o$, this determines the four-momentum 
of the (assumed to exist) massless jet to be 
$$
P_j~=~ (167,126.5,101,0)
\auto\label{pjm}
$$
Taking the directly measured $Q_e$ to be correct, 
the additional momentum projected along the positron direction is 
$$
Q_e - Q_{DA}~= ~(46.5,45.5,-12,0)
\auto\label{eDA}
$$
Adding this back to $P_j$ to, potentially, obtain the true four momentum
of the produced hadronic state we obtain
$$
P_j + Q_e - Q_{DA}~=~ (213.5,172,89,0)
\auto\label{jeDA}
$$
which has a mass squared of
$$
8,077.25~GeV^2~=~ (89.9~GeV)^2
\auto\label{mjeDA}
$$
suggesting that a massive $Z^0$ jet was indeed produced.
The production angle would have been $\gamma_{Z^0}~=~ 27.4^o$
which is large enough to be detected only because $Q^2$
is so large.

If we compute the momentum transfer $k$ using
(\ref{jeDA}) for the momentum $\hat{P}$ of the $Z^0$ we obtain 
$$
k~=~P_{Z^0} - Q_e ~=~ (566,565.5,12,0) 
\auto\label{tZ0}
$$
implying (more exactly than is surely justified by all the reconstruction involved)
that the squared momentum transfer may have indeed been  
small. Thus allowing the interpretation of the event
as diffractive $Z^0$ production.

\subhead{5.9 Other Events}

It will be very interesting to determine whether the foregoing
analysis can reveal further HERA events that might be consistent with
diffractive $Z^0$ production. Although the H1 events, that were presented in the 
paper\cite{H1} that was contemporaneous with the ZEUS paper, carried large
$Q^2$ they were selected with 
different criteria and were presented from a different
viewpoint. There was an emphasis on the possibility of a large mass intermediate 
state in the electron + jet channel that led to the presentation of the kinematics
of the events in a way that makes it impossible to directly apply our analysis.
Also the search for a large mass intermediate state produced an emphasis on large
$y$, and hence low $Q_{\perp}^2$, that is counter to our purpose. In particular,
the two largest
$Q^2$ events presented (with $Q^2 \sim ~31,000~ GeV^2$) both had relatively 
small $x$ ($\sim 0.45$) and consequently had lower 
$Q_{\perp}^2$ ($\sim~ 80~ GeV$) than we would prefer for our analysis. 

Presumably, both ZEUS and $H1$ have further candidate events from 
runs subsequent to 1997. However, cross-sections 
for $Q^2 > 30,000~ GeV^2$ have yet to be published, as have any corresponding 
event pictures. 

\newpage

\mainhead{6. SEXTET PHYSICS AT FERMILAB }

\subhead{6.1 Single Diffraction}

The interactions shown in Fig.~18 will also take place in a hadron collider 
when a $Z^0$, $W^{\pm}$, or photon is emitted from a quark in a hadron.
Unfortunately, it will be very difficult to isolate these
processes because of the small cross-section involved.
However, as in our discussion of deep-inelastic scattering in the previous Section,
the $t$ dependence of the pomeron/hadron vertex implies there should 
be a ``relatively large'' forward amplitude. In fact, this interaction could  
explain the push towards larger rapidities, that is apparently 
observed\cite{d0} at the Tevatron, when 
a $W^{\pm}$ or $Z^0$ is produced in association with a large $E_T$ jet.

Diffractive production of vector boson pairs might also be possible,
although it is not clear whether the corresponding anomaly pole vertices exist. 
Apparently\cite{UA1}, there is already an anomalously large (non-diffractive) $W$ pair 
cross-section at the energy of the $S\bar{p}pS$ collider. Since, as we discuss
in the next Section, we expect
this cross-section to be really large at the LHC, it seems that an
``anomalous'' (although still relatively small) cross-section 
should surely be observed at the Tevatron. A sextet pion 
coupling might then give an unexpectedly large single diffractive component.
A complication is that detection of
events in which one of the pair decays hadronically is much more difficult 
at the Tevatron than it was at the $S\bar{p}pS$ because of the large background 
from the QCD production of $W$ (or $Z$) plus two jets. In addition, the 
vector boson pairs will be produced with much greater momentum at the Tevatron
(than at the $S\bar{p}pS$) and so the problem of the 
close together decays of longitudinal bosons will be much more significant. 

Other anomalous events, related to the single diffractive interactions,   
may also be observed. In particular, a connection between 
diffractive cross-sections and events with twice the average multiplicity density
(in rapidity) is required by the AGK cutting rules. In addition, 
the Wilson lines attached to sextet quarks 
should also generate higher associated multiplicities than triplet quark lines. 
Anomalously low multiplicity events may anticipate the higher energy 
rapidity gap cross-sections that we expect to appear. 

\subhead{6.2 Double Pomeron Exchange at the Tevatron}

Double pomeron production of $W^{\pm}$ and $Z^0$ pairs which, as we discuss in the
next Section, we expect to be a very clean signal of sextet quark physics 
at the LHC, is (probably) inaccessible kinematically at the Tevatron. 
However, a $Z^0$ can also pair with a photon to give a state
with zero sextet quark flavor. Since double pomeron production of $Z^0 \gamma$
is accessible kinematically, although there is not an obvious anomaly pole vertex,
there could be a significant
(although small, because an electromagnetic coupling is involved)
anomalous cross-section for this process at the Tevatron. Since 
there are two hadron/pomeron couplings there should also be a major
increase of the cross-section at smaller $t$. Assuming that the photon can
simply play the role of introducing sextet quark quantum numbers, 
it need not carry electroweak scale transverse momentum. 

\subhead{6.3 The $\eta_6$, $t\bar{t}$, and Large $E_T$ Jets.}

We turn now to non-diffractive sextet quark physics that might be seen (or may have
already been seen) at the Tevatron.

As illustrated in Fig.~23,
the $\eta_6$ has two anomaly couplings to wee gluons in $CSQCD_S$. 
There is both a $Q\bar{Q}$ and an SU(2) singlet gluon coupling (where 
the gluon has a non-leading helicity). 
Therefore, in $QCD_S$, the $\eta_6$ 
mixes with a pure glue state 
\begin{center}
\epsfxsize=4.5in
\epsffile{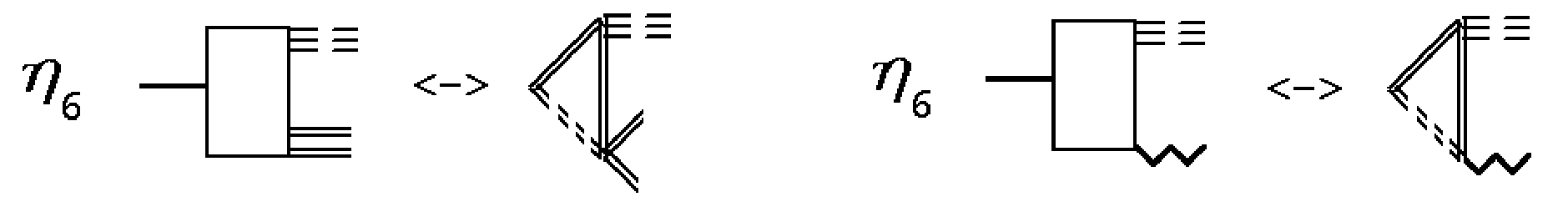}

Fig.~23 Anomaly couplings for the $\eta_6$.
\end{center}
and, as a result, we expect that it
will have an electroweak scale mass, with the sextet quark and 
antiquark carrying electroweak scale constituent masses. The $\eta_6$
will also mix, via the gluon state, with the triplet flavor singlet
(the $\eta_3$) that will be dominated by $t\bar{t}$ at the electroweak scale.

We anticipate, therefore, that the $\eta_6$ has  
an electroweak scale short-distance component which 
carries octet color that is compensated by wee gluons. This short
distance component can be produced via gluon production and, since 
sextet quarks are stable, it will decay, primarily, through 
$t\bar{t}$. Assuming that the major disparity in scales leads to a minimal 
dynamical role for the wee gluons in the process,  
$t\bar{t}$ production at Fermilab could be due to the $\eta_6$, and
could be, essentially, ``perturbatively'' calculable. This would imply, however, that
$m_{\eta_6} \sim $ ``$~2 m_t~$''. 

That top production is due to resonance production would, of course, resolve 
the paradox that the production of a confined, colored, quark can, apparently, be
observed experimentally.
Theoretically, and ``philosophically'', it would surely be attractive if
an electroweak scale mass, i.e. $2 m_t \sim 350~ GeV$, is explained
as the (dynamical) mass of a sextet
quark/antiquark bound state, rather than as (twice the value of) a 
lagrangian parameter of the triplet
quark sector. Whether a well-determined top quark
``mass'' should still be, experimentally, identifiable is not clear. 

Within $QCD_S$, the existence of a non-perturbative 
QCD sector above the ``mass'' of the top quark makes it very unlikely 
that the concept of a perturbative, electroweak scale, current quark 
mass can be well-defined enough to be directly measured. There would be
a large dynamical mass generated above the electroweak scale that, most likely, 
would make 
the concept of the current quark mass very elusive. Alternatively, if we 
identify the $\eta_6$ as responsible for top production then we can identify
$m_t$ as the sextet quark constituent mass scale. This would imply
that the sextet neutron $N_6$ has a relatively low mass of $500-600~GeV$.
As we will discuss in the next Section, this maximises the possibility
that the Cosmic Ray spectrum knee is associated with the appearance of sextet quark
states.

As detailed in Appendix A, the contribution of the sextet quark doublet
to the QCD $\beta$-function is equivalent to the contribution of ten 
triplet quarks. Consequently at the scale where (non-chiral)
sextet quarks enter the dynamics, they will 
halt the evolution of $\alpha_s$ entirely. If the top quark mass is actually
the sextet constituent mass scale, as we have suggested, then the evolution
of $\alpha_s$ will halt at $E_T \sim m_t$. In Fig.~24(a) we show a CDF 
analysis\cite{CDF}
which translates the observed (Run 1) jet excess at large $E_T$ into 
the (non-)evolution of $\alpha_s$.
As can be seen, $\alpha_s$ does indeed stop evolving just at $E_T \sim m_t$.
\begin{center}
\epsfxsize=2.7in
\epsffile{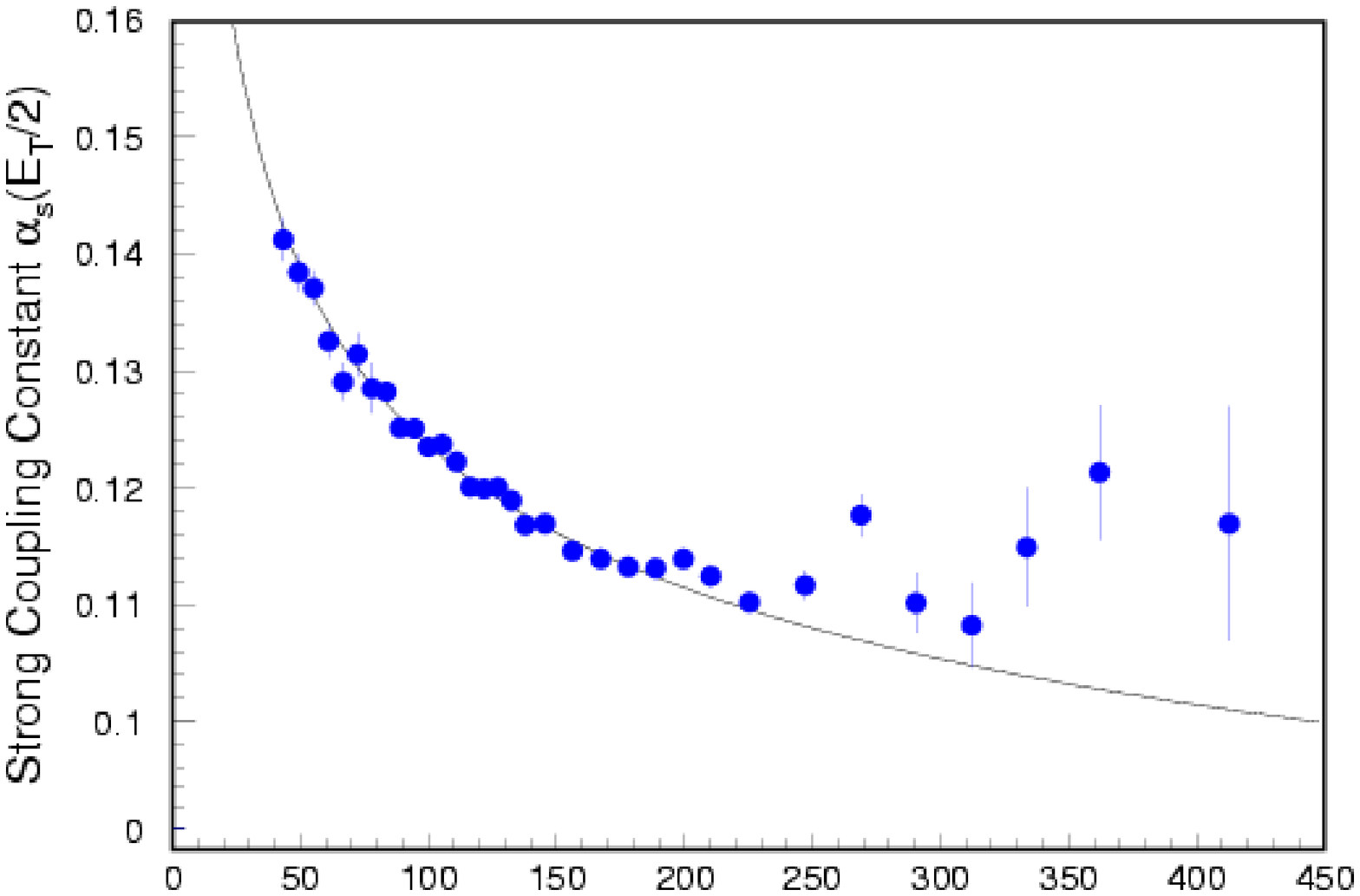}
\hspace{0.2in}
\epsfxsize=2.9in
\epsffile{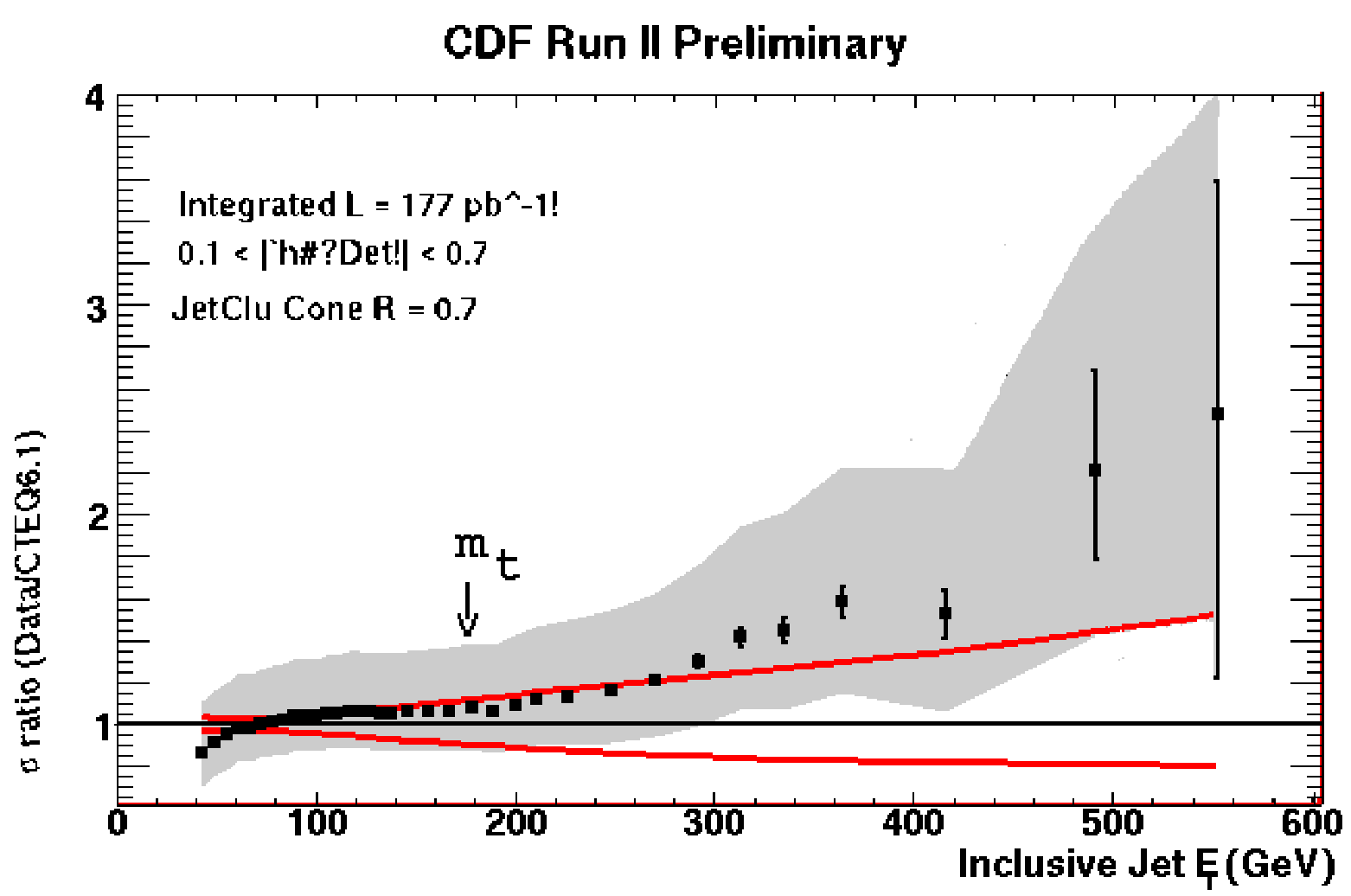}

(a)\hspace{2.7in}(b)

Fig.~24 CDF jet cross-section measurements.
\end{center}

Measurement of the jet cross-section in Run 2 appears, at present, to be entangled
by the very real problem of systematic experimental errors. 
In Fig.~24(b) we show the current
comparison of data with theory\cite{CDFb}. Note that ``theory'' in this case
includes a gluon distribution that was
chosen to best fit the Run 1 excess cross-section.
As can be seen, if we ignore the experimental error problem, the data again
pull away from theory, for $E_T \sim m_t$ upwards, with the effect clearly
growing with $E_T$. 

It seems possible, if not probable, that
above the electroweak scale, QCD jet physics is breaking down in just the manner
that we would expect from $QCD_S$. Indeed,
if the top mass has the significance that we have just discussed,
then the sextet sector has fully entered the theory at this scale.  
In addition to halting the evolution of $\alpha_s$,
the increasing entry of sextet sector states into the dynamics
should imply that the ``excess'' continues to grow as $E_T$ increases.
Indeed, we would expect that in the highest $E_T$ excess region there is an 
enrichment of longitudinal $W^{\pm}$ and $Z^0$
jets with $M_{jet} \approx M_{W/Z}$. As we discuss in the next Section, 
at the LHC such events will have become a major part of the cross-section. 

\subhead{6.4 Non-perturbative Decay Modes} 

If the $\eta_6$ is indeed responsible for $t\bar{t}$ production, then
we would also expect to see ``non-perturbative'' decay modes. 
To discuss these modes, the best we can do
is to exploit the parallel between the \{$\Pi^{\pm},\Pi^0,\eta_6$\} sextet
states, corresponding to \{$W^{\pm},Z^0,\eta_6$\},
and the familiar \{$\pi^{\pm},\pi^0,\eta$\} triplet quark states.
Although the width should be large, if we take 
$~m_{\eta_6} ~\sim~2 m_t~ \sim ~ 350 GeV$, 
the relative couplings and masses of
the vector mesons, and the photon, imply that the 
primary non-perturbative decay mode should be (in parallel with 
$\eta~\to~ \pi^+~\pi^-~\pi^0$) 
$$
\eta_6~~\to~~ W^+~W^-~Z^0 
\auto\label{dk1}
$$
which, when $Z^0 \to b\bar{b}$, would give the same final state as $t\bar{t}$. 
The next most significant mode 
$$
\eta_6~~\to~~ Z^0~Z^0~Z^0 
\auto\label{dk2}
$$
(in parallel with $\eta~\to~ \pi^0~\pi^0~\pi^0$) 
should have a smaller branching ratio, because of the larger $Z^0$ mass. 
In addition, (\ref{dk2}) would  
be indistinguishable from (\ref{dk1}) when the $Z^0$'s decay hadronically, as they
do most of the time. Because the $\eta_6$ mass is so large, 
decay modes involving an electromagnetic coupling, such as
$$
\eta_6~~ \to~~ W^+~W^-~\gamma~, ~~~Z^0 ~Z^0~\gamma ~, ~~~ Z^0 ~\gamma~\gamma~, 
~~~ \gamma~\gamma
\auto\label{dk3}
$$
would be expected to have smaller branching ratios but should 
be present at some level. 

Unfortunately, because the non-perturbative decay modes proceed vis sextet pion 
interactions, the produced vector mesons will be longitudinally
polarized and so, as we discussed in the previous Section, 
when they carry large momentum they will have close together jet
and lepton decay modes that are more difficult to detect.

\newpage

\mainhead{7. DARK MATTER, COSMIC RAY PHENOMENA, AND LARGE CROSS-SECTION 
LHC PHYSICS}

If the sextet sector exists, the LHC will most probably 
be the discovery machine, at least as far as accelerator physics is concerned. 
In the 
next Section, we will give direct theoretical arguments
for effects that should be seen at the LHC. However, we will not
be able to predict, theoretically,
the magnitude of the major phenomena we expect with any great certainty.
However, if there is ``a major change in the strong 
interaction above the electroweak scale'',  it surely should
be visible in Cosmic Ray physics and, more generally, in other cosmic phenomena.
As we now discuss, 
there are candidate phenomena of this kind and, if they are
interpreted as we will suggest, they indicate that 
large cross-section effects are to be expected at the LHC. 
We first give a brief, qualitative, discussion of why
we expect the sextet sector to appear with high-energy cross-sections 
that are larger than hadronic in size and what we expect the major effects of these
cross-sections to be. 

\subhead{7.1 Larger Than Hadronic Size Cross-Sections}

If $QCD_S$ existed in isolation, without the electroweak sector, then, because of 
the larger Casimirs, the sextet sector would constitute
a stronger coupling sector of the theory. Just how significant
the casimir effect is, we do not know. In general, it is clearly present in
perturbation theory but is less significant in conventional non-perturbative
formalisms. In $QCD_S$, because of the
``almost perturbative'' form of confinement that is present,
we expect the effect to be maximal. Most importantly, though, we do not know how 
the wee gluon distribution contributes to the pomeron couplings that determine
the size of asymptotic cross-sections, although sextet couplings should surely
be larger. Therefore, the best we can say is that
sextet pions will be massless and have asymptotic 
cross-sections that are (probably considerably) larger that their triplet
counterpart. The sextet nucleon mass 
scale will be larger than the triplet scale but,
nevertheless, sextet nucleon asymptotic cross-sections should also be larger.
In general, therefore, although 
the asymptotic mass scale will be much larger, the size of asymptotic 
cross-sections, including multi-pomeron cross-sections,
should be larger for the sextet sector, than the triplet sector, in $QCD_S$. 

Adding the electroweak sector
transforms the massless sextet pions into massive vector mesons. Effective
current quark masses also have to be added. At
asymptotic energies neither effect should matter, but such
effects do matter for determining the scale above which
asymptotia sets in. In addition, if we start
(in the real world) with initial triplet states we will only
be able to produce the large cross-section sextet states via multiple gluon
exchange and therefore, to obtain large cross-sections, via the pomeron. 
This does not mean, however, that only
double pomeron production processes can be involved. If the
double pomeron amplitude for the production of a sextet state, such as the
$W^{\pm}$ pair amplitude shown in Fig.~25(a), is large (as we show in the next
Section)
then the associated ``cut-pomeron'' amplitude, shown as the first diagram
in Fig.~25(b), should also be large. (This amplitude is, however, entirely
non-perturbative in that it can not be obtained by an anomaly pole method.) 
\begin{center}
\epsfxsize=1.6in
\epsffile{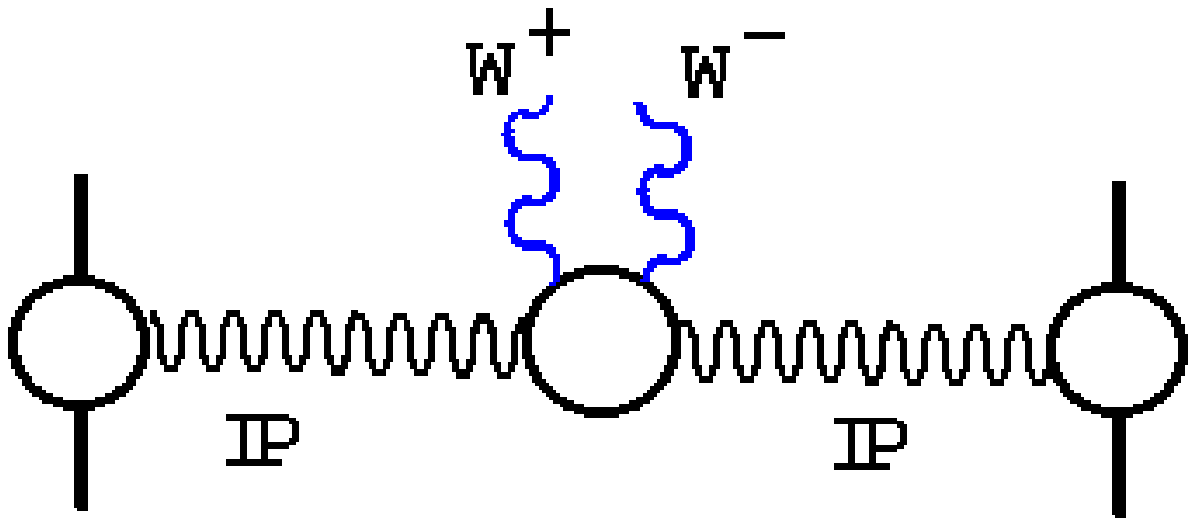}
\hspace{0.5in}
\epsfxsize=3.4in
\epsffile{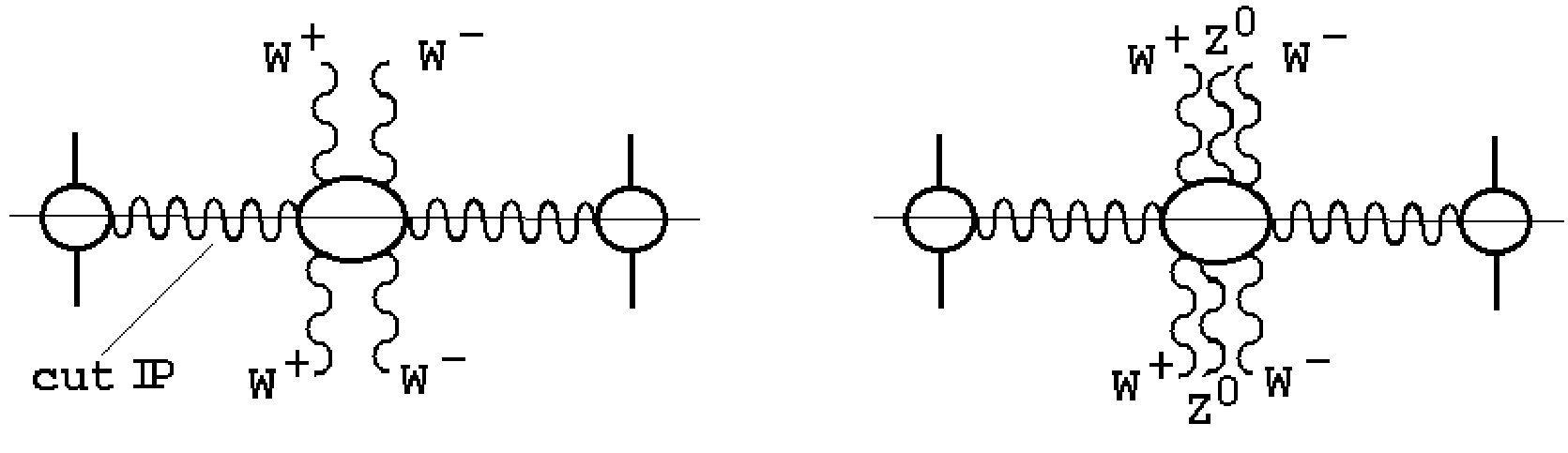}

(a)\hspace{3.2in}(b)$~~~~~~~~~~~~~~~~~$

Fig.~25 (a) The double pomeron $W$ pair amplitude (b) cut-pomeron amplitudes. 
\end{center}

The cut-pomeron amplitude describes the full, 
central region, inclusive cross-section for production of a
$W^{\pm}$ pair. Like the total cross-section, a significant part of the inclusive
cross-section should be describable by pomeron exchange, even when only a 
relatively small rapidity range is effectively available for one cut-pomeron
or the other. Therefore, when the cut-pomeron  amplitude is large it implies that
$W^{\pm}$ pairs (and, similarly, $Z^0$ pairs) will be strongly,
and multiply, produced inclusively across a larger part of the rapidity axis
than is covered by the double pomeron produced state.
The second cut-pomeron amplitude shown in Fig.~25(b)
describes the inclusive production of three boson states which requires, of course,
a bit more energy, and so on for higher cut-pomeron amplitudes. 
Once there is enough energy for cut-pomeron
exchange to begin describing significant production
of the sextet sector then the larger cross-sections
of this sector should imply that sextet states actually come to
dominate the inelastic
(triplet state) hadronic cross-section at a (not too much) higher energy.

We expect, therefore, that the initial 
``major change in the strong interaction 
above the electroweak scale'' will be that multiple vector boson states 
are produced, with large cross-section, across
most (but not all) of the rapidity axis - in close analogy
with pion production at much lower energies. Sextet nucleon production will
set in at higher energies, depending on the mass of
these states. If the pomeron provides the gateway to the 
asymptotically dominant sextet sector, then we could expect that 
to produce a sextet state with mass $M$ requires at least
$\sqrt{S} ~> 10M$
and so if $M \sim~400~GeV$ (to be safely above the threshold
for vector boson pair procution) then we would need
$\sqrt{S} ~> 4~TeV $
- which is just above the Fermilab energy. 

\subhead{7.2 The $N_6$ and Dark Matter}

That the (triplet quark) 
proton is lighter than the neutron is entirely due to the fact 
that the current mass of the $u$ quark is less than that of the $d$ quark. 
Electromagnetic effects, alone, 
would make the proton heavier. 
Because of the absence of hybrid triplet/sextet states, 
the lightest of the sextet nucleons will be stable. However,
sextet quark current masses must be zero. If not, sextet pions would 
be massive and could not mix with the massless $W$ and $Z$ states to give 
them masses, as discussed in Section 5. 
Therefore, the sextet nucleon mass difference 
has to be entirely electromagnetic in origin, and so  
it is the $N_6$ that is stable. If the sextet
quark dynamical mass is given by the top quark mass, as discussed in the last
Section, then the $N_6$ mass should be  $\approx  500~ GeV$ and the $P_6$
mass should be just a little higher. Since triplet and sextet quarks do not
combine to form bound states it is, presumably, reasonable to assume that
sextet nucleons also do not form bound states with triplet nucleons. More 
particularly, perhaps, if pion exchange provides the binding force for nucleons
to form nuclei, the distinct quark content of sextet and triplet nucleons
implies that there is no common ``pion'' that can bind them. 

The $N_6$ is, therefore, neutral, stable, and (because of the dominance
of sextet states) will be the
dominant, heavy, stable state produced in high energy cross-sections. Consequently,
it will be dominantly produced in the high energy interactions that 
are believed to have been responsible for the formation of the early universe.
If it does not form bound states with normal quark matter it will 
abundantly form cold 
dark matter, in the form of (sextet) nuclei, clumps, etc. (Perhaps
sextet pions can exist inside sextet nuclei and provide the binding force.) As 
a result, the existence of the sextet nucleon sector provides a natural
explanation for the dominance of dark matter in the universe. Conversely, once
we establish that the $N_6$ will form dark matter, the dominance of dark matter
in the universe can be regarded as 
evidence confirming that sextet quark states dominate high energy cross-sections.

The dominance of dark matter in the universe does not  
tell us at what energy scale this
dominance appears in total cross-sections.
Specific evidence for the relevant scale
is, however, provided by the cosmic phenomenon that we discuss next.

\subhead{7.3 The Knee in the Cosmic Ray Spectrum}

The ``knee'' in the cosmic ray spectrum is an extraordinary, well-established
and very well-known, phenomenon. As shown in Fig.~26(a), it appears as a break in 
the slope of the spectrum that stands out, as a distinctive feature, as the 
energy increases over some ten orders of magnitude and the flux decreases
by thirty orders of magnitude. In first approximation, there is
one single slope as the energy increases 
from $10^{10}~eV$ to $10^{16}~eV$ and a second slope as the energy increases
from $10^{16}~eV$ to $10^{20}~eV$. It is called the knee because, as is
clear from the larger scale plot shown in Fig.~26(b) (normalized by the low-energy
slope), it is not simply a break 
in slope but rather a ``bump'' in which, for a short energy range, it looks 
like the slope has decreased before it settles at an increased value.

It is widely believed by cosmic ray physicists that the origin of the
knee is cosmic, even though there is no consensus on what the cause
could be. 
\begin{center}\parbox{2.9in}{
\begin{center}
\epsfxsize=2.3in 
\epsffile{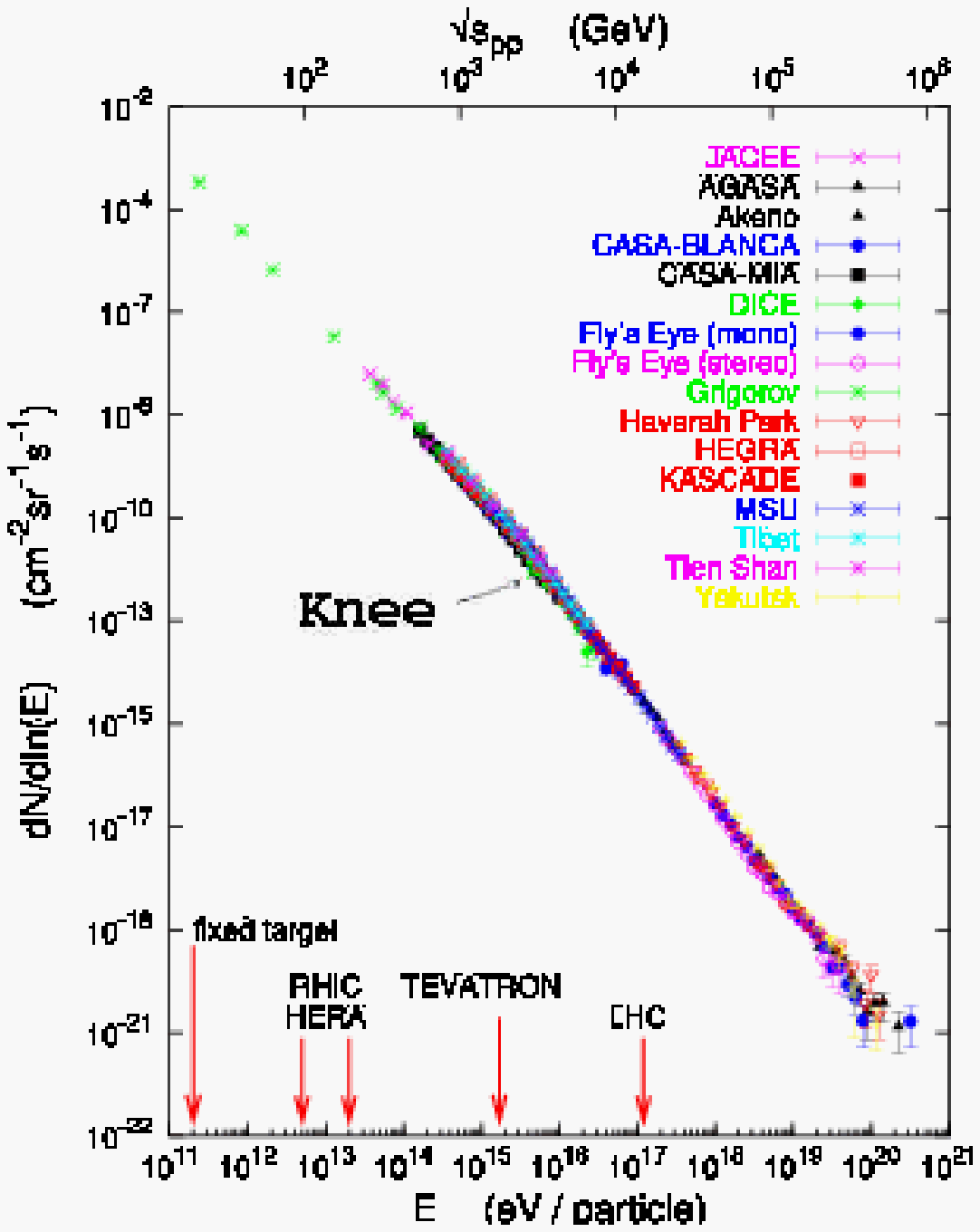}

(a)
\end{center}}
\parbox{2.9in}{
\begin{center}
\epsfxsize=2.5in 
\epsffile{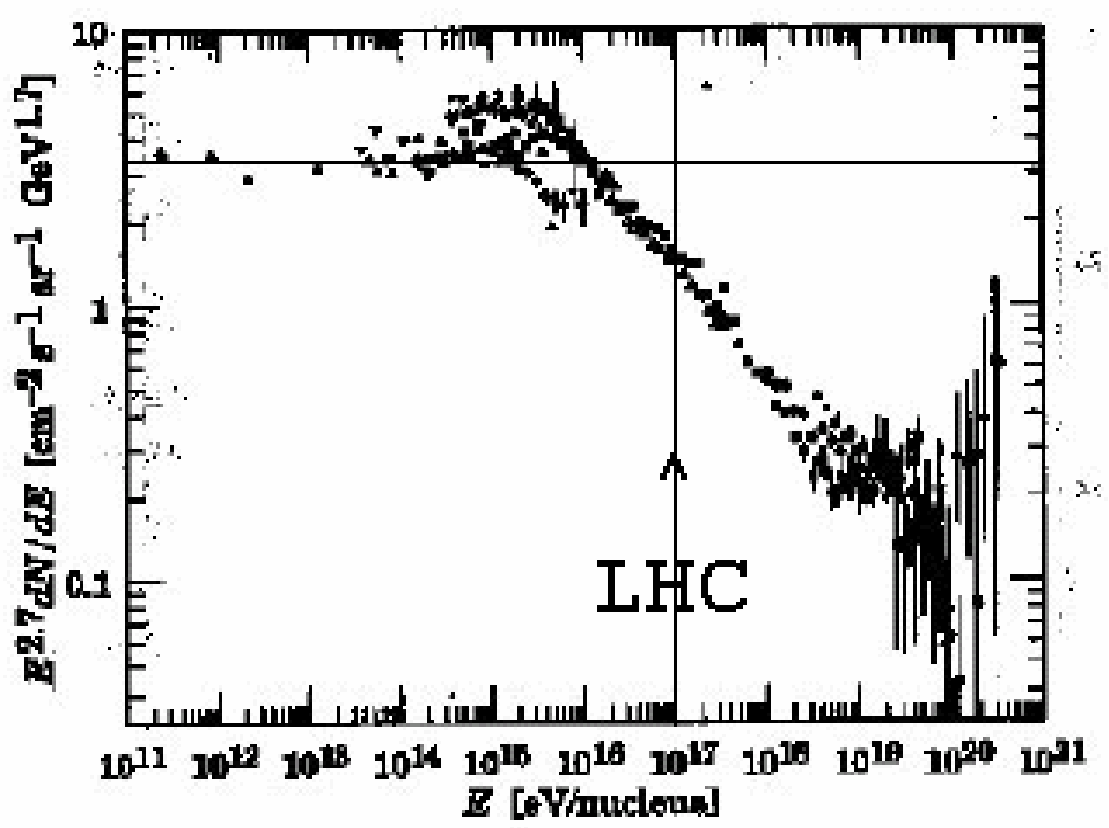}

(b)
\end{center}}

Fig.~26 (a) The full cosmic ray spectrum (b) The knee.

\end{center}
 A priori, it seems almost inconceivable that a conspiracy of
external phenomena could produce such a pronounced local effect, in a spectrum that 
(naively at least) is arriving from all directions and all distances 
of the universe. It seems far more plausible that the cause of the effect is in the 
atmospheric interaction. Indeed, right from it's earliest discovery, it was 
suggested\cite{nik} that the knee could be the threshold for
a new interaction that produces (stable or relatively stable)
neutral particles which are not observed in the ground level detectors.
This would produce an underestimation of the shower energy above the threshold
and would lead to a pile-up of events below the threshold energy which,
together with a depletion of the spectrum above the threshold, would be 
observed as a ``knee''. However, 
the major part of the cross-section has to be affected by
this threshold. Since there was no serious idea what the neutral particle(s) could
be and there was no reason to expect 
such a dramatic effect in the strong interaction,
particularly after the discovery that this interaction is described by QCD, 
there was no general acceptance of the proposal.

We first proposed that the sextet threshold could be responsible for the knee
some time ago\cite{arw94}. In the
meantime, other authors have emphasized the difficulty of explaining the knee as 
an effect of cosmic physics and have made various proposals\cite{nik,kn,bk}
for a threshold effect that could be involved. 
That a large fraction of the cross-section (increasing
as the energy increases) has to be involved, is a problem for all threshold
proposals. 
Because the data from different experiments do not agree about the absolute value 
of the flux and also cover different energy ranges, it is difficult to be sure 
exactly where the threshold should be and how 
much of the cross-section has to be involved. It is clear, however, that the 
threshold should be below the LHC energy and that
the physics involved should be visible at the LHC as a significant part of
the hadronic cross-section ($\sim$ 10-20\% ).
As the discussion in 
{\bf 7.1} shows, the sextet sector threshold has (perhaps uniquely) the potential,
at least, to play this role.

The the prolific production of vector bosons, 
will increase the average transverse momentum of events enormously and lead to
such an increase of the shower spread that a much greater fraction (than expected)
of the shower particles will be undetected. At the LHC, ten or more vector bosons 
could be produced, kinematically, via the cut-pomeron cross-section.
The major consequence will, of course, be a huge increase of the large
$E_T$ jet cross-section. (The effective increase 
due to the non-evolution of $\alpha_s$ will be just a small part of this effect).
There will also be marked changes in the distributions of
leptons produced. In particular, there will be
a much larger fraction of (undetected) neutrinos in the ground level particles.
The production of ``dark matter'' (sextet 
neutron/antineutron pairs) will
straightforwardly take away undetected energy and the effect will be maximal
if the sextet neutron mass is as low as possible.  
At higher energies the inclusive production of $N_6$ pairs will surely
become more and more significant and, necessarily, be a major contribution 
to the loss of detected energy by most of the total cross-section.

It is interesting that 
the production of $N_6$ pairs is not so different from the original 
proposal\cite{nik} of the production of neutrals 
to explain the knee. Of course, the existence of dark matter was unknown
and the link between the two phenomena, that we are proposing,
could not have been imagined.

\subhead{7.4 Cosmic Ray Dijets and Ultra High Energy Events}

There are a number of distinct effects that have been seen in cosmic ray showers
with energies above the knee, for example those discussed
in \cite{nik}. Collectively, they all suggest that
new physics appears above the knee. We catalogued the effects,
and offered explanations of the phenomena involved, in \cite{arw94}. However,
in most cases, the explanations we offered would surely be modified by our current
understanding. In addition, other effects have been discovered since. We will not
attempt a recataloguing here, but instead will concentrate on one of the,
by now, most well-established effects and will also discuss what has since
become one of the most interesting phenomena.

There are very significant anomalies in the rate of high $E_T$ jets (``cores") in 
experiments such as Chacaltaya and Kanbala~\cite{cores}. A QCD Monte Carlo 
was tuned to jet data at fixed
target and collider energies (including the $Sp\bar{p}S$ and Tevatron). 
The prediction for $\chi_{12}$, which is basically the
product of the jets' $E_T$ and the jet-pair separation $R_{12}$,
was then compared with the cosmic ray data. As shown in Fig.~27, for 
energies above $\sqrt{s} \approx $ 5 TeV (i.e. above the knee)
the jet rate for $\chi_{12}~ \centerunder{\raisebox{1mm}{$>$}}{$\sim$} $ 1000
TeV.cm exceeds the QCD expectation
by as much as two orders of magnitude.
If we interpret this is an extension, to higher energies,
of the large $E_T$ jet excess observed at Fermilab, then 
it shows that there is an (orders of magnitude)
increase of just the kind that we expect.
\begin{center}
\epsfxsize=3.5in 
\epsffile{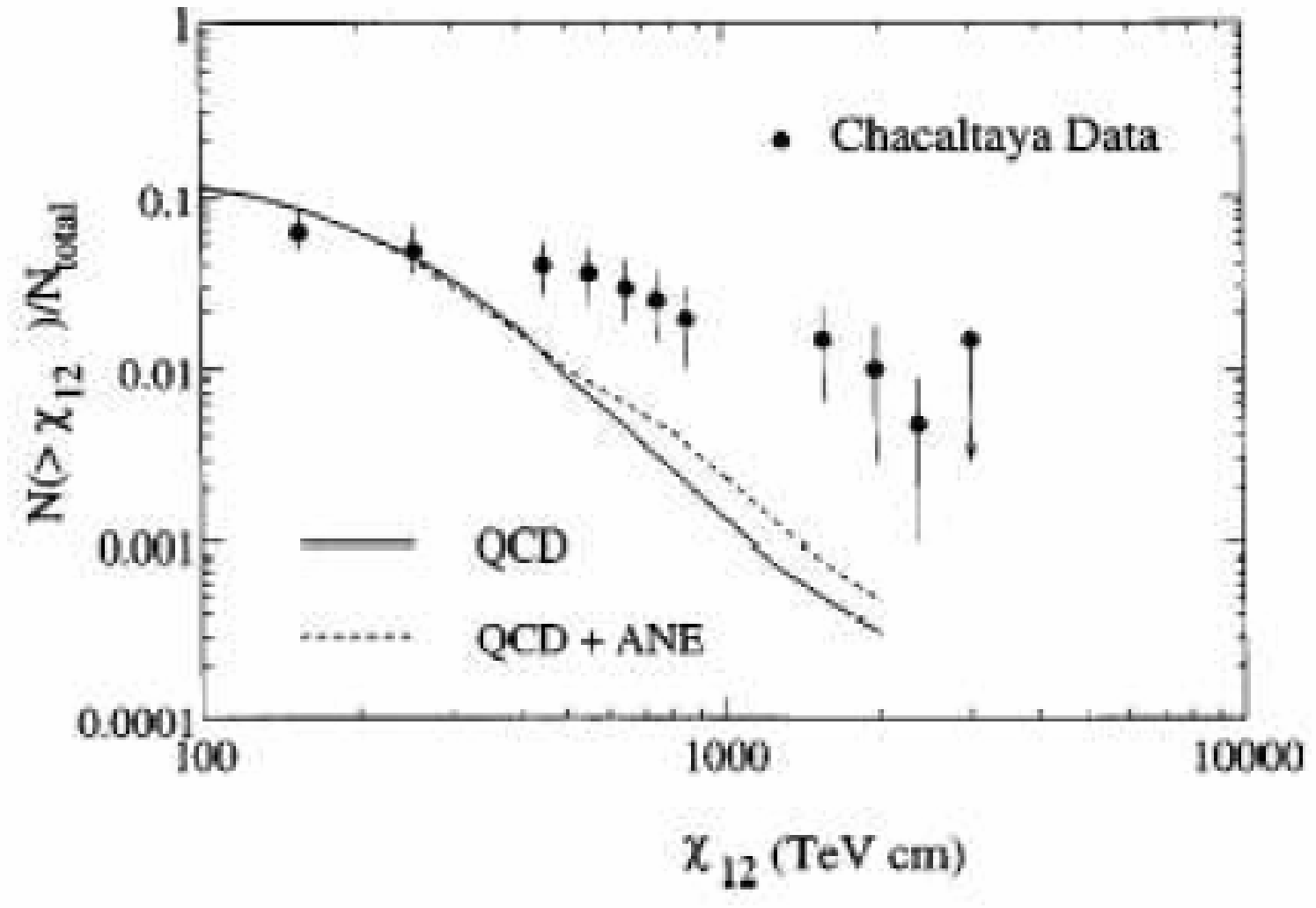}

Fig.~27 Excess large $E_T$ dijets (``cores'').
\end{center}

Ultra high-energy cosmic ray events, with $E_0~ 
\centerunder{\raisebox{0.5mm}{$>$}}{$\sim$}~
10^{20}~eV$,  have attracted great attention because the energy
exceeds the GZK cut-off produced by the interaction of a proton
with the cosmic background radiation. This suggests, of course, that 
the particles involved can not be protons. On the other hand, 
it is believed that the high
velocities involved imply the particles must have traveled a long way and so
should be stable. Within the Standard Model there is no other candidate
besides the proton. As a result, both the origin and the nature, of the 
high-energy events is regarded as a mystery.

Within the sextet sector, there is an obvious candidate for producing the
ultra high-energy events. Sextet neutrons are stable and will 
avoid the GZK cut-off, both because they are neutral and because they are
massive. Also (because they have a large coupling to the pomeron) they will
have a large high-energy hadronic cross-section. Clearly they could be
responsible for the ultra high-energy cosmic rays. Indeed, they are probably 
responsible for an increasing fraction of the spectrum from energies
lower than $10^{20}eV$ upwards. Since they would simply be
very high energy dark matter, which is omnipresent in the universe,
their origin would (presumably) not be a mystery.

To the extent that the existence of the 
ultra high-energy events is evidence for a stable, massive,
particles that are strongly interacting (and preferably neutral), they
could actually be regarded as evidence that dark matter is strongly interacting.
 
\newpage

\mainhead{8. WHAT SHOULD BE SEEN AT THE LHC ?}

Major evidence for the sextet sector, in the high luminosity mode of the LHC, 
will be the much larger than expected
multiple vector boson and large $E_T$ jet cross-sections discussed in 
the previous Section. Because large momentum longitudinal bosons
(that preferentially decay to jet or lepton configurations with isolation
problems) will be
dominantly produced, the full size of the diboson cross-section may 
not be immediately recognized. Instead the major, observed, effect of this cross-section
may be to contribute to the increased magnitude of large $E_T$ jet cross-sections. 
Quite possibly, this increase will not be immediately identified as due to
a sextet quark sector.

A priori, the neutral $N_6$ will also be quite difficult to detect,
since missing energies of several hundred $GeV$ will be common. The $P_6$,
assuming it is not too unstable, should be seen.
Although a massive, charged, particle with a large production cross-section
will surely cause much general interest, it also may not be 
immediately identified with the sextet sector. Instead, 
the double pomeron cross-section may well be 
the most definitive early evidence for the sextet sector.

\subhead{8.1 Double Pomeron Exchange.}

Vector bosons can be pair-produced directly in double pomeron exchange, 
via the sextet pion anomaly mechanism, as illustrated in Fig.~28.
\begin{center}
\epsfxsize=3in
\epsffile{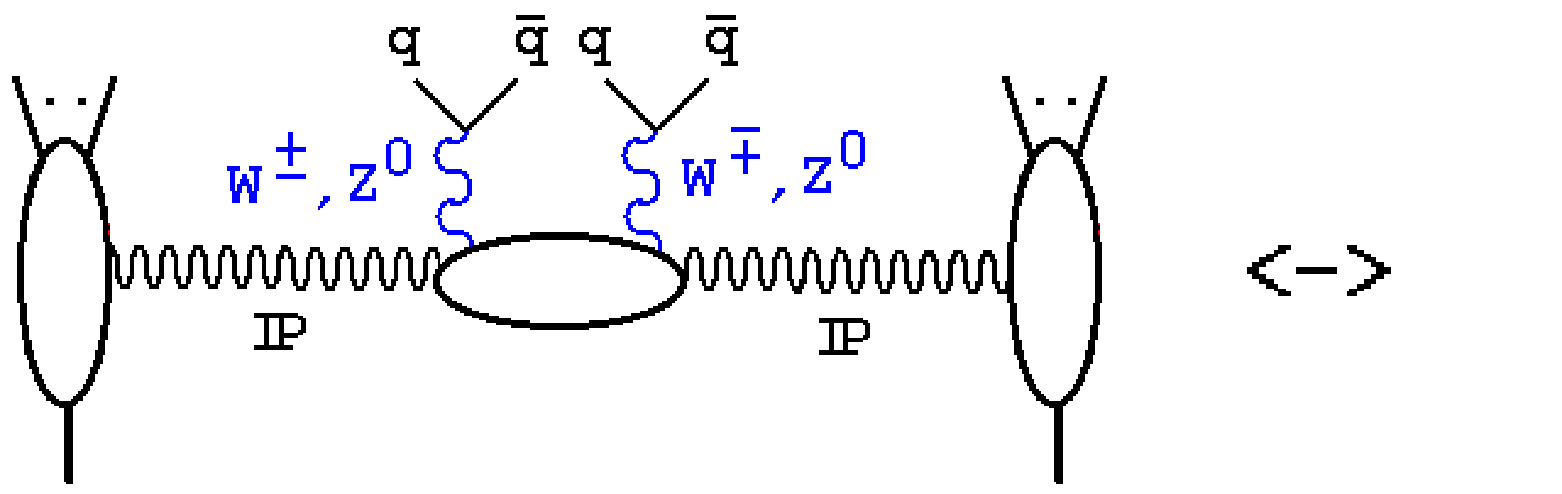}
\epsfxsize=2.4in
\epsffile{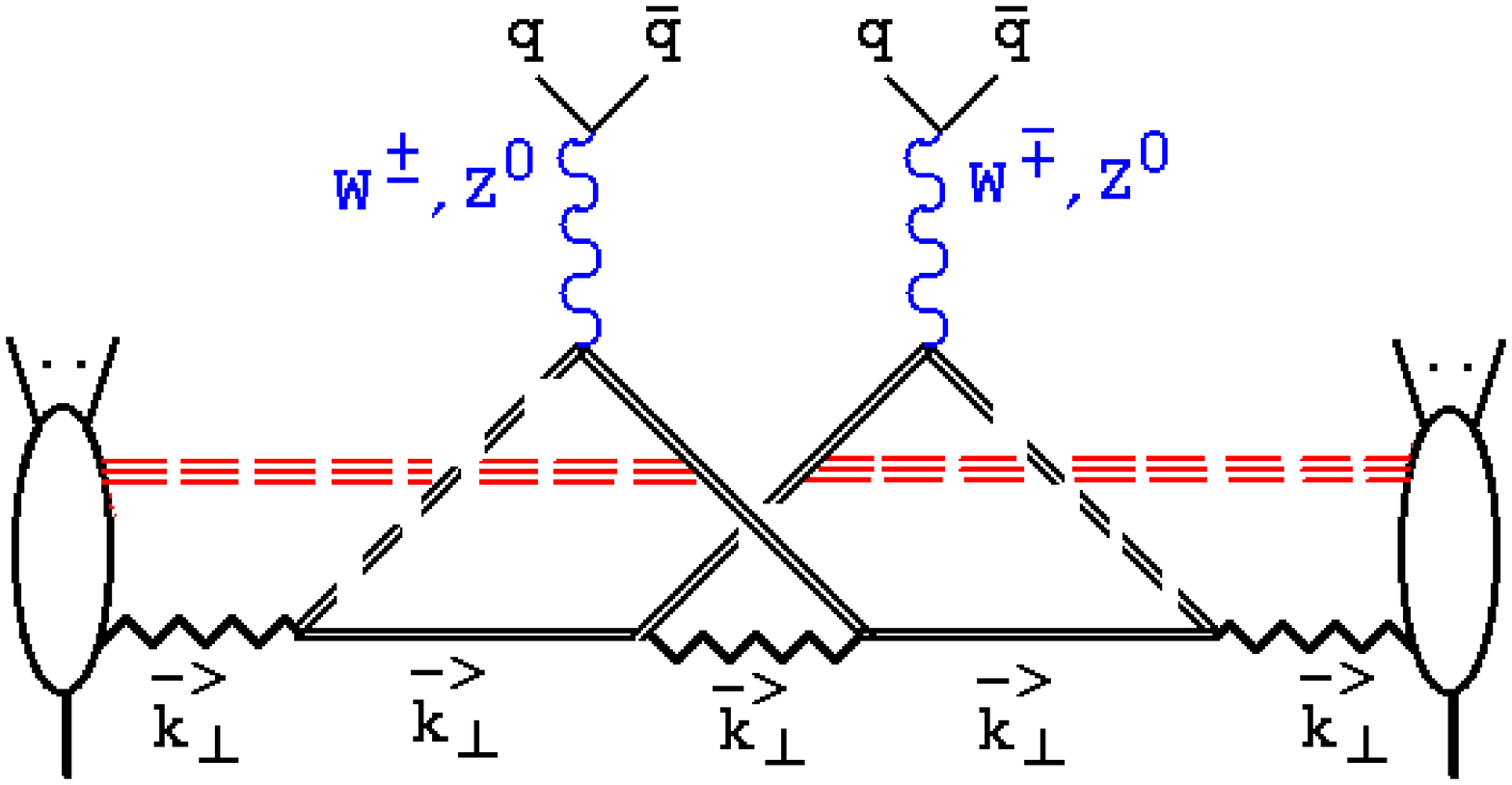}

Fig.~28 Double pomeron production of $W$ and $Z$ pairs via sextet pions.
\end{center}
The kinematics needed for the derivation of this amplitude, as a
straightforward extension of the argument of {\bf 5.2}, are easily satisfied
at the LHC. A parallel argument to that of {\bf 5.4} and {\bf 6.2} can then
be used to obtain
an order of magnitude estimate for the cross-section. 
The jet amplitude, analagous to Fig.~18(a), that has, apart from the anomaly
loops, the same propagators and couplings as Fig.~28 is shown in Fig.~29(a).
When the transverse momentum
is electroweak scale, i.e. $|k_{\perp}| 
\centerunder{\raisebox{1mm}{$>$}}{$\sim$} ~100 ~GeV$, the cross-sections
given by Fig.~28 and Fig.~29(a) are comparable. That is to say,
at large $k_{\perp}$, the double pomeron production of 
$W^+W^-$ and $Z^0Z^0$ pairs will give jet
cross-sections that are as large as those predicted by standard QCD.
\begin{center}
\epsfxsize=2in
\epsffile{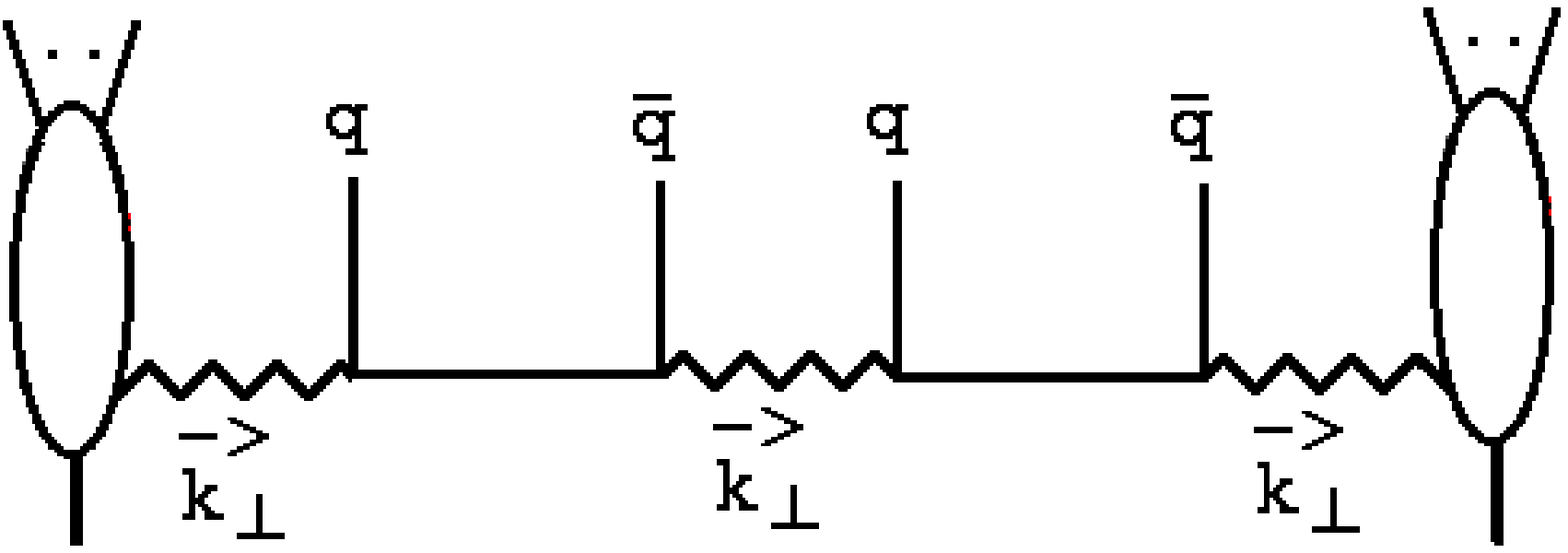}
\hspace{0.6in}
\epsfxsize=2.3in
\epsffile{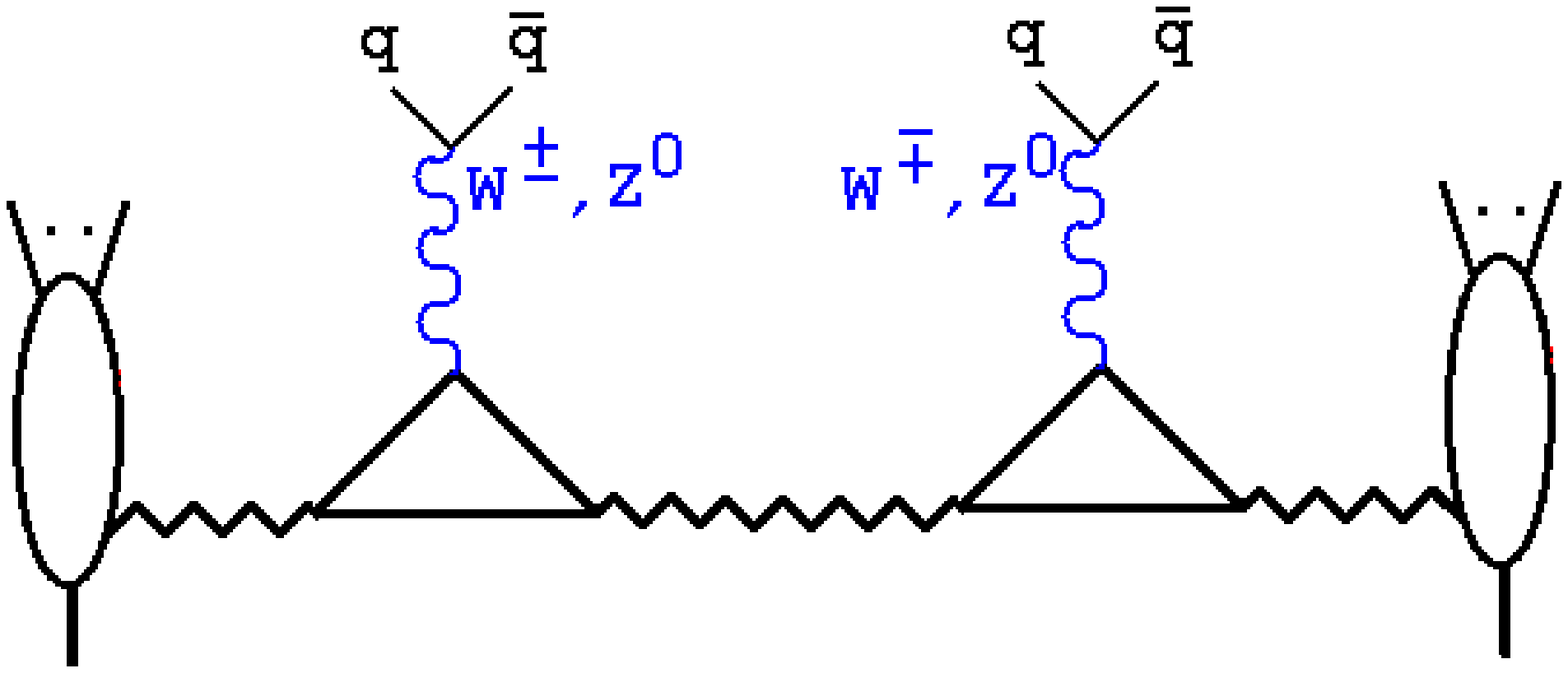}

(a)\hspace{3in}(b)

Fig.~29 (a) The comparable jet amplitude (b) a triplet sector amlitude 
\end{center}
In Fig.~29(b) we show the
lowest-order triplet sector amplitude that would comparably produce 
a vector boson pair decaying to jets, and would also involve the
gluon exchanges necessary for pomeron exchange. Extending the argument of 
{\bf 5.4}, since there are two sextet pions involved, and therefore two 
factors of $F_{\Pi}$,
the cross-section given by Fig.~29(b) would be smaller by a factor
of $~ \centerunder{\raisebox{0.5mm}{$ >$}}{$\sim$}
 ~O(10^{12})$. 

Repeating the argument of {\bf 5.6} in the present context, 
the central double pomeron vertex of Fig.~28 
should vary only slowly with $k_{\perp}$ (with an electroweak scale), 
while the external hadron/pomeron vertices  
will have strong $ k_{\perp}~$- dependence and give a large 
increase as $|t|$ decreases. As was the case in our discussion
of HERA events, we will obtain the maximum
increase if the scattering protons are not diffractively excited.
In this case, the increase will be given by the 
same product of hadron/pomeron couplings that is present in the 
elastic cross-section. When combined with a large $|t|$ amplitude
that is larger than it's triplet sector counterpart 
by  $~ \centerunder{\raisebox{0.5mm}{$ >$}}{$\sim$} ~O(10^{12})$, this 
should imply a large double pomeron cross-section when $|t|$
is at the minimum kinematically allowed value, in agreement with the general
argument of {\bf 7.1}.

\subhead{8.2 LHC Kinematics}

If we consider
the (symmetric) central region production of a state with
mass $4M^2$ by colliding proton beams with momentum $\sqrt{S}/2$,
then this corresponds to a minimum $t$ value of 
$$
t~\sim~ -  \frac{4 M^2 m_p^2}{S}
\auto\label{syt}
$$
If we consider $W^{\pm}$ (or $Z^0$) pair production then, in analogy
with the double pomeron cross-section for pion pairs we might expect
the maximal cross-section to be at 
$M \sim 2 M_W \sim 170~GeV$. In this case, with
 $\sqrt{S}= 14~ TeV$ and the proton mass set to $1~ GeV$, 
we obtain from (\ref{syt})  
$$
t~\sim~~ 4 ~\bigl(~\frac {0.17}{14}~\bigr)^2  ~~\sim ~5 \times 10^{-4}~GeV^2
\auto\label{tsyv}
$$
which is close to the minimum value that is expected to be achieved by TOTEM
in the initial low luminosity running at the LHC. Therefore, it should be possible
to detect the cross-section very close to it's maximal value.

If the CMS central detector is operational
during the initial ``soft physics'' running period, then it should be
straightforward to look for central $W^{\pm}$ and $Z^0$ pairs, 
in combination with very forward scattered protons 
in the TOTEM Roman pots. The cross-section will be maximal when $t$ is at it's minimum
but should, of course, be observable over a range of $t$ values. However,
it will also be preferable to be
as close as possible to threshold to minimize problems caused by the 
asymmetric decays of high momentum 
longitudinal vector bosons. In fact, with this in mind, it
could be that at larger $|t|$ values multiple $W^{\pm}$ and $Z^0$ pairs
will be easier to detect because they are closer to threshold. 
Perhaps, since we expect the cross-sections to be so large, there 
could be spectacular events 
in which the far-forward protons are tagged and only (a multitude of)
large $E_T$ leptons are seen in the central detector! 

A large double pomeron cross-section for $W^{\pm}$ and $Z^0$ pairs
immediately implies that the longitudinal components have direct
strong interactions. Therefore, the observation of such a cross-section
would be an immediate confirmation of the existence of the sextet sector
and the sextet higgs mechanism. In addition, this cross-section can be looked for
as soon as the LHC turns on. Consequently, we regard it as the top signature
to be looked for as evidence for the sextet sector.

If we consider sextet neutrons and, to be appropriately above threshold,
set $M = 1~ TeV$ in (\ref{syt}) we obtain a minimum $|t|$ value of
$$
t~\sim~~ 4 ~\bigl(~\frac {1}{14}~\bigr)^2  ~~\sim ~2 \times 10^{-2}~GeV^2
\auto\label{tsy1}
$$
which will be detectable, if the cross-section is large enough. It would also
be a spectacular process. The deflection of the tagged protons would determine
that a very massive state was produced, while no charged particles would be seen
in any of the detectors. Comparison with charged lepton production should allow 
a clear separation between this process and the multiple production of neutrinos
by $Z^0$'s. 

If the sextet nucleon double pomeron
cross-section is extraordinarily large, it might be detectable in the low
luminosity run of the LHC. If not, 
it might be seen by the high luminosity detector
that will look for double pomeron production of the 
Standard Model Higgs particle. 

\subhead{8.3 Inclusive Cross-Sections for Sextet States.}

As we noted in the previous Section, we expect cut-pomeron amplitudes
of the form of 
Fig.~31(b) to be responsible for the inclusive production of mutiple vector 
bosons across most of the rapidity axis. This should be a 
major effect when the LHC is in high luminosity mode. However,
as we already implied above, it is likely to 
take time to determine that this phenomenon is indeed taking place, since 
the main effect will  
be the contribution to jet cross-sections. Note that since we showed above
that the double
pomeron production of jets via vector bosons will be comparable with the total jet
rates that are expected, we would obviously
expect the inclusive production of jets via vector
bosons to be orders of magnitude larger (as we argued for, more generally,
in the previous Section).
 
For the leptonic decays, the association
of multiple leptons and missing $E_t$ to multiple $W$'s will have obvious 
problems, which the close together 
decays of longitudinal bosons will only make worse.
Multiple $Z^0$ states should be the easiest to detect, but the close together
decays will also be a problem. Nevertheless,
since the cross-sections should be so dramatically
large, they should eventually provide emphatic evidence for the sextet sector.

If the $N_6$ and $P^{\pm}_6$ pairs (and also 
$\eta_6$ pairs) are indeed too massive to be seen in double pomeron
exchange, the central region inclusive cross-sections could
(as we have already said) still be relatively large
via cut-pomeron amplitudes. The problem then becomes how to detect such states.

\subhead{8.4 Large Jet Cross-Sections and the Evolution of $\alpha_s$}

If the influence of the sextet sector on $\alpha_s$ is that the 
evolution stops at $E_T \sim m_t$, as we discussed in Section 6, then the 
LHC jet cross-sections will be further enhanced.
Even according to conventional QCD
predictions, the LHC jet cross-section persists for an order of magnitude
in $E_T$ beyond $E_T > m_t$ and so the lack of 
evolution would be straightforwardly evident, if it were the only 
phenomenon involved.
In effect, Fig.~27, together with the large $E_T$ jet excess
at the Tevatron, are existing experimental evidence that
the jet cross-section will be much larger than anticipated from conventional 
QCD, for almost all of the $E_T$ range. If this enhancement is as big
as we are anticipating, it will be very hard to imagine an alternative
explanation besides the existence of the sextet quark sector.

\vspace{0.2in}

\noindent{\bf ACKNOWLEDGEMENTS}

I am grateful to Mike Albrow, Malcolm Derrick and Geoff Bodwin for 
valuable discussions

\newpage 

\renewcommand{\theequation}{A.\arabic{equation}}
\setcounter{equation}{0}
\vskip 1cm \noindent
\noindent {\large\bf Appendix A. $~\beta$-function Properties of 
$QCD_S$ and $CSQCD_S$ }
\vskip 3mm \noindent

\subhead{A.1 The Infra-red Fixed-Point in $QCD_S$ }

We write the $QCD_S$ $\beta$-function in the form 
$$
\beta(\alpha_s) = - ~\frac{g^4}{(4\pi)^2}~\beta_0 ~ -~ 
~\frac{g^6}{(4\pi)^4}~\beta_1 ~-~ \frac{g^8}{(4\pi)^6}~\beta_2 ~+~ \cdots 
\auto\label{beta}
$$
The three loop calculation of \cite{tar} gives, for $n_f$ triplet flavors,
$$
\beta_0 = 11 - \frac{2}{3}n_f~ ,~~~\beta_1 = 102 - \frac{38}{3}n_f~, ~~~
\beta_2 = \frac{2857}{2} - \frac{5033}{3}n_f + \frac{325}{54}n_f^2
\auto\label{3lp}
$$
When $n_f = 6$, we obtain  
$$
\beta_0 = 7,~~~\beta_1 = 26 ~.
\auto\label{six}
$$
When the two sextet flavors are included we obtain\cite{tar}
$$
\beta_0 = 7 - 4T(R)n^6_f/3~ = 7 - 4(\frac{5}{2})2/3~ = 1/3,
\auto\label{sex1}
$$
and
$$
\beta_1 = ~26 - 20T(R)n^6_f - 4C_2(R)T(R)n^6_f~ =~26 - 100 -66\frac{2}{3} 
~=~-140\frac{2}{3}   
\auto\label{sex2}
$$
where $T(R) = C(R)/C(3) = 5/2$ and $C_2(R) = 10/3$ for sextet quarks.
Therefore, $QCD_S$  is (just) asymptotically-free and also has an
infra-red fixed point at 
$$
\alpha_s~\approx ~ \frac{1}{34}
\auto\label{as}
$$
(There is a sense in which this can be argued to be present to all 
orders\cite{bz}). In addition, 
between the ultra-violet and infra-red fixed points the $\beta$-function
remains very small ($ <~10^{-6}$). As a result the massless theory
evolves only very slowly and is 
almost scale-invariant. 

\subhead{A.2 Asymptotic Freedom in $CSQCD_S$ }

As in the body of the paper, we use $CSQCD_S$ to denote the 
``color superconducting'' version of $QCD_S$
obtained by adding a scalar field and using the usual higgs mechanism. 
(Note that, in this context, the ``higgs mechanism'' is a technical manipulation
that has nothing to do with electroweak symmetry breaking.)
It is a special property of $QCD_S$ that a (complex)
color-triplet Higgs scalar sector can be added\cite{gw,cel} 
- with both the gauge-coupling
{\em and} the Higgs self-coupling asymptotically free. We can illustrate this 
as follows.

Let $g(t)$ and
$h(t)$ be the respective scale-dependent couplings, then 
$$
\frac{dg}{dt} = \beta(g,h)~
= -{1\over 2}b_0t^3 + \cdots
\auto
$$
where, now,
$$
b_0 = {1\over {8\pi^2}} \left[\beta_0 -{1\over 6}\right]
\auto 
$$
$\beta_0$ is calculated from the quark content, as above, 
and the 1/6 is due to the triplet scalar. Similarly
$$
\frac{dh}{dt} = \tilde{\beta} (g,h)~
= Ah^2 + Bg^2 + Cg^4 + \cdots 
\auto
$$
where 
$$
A = {7\over {8\pi^2}},\ B = -{1\over {\pi^2}}~~~ and~~ 
C = {{13}\over {48\pi^2}}.
\auto 
$$
We can have $h \rightarrow 0$ consistently in (3.5) if 
$h = xg^2 + 0(g^3)$.
 This gives a stability equation for $x$, that is
$$
\frac{dx}{dt} = g^2 \left( Ax^2 + B^\prime x + C \right)
\auto
$$
where $B^\prime = B + b_0$.  When the stability condition 
$(B^\prime)^2 > 4AC$ is
satisfied there are two fixed-points of (3.7) and the smaller is stable for 
$t \rightarrow \infty$.  The stability condition gives
$$
\left( 1-\pi^2b_0\right) ^2 > {{91}\over {96}}
\auto 
$$
which for $b_0$ small gives
$$
{5\over {24}} > 8\pi^2 b_0
\auto
$$
If there are 16 color triplet quarks, or six color triplets and two sextets,
then
$$
8\pi^2 b_0 = {1\over6} < {5\over{24}}
\auto 
$$
For comparison, if there are 15 color triplet quarks then
$$
8\pi^2 b_0 = {5\over6} > {5\over{24}}
\auto 
$$
We conclude that, only when the number of quark flavors is ``saturated'',
as in $QCD_S$, can we use
the Higgs mechanism to break the $SU(3)$ gauge symmetry to $SU(2)$, and so
smoothly introduce a (single) massive vector into the theory, while 
{\em maintaining} the short-distance asymptotic freedom of the theory.

\newpage

\renewcommand{\theequation}{B.\arabic{equation}}
\setcounter{equation}{0}
\vskip 1cm \noindent
\noindent {\large\bf Appendix B. ~Properties of the Triangle Anomaly}
\vskip 3mm \noindent

In this Appendix we summarize the various properties of the triangle 
diagram that are used in the paper.
We consider the contribution of the massless fermion loop, 
shown in Fig.~B1, 
\begin{center}
\leavevmode
\epsfxsize=2in
\epsffile{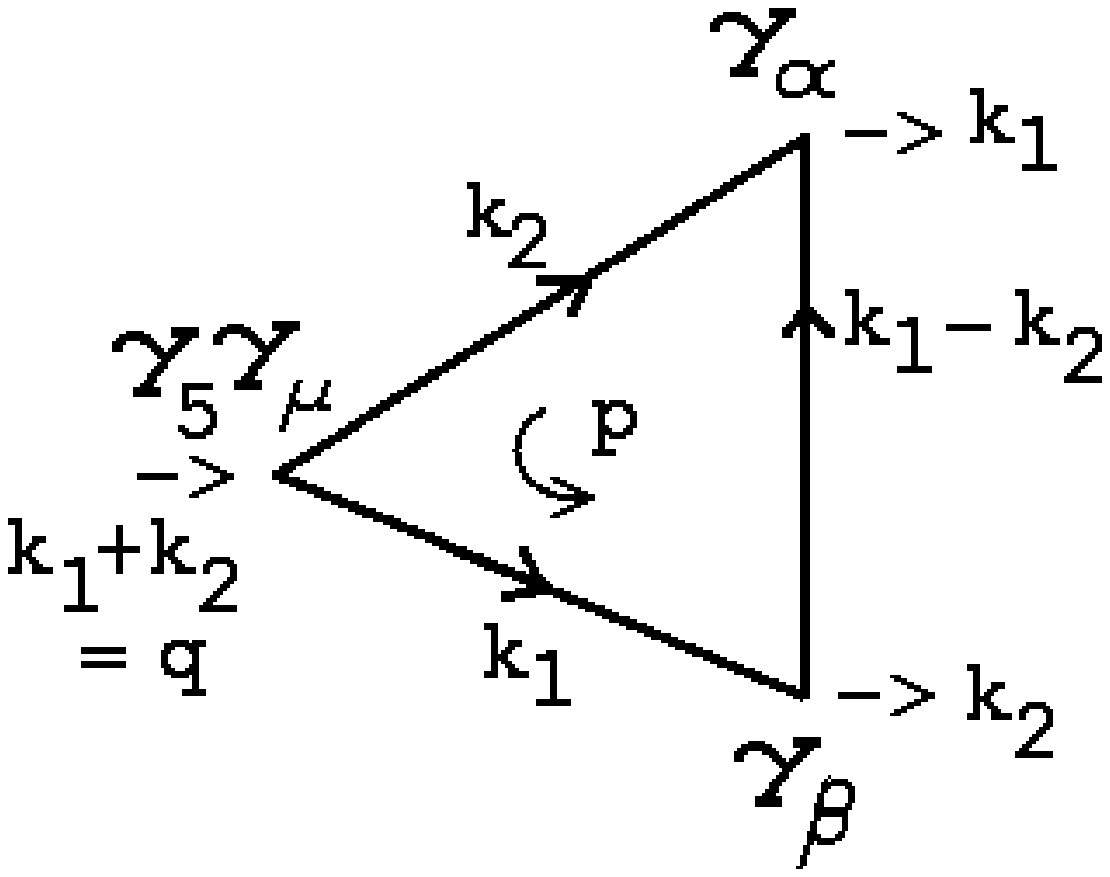}

Fig.~B1 The Fermion Loop Contribution to $T_{\mu \alpha \beta}(k_1,k_2)$

\end{center}
to an axial-vector/two-vector three current vertex, i.e.
$$
\Gamma_{\mu \alpha \beta}(k_1,k_2)~~
=~ {i \over (2 \pi)^4} \int {  d^4 p~ Tr \{ \gamma_5
\gamma_{\mu} ~ (\st{k}_1
+ \st{p})~  \gamma_{\alpha}~ ~\st{p}~
\gamma_{\beta}~ (-\st{k}_2 + \st{p} ) \} 
\over  p^2  (k_1 + p)^2 
 (p - k_2)^2 }
$$
A general decomposition of the symmetrized vertex
$$
T_{\mu \alpha \beta}(k_1,k_2)~=~
\Gamma_{\mu \alpha \beta}(k_1,k_2)~+~
\Gamma_{\mu \beta \alpha }(k_2,k_1)
\auto\label{syvt}
$$
into invariant amplitudes is 
$$
\eqalign{T_{\mu \alpha \beta}(k_1,k_2) ~&= ~ A_1~
{\hbox{\large $\epsilon$}}_{\sigma\alpha\beta\mu}~ k_1^{\sigma}  ~+~ A_2~ 
{\hbox{\large $\epsilon$}}_{\sigma\alpha\beta\mu} ~k_2^{\sigma} 
~+~A_3~
{\hbox{\large $\epsilon$}}_{\delta \sigma\alpha\mu}~ 
k_{1\beta}k_1^{\delta} k_2^{\sigma}  \cr
~~~& +~A_4~  {\hbox{\large $\epsilon$}}_{\delta \sigma\alpha\mu}
~ k_{2\beta}k_1^{\delta}
k_2^{\sigma}~+~A_5~  {\hbox{\large $\epsilon$}}_{\delta \sigma\beta\mu}
~k_{1\alpha}k_1^{\delta}
k_2^{\sigma}~+~A_6~ {\hbox{\large $\epsilon$}}_{\delta \sigma\beta\mu} 
~ k_{2\alpha}k_1^{\delta}
k_2^{\sigma} }
\auto\label{inde}
$$
with
$$
\eqalign{A_1(k_1,k_2)~&=~-A_2(k_2,k_1) \cr 
A_3(k_1,k_2)~&=~-A_6(k_2,k_1) \cr 
A_4(k_1,k_2)~&=~-A_5(k_2,k_1) } 
\auto\label{bsym}
$$

The large momentum region ``anomaly'' contribution to $A_1$ and $A_2$ gives
$$
A_1~=~ \frac{1}{4\pi^2}~+~ \cdots~, ~~~~~~
A_2~=~ \frac{-1}{4\pi^2}~+~ \cdots~, ~~~~~~
\auto\label{uvco}
$$ 
leading to the ``anomalous'' divergence equation 
$$
(k_1 + k_2)^{\mu}~T_{\mu \alpha \beta}~=~
\frac{1}{ 2 {\pi}^2 }~{\hbox{\Large $\epsilon$}}_{\delta\sigma\alpha\beta} 
~k_1^{\delta} k_2^{\sigma}
\auto\label{awi}
$$
It is well-known\cite{gr1} that (\ref{uvco})
can be understood as the consequence of a large momentum shift of
the Dirac sea, during the interaction, that does not conserve axial charge.

As is also well-known, the numerical value of (\ref{uvco}) is fixed
by requiring that the vector Ward identities hold, i.e.
$$
k_1^{\alpha}~\Gamma_{\mu \alpha \beta}~=0 ~,~~~
k_2^{\beta}~\Gamma_{\mu \alpha \beta}~=0
\auto \label{vwi}
$$
and so vector charge is conserved.
For the invariant amplitudes $A_i$, the Ward identities require that
$$
A_2~=~k_1^2~A_5 ~+~k_1\cdot k_2 ~A_6
\auto\label{vwi1}
$$
and
$$
A_1~=~k_2^2~A_4 ~+~k_1\cdot k_2 ~A_3
\auto\label{vwi2}
$$
These identities imply, in turn, an inter-relation between the
ultra-violet anomaly contribution and the infra-red structure of the other $A_i$.
For example, when $k_1^2=0$,  (\ref{vwi1}) becomes
$$
A_2~=~k_1 \cdot k_2~A_6 ~=~\frac{q^2 - k_2^2}{2} ~A_6
\auto\label{vwi11}
$$
suggesting that there is a pole in $A_6$. 
In particular, if we insert the ultra-violet
anomaly term (\ref{uvco}) into (\ref{vwi11}), we obtain  
$$
A_6~\centerunder{$\sim$}{\raisebox{-5mm}{$k_1^2 \to 0$}}
~\frac{1}{2\pi^2~(k_2^2 - q^2)}~+~ \cdots
\auto\label{apol}
$$
which appears to determine that, when $k_1^2=0$, there is a pole in $A_6$ at 
$k_2^2=q^2$.

In fact, explicit expressions for the $A_i$ can be given 
when $k_1^2=0$ (references to the original calculations 
can be found in \cite{arw02}), i.e.
$$
\eqalign{~~~~A_1~&=~{1\over 4{\pi}^2} \biggl({k_2^2 \over k_2^2 -q^2 }~ln{k_2^2
\over q^2} ~+~1 \biggr)~~~~~~~~~~~~~~~~~~~~~~~~~~~~~~~~~~~~~~\cr 
A_2~&=~{1\over 4{\pi}^2} \biggl({k_2^2 \over k_2^2 -q^2 }~ln{k_2^2
\over q^2} ~-~1 \biggr) \cr
A_3~&=-A_6~=~{1 \over 2{\pi}^2 }{1 \over k_2^2 -q^2}
 \biggl({k_2^2 \over k_2^2 - q^2}~ln{k_2^2
\over q^2} ~-~1 \biggr) }
\auto\label{k1m0}
$$
(While $A_4$ can be obtained from (\ref{vwi2}), $A_5$ is undetermined
by (\ref{vwi1}) and is considerably more complicated.)
Both (\ref{uvco}) and (\ref{apol}) are clearly present in (\ref{k1m0}).  
However, it can easily be checked that 
there is no pole at $k_2^2=q^2$ in $A_6$. The logarithms of $k_2^2$
and $q^2$ are due to the ``normal thresholds'' in these channels, while the pole 
at $k_2^2=q^2$ is a (triangle diagram) ``anomalous threhold''. In general
anomalous thresholds are hidden by normal thresholds. Consistent with
this, the pole at $k_2^2=q^2$
is present only if the expressions in (\ref{k1m0}) are continued to 
unphysical sheets of the logarithms.

In special kinematic configurations, the
 ``anomaly pole'' does appear on the physical sheet. In particular, 
with $k_1^2$ already set to zero, 
$$
k_2^2~=0 ~~~\implies~~~ 
A_3~=-A_6~=~\frac{1}{ 2{\pi}^2~q^2 }
\auto\label{apol1}
$$
while
$$
k_1=0~\equiv~k_2^2=q^2 ~~~\implies ~~~
A_3~=-A_6~=~\frac{1}{ 4{\pi}^2~q^2 }
\auto\label{apol2}
$$
In both of these kinematic
configurations the invariant functions containing
the anomaly pole reduce to just the pole term with the residue determined
entirely by the anomaly. In (\ref{apol2}) the thresholds actually
produce a partial 
cancelation of the pole. This partial cancelation is related to the property
that, if $q^2$ is integrated over,  
the real part of the anomaly pole
cancels and only the imaginary part $\delta$-function remains.
As we discuss in Section 2,  
this is important for the contribution of the U(1) anomaly in pomeron vertices.

If the massless fermions are actually confined, the anomaly pole can be 
interpreted as a Goldstone boson pole signaling chiral symmetry breaking.
As we showed explicitly in \cite{arw02}, and briefly describe below,
the pole is generated in the infra-red internal momentum 
region. Consequently, the Ward identities (\ref{vwi1}) 
and (\ref{vwi2}) involve a direct cancelation between the large internal momentum
region generating 
anomaly contributions of the form (\ref{uvco}) and the small internal momentum
region giving the anomaly pole contribution.
In effect, there are two distinct
consequences of the presence of the ultra-violet anomaly (\ref{uvco}). 
The first is the
anomalous Ward identity (\ref{awi}). The second is that, for general momenta,
the vector
Ward identities require a cancelation between separate contributions
(with different kinematic structure) from large and 
small internal momentum regions. If an internal large transverse momentum  
cut-off is introduced, (\ref{uvco}) will be 
modified and the vector Ward identities will no longer hold. The contribution,
to the vector current divergences of the anomaly pole terms 
will survive, however, since they are generated in the infra-red transverse
momentum region\cite{arw02}. 

If we keep just the anomaly pole contributions 
of $A_3$ and $A_6$ to $T_{\mu \alpha \beta}$ we can write
$$
T_{\mu \alpha \beta}(k_1,k_2) ~=~-~{1 \over 2 {\pi}^2 } ~
{({\hbox{\Large $\epsilon$}}_{\delta \sigma\alpha\mu}k_{1\beta}
 ~-~{\hbox{\Large $\epsilon$}}_{\delta \sigma\beta\mu} 
~ k_{2\alpha})~ k_1^{\delta}
k_2^{\sigma} \over (k_1+k_2)^2} ~~~ +~\cdots 
\auto\label{pipo}
$$
This expression does not satisfy the vector Ward identities and does
not have the axial current anomaly.
When $k_1^2= k_2^2= 0$, and $q^2 \to 0$, we can rewrite (\ref{pipo}) as
$$  
T_{\mu \alpha \beta}(k_1,k_2) ~\sim ~{1 \over 2 {\pi}^2 }~{
- q_{\mu}~ [~{\epsilon}_{\delta\sigma\alpha\beta}
~ k_1^{\delta}
k_2^{\sigma}~]  \over q^2}
\auto\label{sifo}
$$
which now satisfies both vector Ward identities and also gives
the anomalous divergence (\ref{awi}). We conclude that, by itself, the anomaly
pole contribution violates the vector Ward identities, except at 
$k_1^2= k_2^2= 0$. 

The ultra-violet anomaly contribution (\ref{uvco}) is
absent in (\ref{sifo}) and yet the anomaly is present.
To understand how the anomalous
divergence can be due to the anomaly pole we must first 
discuss the intermal momentum configuration that generates the pole.
The analysis of \cite{arw02} shows that, if external light-like momenta
$k^+$ and $k^-$ are directed as shown in Fig.~B2, and $p$ is the internal 
loop momentum, the pole is generated at $p=0$. 
\begin{center}
\epsfxsize=2in
\epsffile{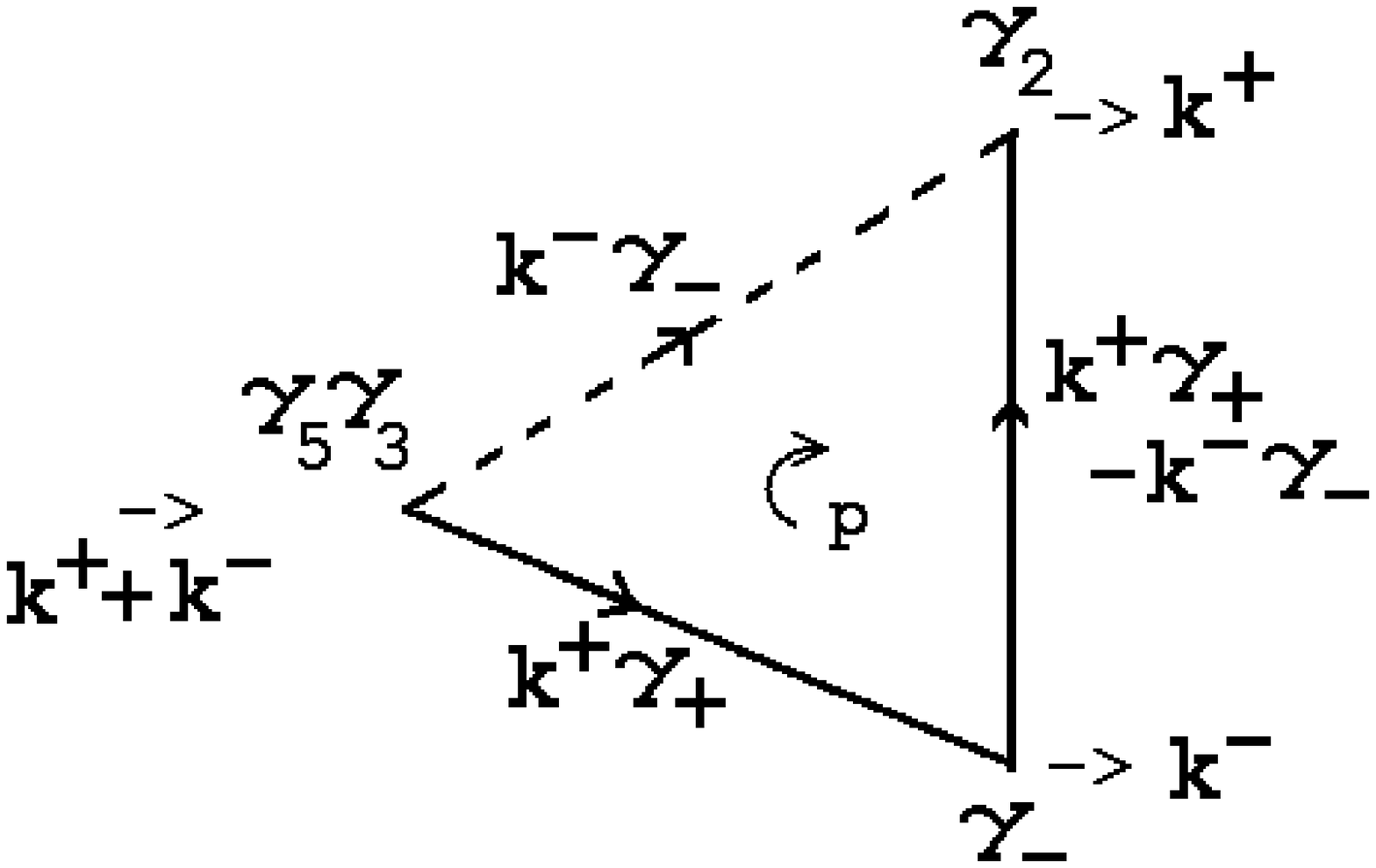}

Fig.~B2 Generation of the anomaly pole.
\end{center}
The $\gamma$ matrices shown give an anomaly pole term
$$
\Gamma_{32-}~=~-~  \frac{[{\epsilon}_{\sigma\delta 32}~k_1^{\sigma} 
k_2^{\delta}]~k_{1-} }{2 \pi^2 q^2} 
~~=~-~{k_+^2 k_- \over 2 \pi^2 q^2} 
\auto\label{llm}
$$
If $k^- \to 0$ then $q^2 = 2k^+k^- ~\to 0$. The finite light-like
momentum $k^+$ then flows along two 
of the internal lines while the third, the dashed line in Fig.~B2, carries
zero momentum. Because both poles of the zero momentum propagator
participate in generating the anomaly pole, there is effectively 
a particle/antiparticle ``chirality transition'', that is equivalent
to an infra-red shift of the Dirac sea, during the interaction. This
is how the anomaly pole produces an infra-red non-conservation of axial charge
that parallels that produced by the ultra-violet Dirac sea shift.
The infra-red effect is present only when the fermions involved are massless
and then only in the infra-red region where (\ref{sifo}) is valid.

If the $1/q^2$ factor in (\ref{sifo}) is to be 
interpreted as a Goldstone boson particle pole then
$q_{\mu}$ must provide the coupling to the axial current
while $[~{\epsilon}_{\delta\sigma\alpha\beta} k_1^{\delta} k_2^{\sigma}~]$ 
provides, potentially, a coupling to physical currents.
There is a problem, however, in that $k_1$ and $k_2$ are both light-like  
and so $q^2=0$ implies that they are also parallel. Therefore, because of the  
$\epsilon$-tensor, the pole residue vanishes, as is seen explicitly in
(\ref{llm}). This should be expected, of course.
Otherwise we would be able to obtain a coupling of a pion to finite momentum
gluons. Nevertheless, we can avoid the vanishing of the pole residue if,
as we go to the pole, we simultaneously go to an infinite-momentum frame. 
If we make a boost $a_3(\zeta)$ along the $3$-axis and consider
$\Gamma_{32-}$ defined in the new frame, we can use either (\ref{sifo})
or (\ref{llm}) to obtain 
$$
\Gamma_{32- } ~\sim ~ \frac{k_+\cosh{\zeta}~\bigl[-~k_+ k_- \sinh{\zeta} ~\bigr]}
{ \sqrt{2} ~q^2} 
\auto\label{imf}
$$
Since we still have $q^2=2k_+k_-$, the limit $k_- \to 0$ again gives $q^2 \to 0$.
However, the coupling 
$[~k_+ k_- \sinh{\zeta}~] $ is finite if 
$ k_- \cosh{\zeta}~$ is kept finite, i.e. if the mass-shell and
infinite momentum limits are combined. As discussed in Section 2, the anomaly
then provides a coupling to infinite momentum wee gluons. This is very important
because, on general grounds, we expect to see 
wee-partons carry vacuum properties in the infinite momentum frame !! 

\newpage

\renewcommand{\theequation}{C.\arabic{equation}}
\setcounter{equation}{0}
\vskip 1cm \noindent 
\noindent {\large\bf Appendix C. ~ The Multi-Regge Program}
\vskip 3mm \noindent

In this Appendix we provide a general description of the multi-regge program
that we have formulated over the years which, as we note often in the main
body of the paper, should ultimately provide the best framework for a complete
derivation of the high-energy solution of $QCD_S$.
We include some minimal historical background in order
to explain the motivation
for the program and to show why we have
been led to connect $QCD_S$ to the Critical Pomeron. 
More technical descriptions of the arguments we give can, for the most part,
be found in our recent papers. 

We will assume that the reader has a basic knowledge of reggeon
diagrams.
A review of elastic scattering diagrams and the transverse
momentum kernels that appear in Section 2 can be found in 
Section IIIB of \cite{arw02}. We will also make considerable use of multi-reggeon 
diagrams\cite{arw98} 
that are the extension to multiparticle amplitudes of the elastic diagrams
described in \cite{arw02}. For our present purposes,
it will be sufficient to understand firstly that, in the multi-reggeon diagrams,
there are several distinct reggeon channels 
in which the elastic scattering kernels again appear - with
all the same properties. Secondly, and very importantly, 
the vertices which couple the distinct
reggeon channels contain anomalies that are not present in
the (vector) gluon reggeon interactions appearing in elastic 
diagrams. We give details of these vertices, and the anomalies that occur, 
in the context of the discussion.

To begin with, we note that the asymptotic freedom of QCD almost certainly implies 
that total cross-sections
must rise asymptotically if perturbation theory is to have any validity.  
The Critical Pomeron description of rising cross-sections  
was discovered\cite{cri} thirty years ago. 
While it's derivation as a renormalization group solution of Reggeon Field 
Theory (RFT) implied that it satisfied full multiparticle
$t$-channel unitarity\cite{gpt,arw00}, it was soon established
that it also satisfies all known $s$-channel unitarity constraints\cite{mm}. 
It remains today the only known rising cross-section solution 
of unitarity in the regge limit. In a sense, it is a fixed point solution
of the regge limit (expressed in terms of physical degrees of
freedom) analagous to the asymptotically-free fixed-point
solution of the short-distance limit. The Critical Pomeron is, however, much 
more difficult to realize in a physical theory.

The critical solution of RFT was found via the well understood 
sub-critical theory (essentially the multiperipheral
model plus unitarity corrections). The physical significance of the
supercritical theory was the subject of much dispute
and conflicting proposals were put forward. The
solution we proposed\cite{arw77,arw78,arw91} 
has the advantage that it is described by an explicit
diagrammatic expansion that clearly satisfies reggeon unitarity. 
The supercritical diagrams are generated (as in a normal
supercritical phase) by introducing a pomeron condensate in the 
critical RFT lagrangian.
The condensate generates new classes of RFT diagrams, a simple example of 
which is shown in Fig.~C1. 
The two pomeron propagators produced by the condensate  
give $k_{\perp}$ poles that have to be interpreted as
particle poles, implying that there is a pomeron
transition to a two vector reggeon state as shown.
\begin{center}
\leavevmode
\epsfxsize=3.8in
\epsffile{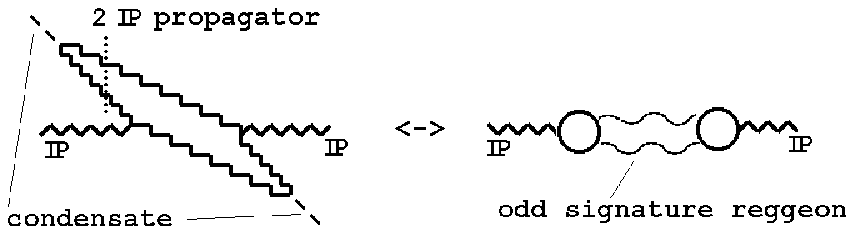}

Fig.~C1 A new RFT diagram generated by the pomeron condensate
\end{center}

Reggeon states involving many vector particle poles  
similarly appear in higher-order diagrams. Consequently, the RFT phase transition
can be described by saying that 
divergences in rapidity produced in the subcritical 
graphical expansion (because the bare pomeron intercept is above one) are 
converted to vector particle divergences in $k_{\perp}$ in the supercritical
expansion. That is, the supercritical phase is characterized by the 
``deconfinement of a vector particle on the pomeron trajectory''.
 
We soon realized\cite{arw78} that the deconfinement of a vector
particle is exactly what should happen if it is possible 
to make a smooth transition from QCD to CSQCD, suggesting 
that the Critical Pomeron occurs, in particular circumstances, 
in QCD. It also suggested that a transition from perturbative reggeized gluon 
diagrams, describing a spontaneously-broken gauge theory, to
non-perturbative pomeron diagrams describing a confining theory,
could be followed within RFT. Confinement
would have to be produced by the infra-red divergences of reggeized gluon diagrams.

We proposed\cite{arw81}, therefore, starting with the gluon and quark
reggeon diagrams of QCD, but with the gauge symmetry 
competely broken so that all gluons are massive. 
The aim was to first restore the gauge symmetry
to SU(2), to obtain CSQCD, and to show that the diagrams obtained could 
be identified with those of
supercritical pomeron RFT. We anticipated that infra-red divergences
would produce confinement of SU(2)
color and a pomeron, while the broken part of the gauge group 
would provide the accompanying massive vector meson. 
Restoring the symmetry to SU(3) would then be done within
RFT and the result would be the Critical Pomeron. We gave arguments, based on
complimentarity, that with a transverse momentum cut-off imposed the symmetry 
restorations should take place smoothly and,
over the years\cite{arw98,arw01}, made a number of attempts to 
implement our proposal. 

The derivation of our supercritical RFT solution involved many subtleties\cite{arw91}
that we eventually realized implied that 
the nature of the scattering hadron states has to be 
closely related to that of the pomeron. In particular, the 
``pomeron condensate'' that defines the supercritical phase has to be associated
with a (``wee parton'') component of the scattering hadrons. To derive the 
solution in a well-defined way, 
it had proved necessary\cite{arw91} to consider a multi-regge 
amplitude in which regge pole hadrons scatter via pomeron exchange.
As a result we believed we should consider an analagous amplitude in QCD.
That is, we should consider a ``many-body'' scattering amplitude, in an 
appropriate multi-regge region\cite{arw98} of phase-space, 
in which regge pole pions and the pomeron could emerge together,
as illustrated schematically in Fig.~C2.
Starting with the appropriate multi-regge 
perturbative reggeon diagrams, we would look for  
infra-red divergences that could produce pions and the pomeron
as the gluons become massless.
That, a priori, very complicated diagrams were to be considered
is not as bad as it seems because the 
general structure of high-order diagrams is determined\cite{arw98} 
from that of lower-order diagrams by reggeon unitarity.
\begin{center}
\epsfxsize0.4in
\epsffile{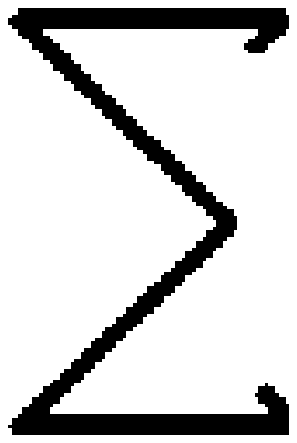}
\epsfxsize=2.8in
\epsffile{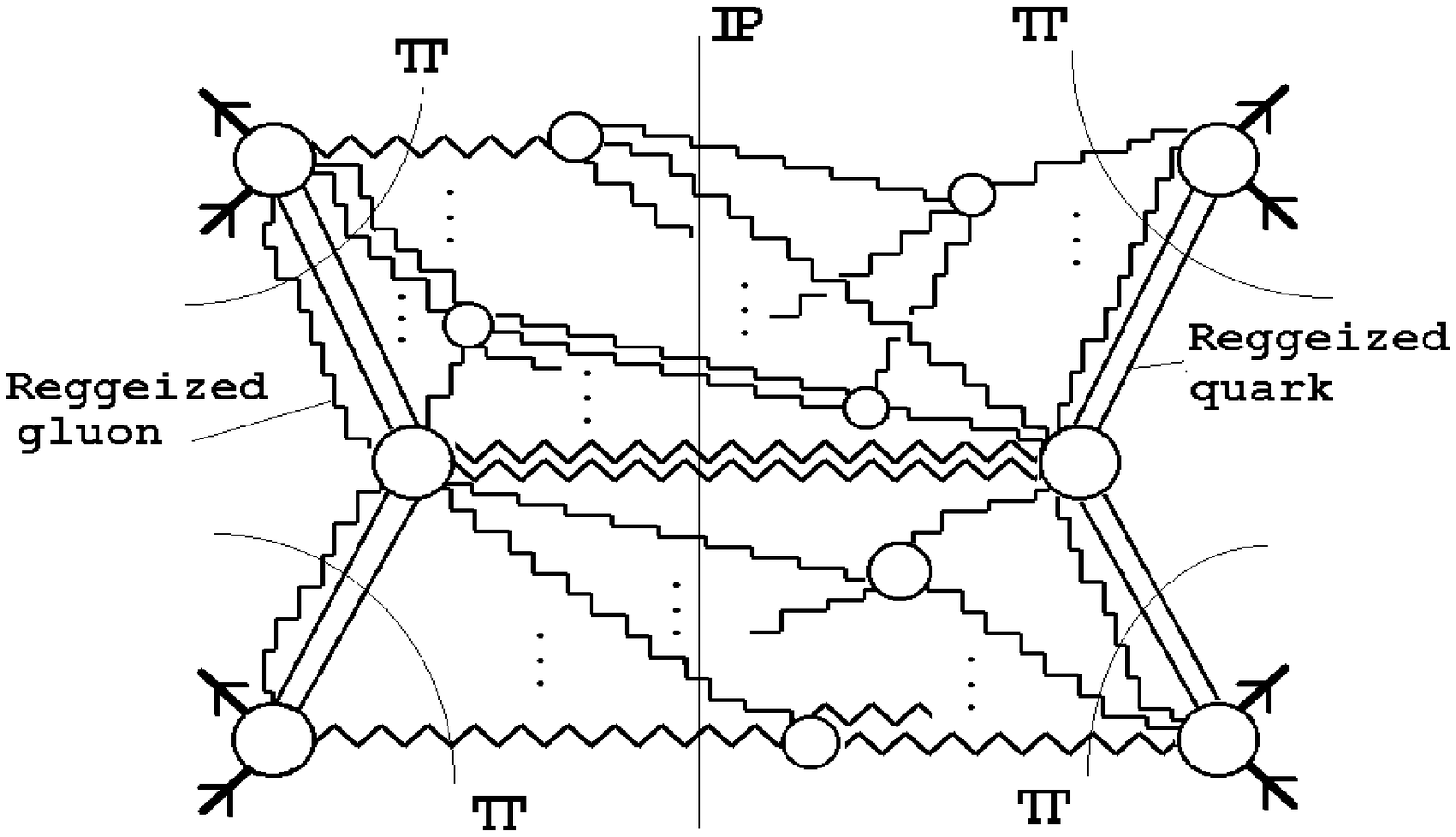}
\epsfxsize=2.3in
\epsffile{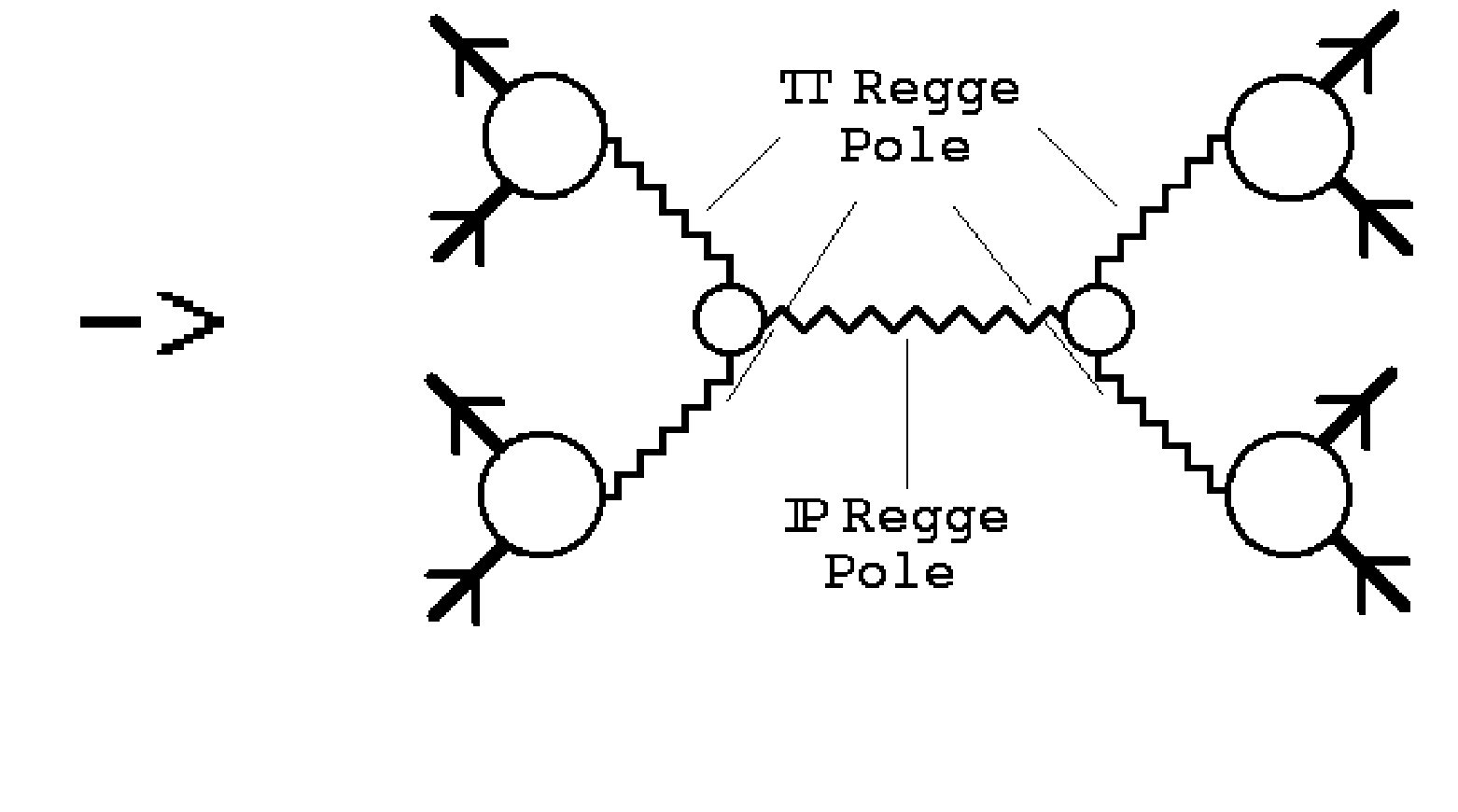}

Fig.~C2 The transition from perturbative reggeon diagrams to reggeized 
pions scattering via pomeron exchange. 
\end{center}

In the sub-critical theory it was apparent that
the criticality of the pomeron should depend on the number of 
hadron states and, therefore, on the number of quark flavors. Consequently,
``saturating'' the asymptotic freedom constraint, as in $QCD_S$,
would be most likely to produce the critical behavior. The only physically
realistic way to do this was\cite{arw81} to add two sextet flavors and have the
sextet higgs mechanism operative. 
It also became clear, rather quickly, that 
the special infra-red scaling properties of
reggeon interaction kernels in $QCD_S$, which follow from the presence 
of an infra-red fixed-point, would have to be an essential ingredient
of the infra-red divergence structure\cite{arw81}. In addition there would 
have to be interactions (anomaly related?) 
to which divergences produced by the scaling properties
would couple. Finally we realized that, because the Higgs mechanism scalar 
field is asymptotically free in $CSQCD_S$, restoration of SU(3) color 
(which is to give the Critical Pomeron) 
can be carried out without a transverse momentum cut-off. 

It soon seemed, therefore, that if Fig.~C2 was to be 
implemented fully then, most likely, 
we would have to specifically consider $QCD_S$. There was, however, 
a major problem that, for a long time, 
prevented us from systematically developing a program
to implement Fig.~C2. If we consider $QCD_S$ in 
isolation, then we can not find suitable external scattering 
states to provide a perturbatively well-defined starting amplitude 
within which pions and the pomeron could emerge as in Fig.~C1. Without this 
we can not determine, for sure, whether anomaly related divergences 
occur. Consequently, the anticipated 
mapping onto supercritical RFT can not be carried out.
We initially supposed\cite{arw81} that the external states
could be multi-quark states. 
However, as will soon become clear, the pions and pomeron, that we are led to,
do not couple to such states. In \cite{arw98} we assumed the existence
of external couplings with particular properties but, in this case, it was clear 
that the nature of the infra-red divergences that occured depended on these 
assumptions.
 
Only recently\cite{arw03,arw02}, have we understood that adding 
the electroweak vector boson sector of the Standard Model to
$QCD_S$ solves the problem of the external states for Fig.~C2.
As illustrated in Fig.~C3, the desired pion amplitude should 
appear in a multi-regge
limit\cite{arw98} of an amplitude for multiple vector boson scattering.
\begin{center}
\epsfxsize=1.1in
\epsffile{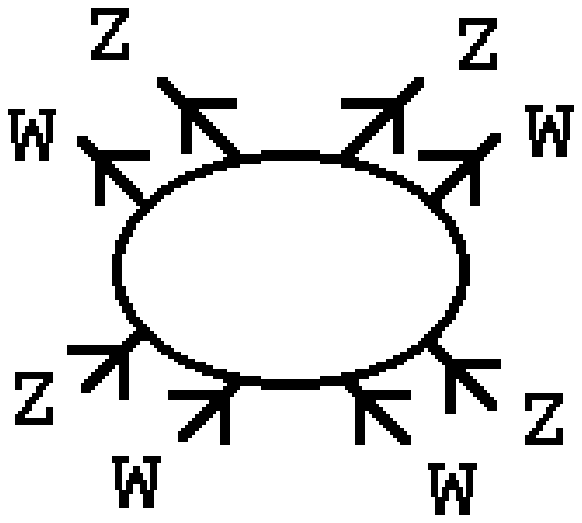}
\epsfxsize=3in
\epsffile{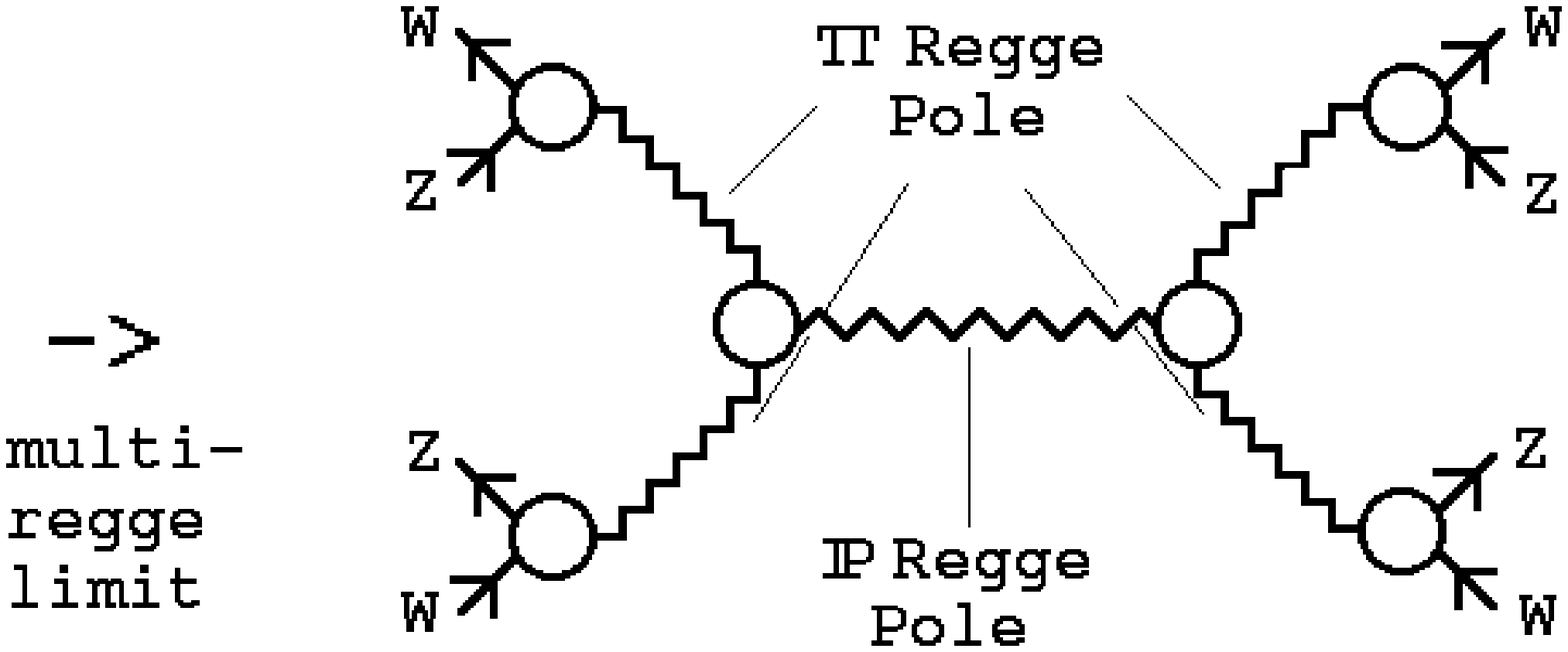}

Fig.~C3 The pion amplitude within a multiparticle vector boson amplitude.
\end{center}  
Conversely, since W's and Z's have explicit perturbative couplings,  
this amplitude also has a perturbative reggeon diagram description. 
Most importantly for our program,
because of the presence of elementary left-handed couplings, the perturbative 
reggeon vertices of the external $W$ and $Z$ states contain\cite{arw03} 
anomalies that can give (with the cut-off manipulation that we discuss below)
the infra-red divergences that we are looking for.
We want, of course, 
to add the vector boson sector of the Standard Model to $QCD_S$
in order to study the sextet higgs mechanism. 
Clearly, the fact that
electroweak vector bosons provide the perfect external states for our 
multi-regge program could be related to the actual
validity of the solution of $QCD_S$ that we find. That is to say, our solution
of $QCD_S$ is effectively induced by the presence of the electroweak vector
bosons and may, perhaps, only be valid in their presence\cite{arw05}.  

For massless gluons, the individual reggeon diagrams in Fig.~C1 have well-known
infra-red divergences (that we will return to later) but, 
if the scattering states are color zero vector
bosons, we expect that these divergences will
cancel in the sum over all diagrams of a given order. 
Therefore, there must be an 
additional divergence phenomenon, if $CSQCD_S$ reggeon diagrams are to be mapped
on to the pomeron diagrams of supercritical RFT via divergences.
In fact, we now understand
well that it is the appearance\cite{arw02,arw021} of chiral anomalies 
in, high order, multi-regge vertices that produces the divergences that 
we are looking for. The anomalies occur because these vertices
contain triangle diagrams that result from the contraction (in the
regge limit) of larger loop feynman diagrams. Even though there are 
no axial vector currents in the elementary QCD interaction, $\gamma_5$
couplings are generated within these vertices 
by products of orthogonal $\gamma$-matrices.
  
We anticipate that, without a transverse momentum
cut-off, the anomalies appear as a large transverse momentum phenomenon that 
produces (non-unitary) power
enhancement of the high energy behavior. We have shown this
explicitly in our analysis\cite{arw03} of elastic vector boson scattering.
The enhancement 
is avoided by the introduction of a cut-off but there is then 
a violation of gauge invariance Ward identities for 
the anomaly generating vertices
(in analogy with our discussion of the elementary triangle diagram in 
Appendix B). As a result, as we discuss more explicitly
below, infra-red transverse momentum
divergences appear which couple directly
to the part of the anomaly diagrams which, because 
the quarks involved are massless,
contain the infra-red ``anomaly pole''. (The anomaly pole
contribution to a triangle diagram is discussed briefly in Appendix B and,
at much greater length, in \cite{arw02}.) In effect, introducing a transverse 
momentum cut-off removes ultra-violet 
chirality violation produced by the anomaly
and replaces it with infra-red chirality violation that produces 
anomaly poles. Since an anomaly pole can be interpreted  
as a Goldstone boson particle pole, this provides a crucial
mechanism for a bound-state, Goldstone boson, spectrum to appear out of
reggeon diagrams via infra-red divergences. Indeed, we will assume that 
anomaly poles survive higher-order corrections only when they are associated
with a chiral symmetry. 

Understanding that the infra-red 
divergence phenomenon that we are looking for should appear
as a consequence of anomalies if a transverse momentum cut-off is imposed,
the first step of our program is to look for this phenomenon within 
the multi-regge diagrams of $CSQCD_S$ obtained 
by setting the mass of an SU(2) subgroup of gluons to zero. As described in
Section 2, $CSQCD_S$ contains an SU(2) triplet of massless (reggeized) gluons,
plus two SU(2) doublets and one singlet of massive (reggeized) gluons.  
The color symmetry breaking can be done, as we have already discussed, 
by adding a scalar field
and using the usual Higgs mechanism (this is a technical manipulation
that has nothing to do with electroweak symmetry breaking).

The main infra-red divergence of massless gluon
reggeon diagrams is that associated with reggeization. Independently 
of the transverse momentum cut-off, this divergence 
exponentiates to zero all amplitudes with non-zero SU(2)
color (in the reggeon channel), while leaving finite color zero amplitudes. 
As described in more detail in \cite{arw02}, an infra-red fixed point implies 
that the interaction kernels of color zero massless
gluons have a crucial infra-red scaling property (the ultra-violet version
of which produces the leading-order BFKL pomeron). 
This scaling is an essential 
component of the anomaly related infra-red divergences that 
we are looking for, as we now discuss. 

We consider reggeon states which contain both an SU(2)
color zero massless gluon component
and an additional SU(2) color zero component - either a massive gluon or a 
quark-antiquark pair. We consider the possibility
of an infra-red divergence from the infra-red region where all
the transverse momenta of the massless gluons scale uniformly to zero. 
If the massless gluons carry, overall, normal color 
parity ($= $ the signature) they will
interact with the additional color zero component and, as a result,
any divergence that occurs  
will be exponentiated via reggeization effects, giving a zero amplitude. 
If, however, the massless gluon component 
carries anomalous ($\neq $ the signature) color parity the divergence 
will not exponentiate. This is because, as explained in \cite{arw02},
a gluon component of this kind can only couple to an anomaly vertex and 
anomalies can not occur in vector reggeon interactions that take place
within a reggeon state. Consequently, 
the massless gluon component will have only self interactions, 
as illustrated in Fig.~C4.
\begin{center}
\epsfxsize=5in
\epsffile{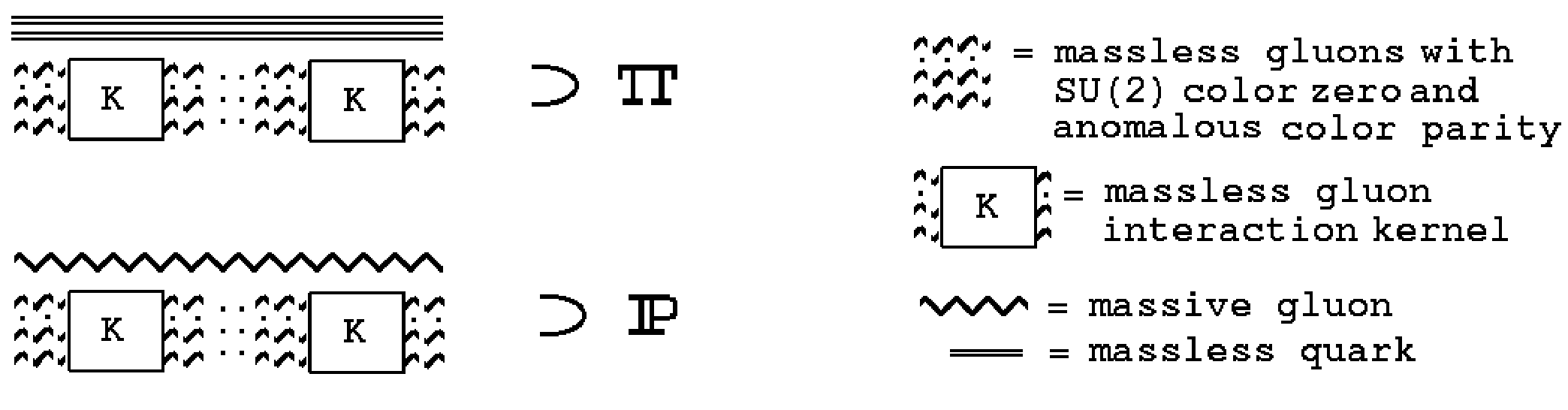}

Fig.~C4 Pomeron and pion reggeon states in $CSQCD_S$.
\end{center}

Provided there are external (to the reggeon state) anomaly vertices to which the
complete reggeon states shown in Fig.~C4 can couple 
the scaling property of the gluon self-interactions will produce a 
divergence (at zero transverse momentum for the gluons). The residue 
of this divergence contains a reggeon state that we can potentially 
identify as either a ``pion'' or a ``pomeron'', as shown.
If we can absorb this divergence into a ``reggeon condensate'', 
this condensate will be an essential, zero transverse momentum, part
of both the pion and the pomeron in $CSQCD_S$. Since 
``anomalous gluons'' with SU(2) color zero necessarily have odd signature
(three is the minimal number),
the pomeron given by Fig.~C4 will be an even signature regge pole that is
exchange degenerate with an odd signature, massive, gluon reggeon.
This, together with the existence of a ``pomeron condensate'', are 
crucial features of supercritical RFT. Also,
since all amplitudes have SU(2) color zero, if pion anomaly poles appear
as we anticipate, we will
have a spectrum with confinement and chiral symmetry breaking.

For the pion and pomeron to appear as in Fig.~C4,
via infra-red divergences, Fig.~C2 has to be realized by
the appearance of a ``lowest-order'' amplitude, of the form shown in Fig.~C5, 
in which an anomaly that can couple the reggeon
states appears in each vertex (as indicated by the $A$).
The notation for Fig.~C5 is the same as that for Fig.~C4 except that a new notation
is introduced to indicate that each of the massless gluons now
carries zero transverse momentum.
\begin{center}
\epsfxsize=3.2in
\epsffile{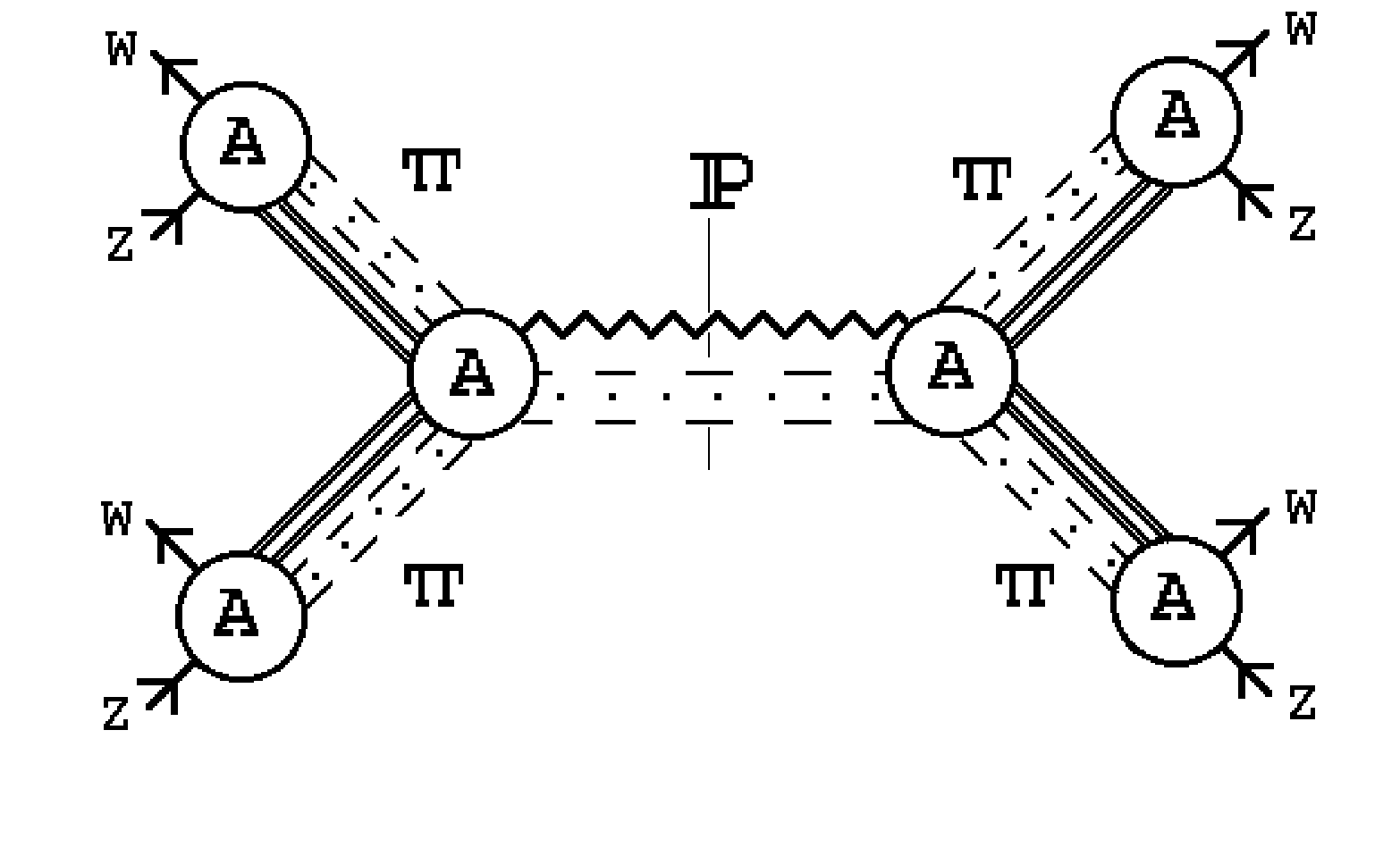}

\epsfxsize=4.3in
\epsffile{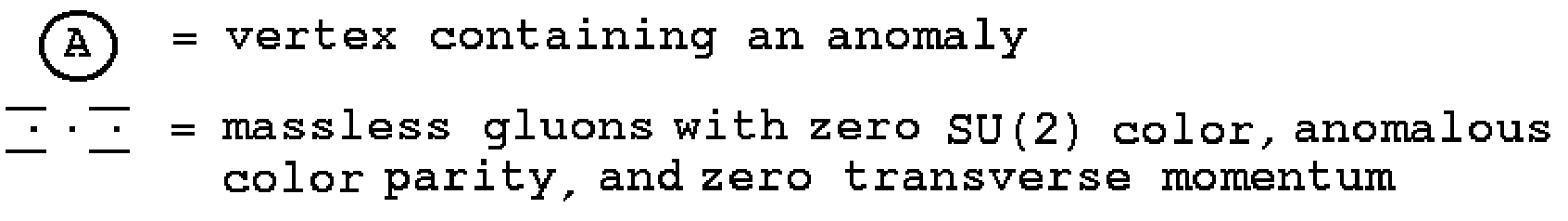}

Fig.~C5 Anomaly vertices that must appear in the pion amplitude.
\end{center}
 
The external anomaly vertices needed for Fig.~C5 are identical to those that
appear in our recent paper\cite{arw03} deriving elastic 
scattering amplitudes of electroweak vector bosons. 
In that paper we demonstrated 
that anomalous color parity gluons have the needed coupling.
In the elastic scattering context, it is very 
clear how the use of a cut-off removes
bad, large transverse momentum based, high-energy behavior produced 
by the anomalies and, instead, introduces anomaly dominated infra-red divergences 
that potentially produce ``non-perturbative'' anomaly pole Goldstone bosons. 
Although we did not discuss the generation of the anomaly pole explicitly
in \cite{arw03}, we did give a brief summary of how
the anticipated infra-red divergences should be mapped onto RFT and the amplitude
for pion exchange obtained.

The anomaly vertices obtained in \cite{arw03}
contain triangle diagrams resulting from the contraction of larger loop
feynman diagrams just as illustrated in Fig.~2. 
As described in Appendix B, an anomaly pole 
is generated in the triangle diagram by a zero momentum quark line 
(the partially broken line in Fig.~2).
It is important that, as illustrtated in Fig.~2, 
it is the longitudinal polarization of the on-shell
massive vector boson that produces the quark/antiquark coupling in the
anomaly triangle diagram. (Note that, 
in the calculation of \cite{arw03}, the on-shell massive gluon in Fig.~2
was replaced by a massive electroweak vector boson.)

The anomaly that occurs in diagrams that contribute to the
pion/pion/pomeron vertex in Fig.~C5 is discussed 
in \cite{arw021} and \cite{arw011}. 
The reduction to a triangle
diagram is as illustrated in Fig.~C6 and it is the U(1) anomaly that is involved.
\begin{center}
\epsfxsize=5in
\epsffile{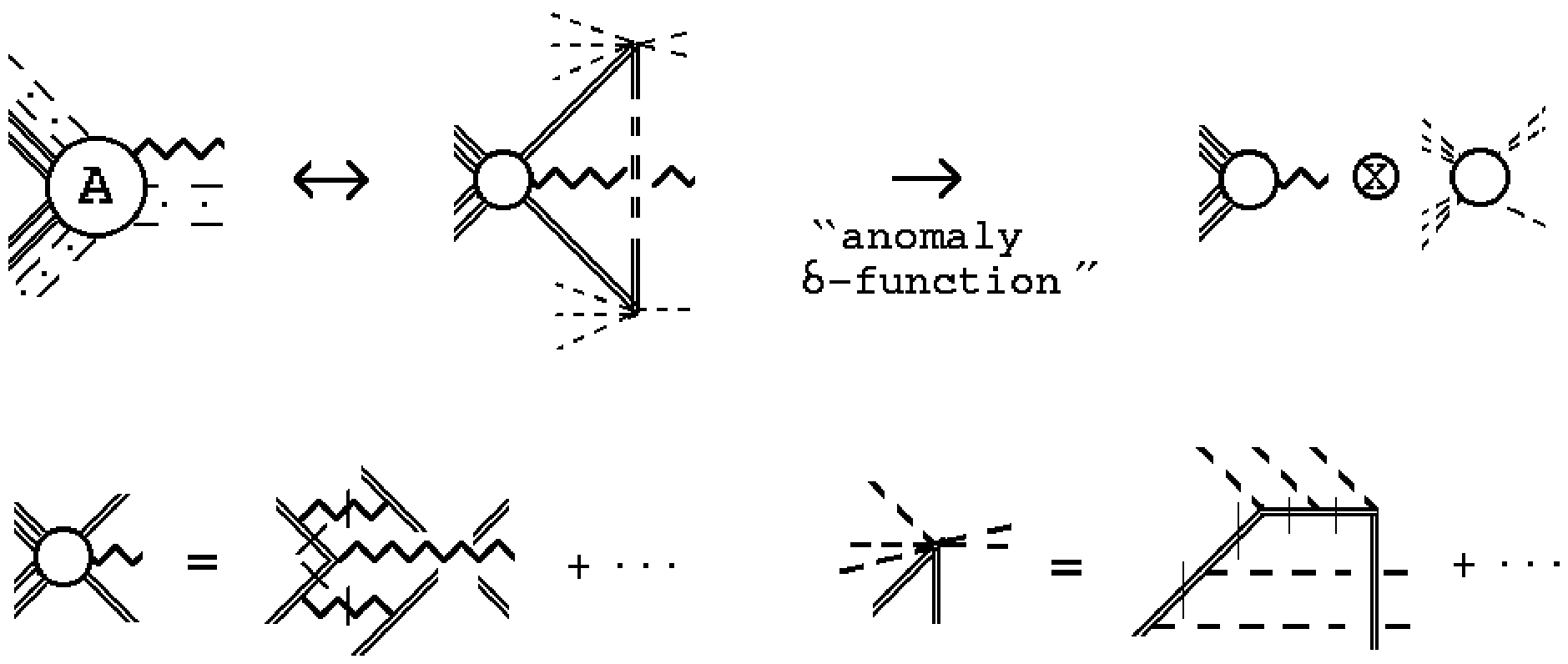}

Fig.~C6 The reduction to a triangle diagram that involves the U(1) anomaly.
\end{center}
(The notation is the same as in previous diagrams.) The anomaly pole is
present but, because it is integrated over,
it contributes as an ``anomaly $\delta$-function'' and plays the remarkable 
role, also illustrated in Fig.~C6,
that it factorizes off the (zero transverse momentum)  
anomalous gluon interaction from the remaining ``hard interaction''. 
The anomaly $\delta$-function is again 
generated by a zero momentum quark line (illustrated by a broken
line in Fig.~C6) which 
undergoes a chirality transition. The hard interaction
production of a massive gluon has an overall axial vector
nature that compensates for this transition.
To produce the axial coupling it is essential that the exchanged
on-shell massive gluons within the hard interaction are longitudinal.
(It is also important that these gluons
carry zero light-cone momentum in a frame in which the pion carries
finite light-cone momentum.)

Because each of the external reduced triangle diagrams that we have discussed
contain anomaly contributions,
imposing a transverse momentum cut-off will lead to a violation of Ward 
identities for the gluons coupling to such vertices. A scaling 
transverse momentum divergence should then appear in each pion channel
and generate Fig.~C5, as we have discussed above. We anticipate that there will
be an overall logarithmic divergence as the transverse momenta in all channels
are scaled uniformly to zero and that this is what we will have to factorize off
to obtain the physical amplitude. However, to be sure of this and
to elaborate the divergence phenomenon in full
requires more details of the calculation than we presently have. We have, so far,
carried out the full analysis only in the situation described in \cite{arw02}
in which we used the pion anomaly pole approximation that 
we describe in Section 2.

For the more general case of the reggeized pion amplitude
appearing in Fig.~C5, we can say the following. 
The divergence is at zero transverse momentum, which (in an appropriate frame)
should be equivalent to zero four momentum for the gluon 
vertex of the effective triangle diagram in Fig.~2. From (\ref{apol2}), 
we see that in this kinematic configuration 
the triangle diagram amplitude reduces to a ``pion'' 
anomaly pole, as illustrated in Fig.~2. Thus, as we have anticipated, 
the anomaly pole should 
provide the mechanism whereby a pion particle pole appears (in 
the residue of the infra-red divergence) as part of the pion reggeon state.

To obtain the multi-regge amplitudes of $QCD_S$ from those of
$CSQCD_S$, via RFT, it is clearly necessary to understand in complete detail
how the full set of divergent $CSQCD_S$ diagrams maps onto super-critical RFT.
In higher-orders we expect to find vertices, of the form shown in Fig.~10, in
which massive gluons are produced by a wee gluon interaction only. 
Interactions of this kind should lead to particle pole interactions within pomeron 
vertices, just as is produced by the pomeron condensate
in the supercritical pomeron phase\cite{arw91}.

Although there is every indication that the reggeon diagrams of $CSQCD_S$ 
can be mapped onto supercritical RFT, it remains 
a major challenge to carry out the mapping in full.
Our hope is that the (relative) simplicity of the external vector 
boson couplings, appearing in Fig.~2, will finally make it feasable. 
It also remains to be 
determined how a pion anomaly pole, which occurs 
at zero transverse momentum, combines with reggeization 
at spacelike momentum transfer. In the anomaly pole
vertex method\cite{arw02}, that we use in Section 2,
we effectively assume that the
on-shell pion couplings can be obtained by an anomaly pole coupling of the 
form shown in Fig.~2. While the above discussion suggests that this should be 
a straightforward outcome of the full multi-regge calculation, 
it remains to be shown.

With the mapping of $CSQCD_S$ onto supercritical RFT established, it should
be straightforward to show that the high-energy 
behavior of $QCD_S$ is that of the Critical Pomeron. 
Critical Pomeron amplitudes
can be, and have been, calculated\cite{mm} 
without reference to QCD. There will, however,
be much to understand about the limiting process involved, particularly
with respect to the formation of baryons. For our present purposes 
we have, in Sections 2 and 3,
concentrated on the underlying physical phenomenon which
describes the pomeron in $QCD_S$.

\end{document}